\documentclass[aps,rmp,preprint,amssymb,graphicx,longbibliography]{revtex4-1}

\usepackage{graphicx}

\usepackage{hyperref}

\def\twiddle{\lower.9ex\rlap{$\kern-.1em\scriptstyle\sim$}}
\def\bigtwiddle{\lower1.ex\rlap{$\sim$}}
\def\gtrsim{\mathrel{\raise.3ex\hbox{$>$\kern-1.05em\lower1ex\hbox{
$\sim$}}}}
\def\lesssim{\mathrel{\raise.3ex\hbox{$<$\kern-1.05em\lower1ex\hbox{
$\sim$}}}}

\begin{document}

\title{Invisible Axion Search Methods}

\author{Pierre Sikivie}

\affiliation{Department of Physics and
Institute for Fundamental Theory,\\ 
University of Florida, Gainesville, FL 32611, USA}

\date{September 29, 2020}

\begin{abstract}

In the late 1970's, the axion was proposed as a solution to the 
Strong CP Problem, {\it i.e.} the puzzle why the strong interactions 
conserve parity P and the product CP of charge conjugation and
parity in spite of the fact that the Standard Model of elementary 
particles as a whole violates those symmetries.  The original axion 
was soon ruled out by laboratory experiments and astrophysical 
considerations, but a new version was invented which is much more 
weakly coupled and which evades the laboratory and astrophysical 
constraints.  It was dubbed the ``invisible" axion. However, the  
axion cannot be arbitrarily weakly coupled because it is overproduced
in the early universe by vacuum realignment in the limit of vanishing 
coupling.  The axions produced by vacuum realignment are a form of 
cold dark matter today.  The axion provides a solution then not 
only to the Strong CP Problem but also to the dark matter problem.  
Various methods have been proposed to search for dark matter axions 
and for axions emitted by the Sun. Their implementation and improvement 
has led to significant constraints on the notion of an invisible axion.  
Even purely laboratory methods may place significant constraints on 
invisible axions or axion-like particles.  This review discusses the 
various methods that have been proposed and provides theoretical 
derivations of their signals.

\end{abstract}

\maketitle



\tableofcontents

\vspace{0.5cm}

\section{Introduction}

During the 1970's, the Standard Model (SM) of elementary particles
\cite{Chen84,Dono14} came to the fore as a correct description of
all fundamental interactions other than gravity.  It has proved
since to be tremendously successful, explaining practically all
relevant data in terms of a small number of parameters. Already
in its early days, however, it was seen to present a puzzle: one
would not expect within the SM that the strong interactions conserve
parity P nor the product CP of charge conjugation C with parity.
The strong interactions and the electromagnetic interactions
are observed to conserve P and CP.  The weak interactions on
the other hand violate P, C and CP.  The trouble with the
SM is that the P and CP violation of the weak interactions
produces P and CP violation in the strong interactions unless
an unexpected cancellation occurs.  This is commonly referred
to as the Strong CP Problem.

The amount of P and CP violation in the strong interactions is 
controlled by a parameter, $\theta_{\rm QCD}$, which appears as 
the coefficient of a P and CP odd term in the action density 
\begin{equation}
{\cal L}_{\rm SM} = ... + \theta_{\rm QCD} {g_s^2 \over 32 \pi ^2}
G^a_{\mu\nu}~\tilde{G}^{a \mu \nu}
\label{Lt}
\end{equation}
where the $G^a_{\mu\nu}$, $a = 1, 2, ... 8$, are the field strengths 
of Quantum Chromodynamics (QCD), $\tilde{G}^{a\mu\nu} \equiv 
{1 \over 2} \epsilon^{\mu\nu\alpha\beta} G^a_{\alpha\beta}$, and 
$g_s$ is the QCD coupling constant.  Unless stated otherwise, we
use units in which $\hbar = c = 1$ \footnote{\label{units} A 
short appendix on units and conventions is included.} and 
conventions in which the Minkowski metric $(\eta_{\mu\nu})$ = 
diag(+1, -1, -1, -1) and $\epsilon^{0123} = +1$. The dots represent 
all the other terms in the SM action density, i.e. the terms that 
lead to its numerous successes.  Eq.~({\ref{Lt}) shows the one term 
that is not a success.  $\theta_{\rm QCD}$ is an angle, i.e. it 
is cyclic with period $2 \pi$.  QCD depends on $\theta_{\rm QCD}$ 
because of the existence in that theory of quantum tunneling events
\cite{'tHo76a,'tHo76b}, called ``instantons", which violate P and CP 
if $\theta_{\rm QCD}$ differs from zero or $\pi$.  Since in actuality 
the strong interactions obey P and CP, as well as can be observed, 
$\theta_{\rm QCD}$ must be close to one of its CP conserving values.  
The best constraint derives from the experimental upper limit on the 
neutron electric dipole moment: $|d_n| < 3 \cdot 10^{-26} e$ cm 
(90\% CL) \cite{Pend15}.  For small $\theta_{\rm QCD}$ the 
contribution of the term shown in Eq.~(\ref{Lt}) to the neutron 
electric dipole moment is of order \cite{Balu79,Crew80}
\begin{equation}
d_n \sim \theta_{\rm QCD} {m_u m_d\over m_u + m_d} {1\over
\Lambda_{\rm QCD}} {e\over m_n}\sim 3\cdot 10^{-16} 
~\theta_{\rm QCD}~e~{\rm cm}~~\ ,
\label{dnest} 
\end{equation}
where $m_u$ and $m_d$ are the up and down quark masses, 
$m_n$ is the neutron mass, and $\Lambda_{\rm QCD}$ the 
QCD scale. $\theta_{\rm QCD}$ should therefore be less than 
of order $10^{-10}$ (mod $\pi)$.  $\theta_{\rm QCD} = 0$ or 
$\pi$ is unexpected in the SM because P and CP are violated 
by the weak interactions.  CP violation is introduced by 
giving apparently random phases to the Yukawa couplings 
that give rise to the quark masses.  The overall phase 
of the quark mass matrix feeds into $\theta_{\rm QCD}$ 
which is therefore generically of order one.  The puzzle 
why $\theta_{\rm QCD}$, expected to be of order one, is in 
fact less than $10^{-10}$ is the Strong CP Problem.

Soon after the Strong CP Problem was recognized, Peccei and
Quinn (PQ) proposed a modification of the SM that offers a
solution \cite{Pecc77a,Pecc77b}.  They postulated a $U_{\rm PQ}(1)$ 
symmetry that 1) is an exact symmetry of the classical action, 
2) is spontaneously broken, and 3) has a color anomaly, i.e. 
it is explicitly broken by the non-perturbative QCD instanton 
effects that make physics depend on the value of $\theta_{\rm QCD}$.
When this recipe is followed, the parameter $\theta_{\rm QCD}$ is 
replaced by $a(x)/f_a$ where $a(x)$ is a dynamical pseudo-scalar 
field and $f_a$ is a quantity with dimension of energy, called the
axion decay constant.  $f_a$ is of order the vacuum expectation 
value that spontaneously breaks $U_{\rm PQ}(1)$ symmetry.   
\footnote{\label{decayconstant} Confusingly, the expression 
``decay constant" has different meaning in nuclear physics than 
in particle physics.  In nuclear physics, ``decay constant''
means what particle physicists term "decay rate".  Eq.~(\ref{lif})
gives the decay rate of the axion to two photons in terms of the 
axion mass and the axion decay constant.} $a(x)$ is the associated 
Nambu-Goldstone boson.   Weinberg and Wilczek (WW) pointed out 
that the non-perturbative instanton effects that make physics 
depend $\theta_{\rm QCD}$ introduce an effective potential for 
$a(x)$ \cite{Wein78,Wilc78}.   The minimum of this effective 
potential was later shown to be at $a(x) = 0$ \cite{Vafa84}.  The 
Strong CP Problem is solved after the $a(x)$ field settles there.

The PQ mechanism modifies the low energy effective theory of 
the SM by the addition of a light pseudo-scalar particle, 
called the ``axion", the quantum of the $a(x)$ field.   The 
properties of the axion depend mainly on the value of the 
axion decay constant $f_a$; see Section 2. The axion mass 
$m_a$ and all its interaction strenghts are inversely proportional 
to $f_a$.  In the original PQWW model, $f_a$ is of order the 
electroweak scale, implying an axion which is relatively strongly 
coupled and heavy, i.e. $m_a$ of order 100 keV.  The PQWW model 
was soon ruled out by a variety of laboratory experiments, 
including unsuccessful searches for axions in beam dumps and 
in rare particle decays such as $K^+ \rightarrow \pi^+ + a$
\cite{Kim87}, and by stellar evolution constraints \cite{Turn90,
Raff90}.  The latter arise because stars emit the weakly coupled 
axions from their cores whereas they emit photons only from their 
surfaces.  If axions exist, stars have an additional energy loss 
mechanism, causing them to evolve faster.  When the negative 
results from accelerator based axion searches are combined 
with the stellar evolution constraints, axion models with 
$f_a \lesssim 10^9$ GeV are generically ruled out.  

Although the original PQWW model is untenable, the general idea 
of Peccei-Quinn symmetry and its concomitant axion are not.  
Jihn E. Kim and others showed that $U_{\rm PQ}(1)$ need 
not be broken at the electroweak scale \cite{Kim79,Shif80,
Zhit80,Dine81}.  It may be broken at an arbitrarily high 
energy, e.g. the hypothetical ``grand unification scale" 
of $10^{15}$ GeV.  When $f_a$ is that large, the axion is 
very light ($m_a \simeq 6 \cdot 10^{-9}$ eV for $f_a = 
10^{15}$ GeV) and extremely weakly coupled: all axion 
production and interaction rates are suppressed by 
approximately 25 orders of magnitude compared to those 
of the PQWW axion.  Thus was born the idea of the 
``invisible axion", a solution to the Strong CP Problem 
that conveniently avoids all constraints from laboratory 
searches and stellar evolution, by making $f_a$ arbitrarily 
large.  

Fortunately cosmology came to the rescue.  Indeed, for 
$a(x)$ to relax to zero, the axion field oscillations 
must commence sufficiently early in the history of the 
univere (today is too late!) and for this the axion must 
be sufficiently heavy \cite{Pres83,Abbo83,Dine83} since 
the oscillation period is $2\pi/m_a$.  The finite age of 
the universe implies a limit on how small $m_a$, or 
equivalently how large $f_a$, can be.  

Unlike most other particles, relic axions are produced in 
the early universe in two different populations, which we 
call ``hot" and ``cold".   The hot axions are thermally 
produced in the primordial plasma.  Like relic photons 
and neutrinos, they have a temperature of order a couple 
of degrees Kelvin today.  Hot axions move too fast to 
gather in galactic halos and, for this reason, are not 
a good candidate for the dark matter observed in galactic 
halos and in clusters of galaxies. Like relic SM neutrinos 
they are a form of ``hot dark matter".  There is no known 
technique to detect hot relic axions in the laboratory.

The cold axion population is produced in the process of 
axion field relaxation, usually referred to as ``vacuum 
realignment", mentioned in the paragraph previous to last.   
The vacuum realignment process is specific to Bose fields,
such as axions or axion-like particles, that are both very 
light and very weakly coupled.  The key point is that when 
the axion mass becomes larger than the inverse age of the 
universe at that time, the axion field is not initially at 
the minimum of its effective potential (because it has no 
reason to). It begins to oscillate then and, because the
axion is very weakly coupled, these oscillations do not 
dissipate into other forms of energy.  The energy density 
in relic axion field oscillations is a form of cold dark 
matter \cite{Ipse83}.  Indeed, among all the widely 
considered dark matter candidates, axions are the coldest.

As was implied above, the cold axion cosmological energy 
density is an increasing function of $f_a$, and therefore a 
decreasing function of the axion mass.  The axion mass for 
which, in the simplest scenarios, the cold axion density equals 
that of dark matter is of order $10^{-5}$ eV.  There are however 
large uncertainties.  The largest source of uncertainty is 
whether inflation homogenizes the axion field.  If inflation 
takes place after the phase transition in which $U_{\rm PQ}(1)$ 
is spontaneously broken, the value of $a(x)/f_a$ before the 
axion field oscillations begin, called the initial misalignment 
angle $\theta_{\rm in}$, is the same throughout the observable 
universe \cite{Pi84}.  Because the cold axion cosmological 
energy density is proportional to $\theta_{\rm in}^2$ (for 
small $\theta_{\rm in}$), there is a 10\% chance that the 
axion density is suppressed by a factor of order $10^{-2}$, in 
which case the axion mass for which the axion density equals 
that of cold dark matter is approximately 100 times smaller, 
$10^{-7}$ eV instead of $10^{-5}$ eV.  Likewise, there is a 
1\% chance that it is suppressed by a factor $10^{-4}$, with 
the cosmologically interesting axion mass most likely near 
$10^{-9}$ eV, and so on.  There are additional sources of 
uncertainty, including: the contribution to the cold axion 
energy density from the decay of topological defects (axion 
strings and domain walls), the precise temperature dependence 
of the axion mass, and the amount of entropy produced during 
the QCD phase transition.  Finally, we do not know what fraction 
of dark matter is axions, in case dark matter is composed of 
several species.  These and other topics in axion cosmology 
are reviewed in refs. \cite{Siki08,Mars16}.

Various methods have been proposed to detect ``invisible" axions. 
Most methods do not attempt to produce and detect axions but attempt 
instead to detect axions that are already in the laboratory either 
as dark matter or as particles emitted by the Sun.  Indeed 
experiments that attempt to both produce and detect axions 
pay twice the price of very weak coupling and for this reason 
have extremely low event rates.  On the other hand such 
experiments make fewer assumptions and have better control 
over experimental variables.  The goal of this review is to 
discuss the various methods that have been proposed and to 
provide theoretical derivations of their signal strengths.  
In a number of cases, noise and backgrounds are discussed 
as well.  Previous reviews, with greater emphasis on 
experimental techniques, can be found in refs. \cite{Rose00,
Brad03,Aszt06,Iras18}.

QCD axions are very well motivated because they solve the 
Strong CP Problem and they are a good dark matter candidate.
Their allowed mass range is $10^{-13}$ to $10^{-2}$ eV, where 
the lower bound is from the assumption that the scale of PQ 
symmetry breaking is smaller than the Planck scale and the 
upper bound is from stellar evolution arguments.  For QCD 
axions there is a definite relationship between mass and 
interaction strength.  They are proportional to each other.  
The residual model dependence is relatively small, except 
perhaps for the coupling of the axion to electrons.  See 
Section 2.  QCD axions appear in many theories of physics 
beyond the SM, including supersymmetric extensions and string 
theory \cite{Svrc06,Aria12}.  In fact such theories often 
predict additional axion-like particles (ALPs), distinct 
from the QCD axion but with similar properties.  Let us 
define an ALP as a light pseudo-scalar particle with 
couplings to ordinary particles like those of the QCD 
axion but without any a-priori relationship between 
coupling strength and mass.  Many QCD axion search 
techniques are relevant to ALPs as well.  In such 
cases it will be natural to include ALPs in the 
discussion.  For the sake of definiteness, ALPs 
outside the allowed mass range of QCD axions 
($10^{-13}$ to $10^{-2}$ eV) are not considered.

Finally, let us mention that an argument has been made that 
the dark matter {\it is} axions, or ALPs, at least in part.
The argument is based on the observation that cold dark matter
axions thermalize through their gravitational self-interactions
and, as a result, form a Bose-Einstein condensate \cite{Siki09}.  
A thermalizing or rethermalizing Bose-Einstein condensate has 
properties different from ordinary cold dark matter \cite{Erke12}, 
and it has been found that observations support the hypothesis 
that the dark matter is a rethermalizing Bose-Einstein condensate
\cite{Siki11}.

\section{Axion properties}

This section provides basic information on axions, including 
formulae for the axion mass and for its couplings to ordinary 
particles, limits on axion properties from astrophysics and 
cosmology, an estimate of the flux of axions from the Sun, 
and two proposals for the local distribution of dark matter 
axions.   Axion models are reviewed in ref. \cite{DiLu20}.

\subsubsection{Axion mass}

In terms of the decay constant $f_a$, the axion mass is given by 
\cite{Wein78}
\begin{equation}
m_a \simeq {\sqrt{m_u m_d} \over m_u + m_d} {f_\pi m_\pi \over f_a} 
\simeq 6 \cdot 10^{-6}~{\rm eV} 
\left({10^{12}~{\rm GeV} \over f_a}\right)~~\ ,
\label{mass}
\end{equation}
where $m_\pi$ is the pion mass, and $f_\pi \simeq 93$ MeV the pion 
decay constant.  

Formulae for the axion coulings in the PQWW model were derived 
in refs. 
\cite{Wein78,Wilc78,Bard78a,Gold78,Kanda78,Elli78,Trei78,Donn78}.  The 
relevant formulae for the invisible axion models can be found in 
the original papers \cite{Kim79,Shif80,Zhit80,Dine81} on these 
models.  More general discussions of the axion couplings can be 
found in refs. \cite{Kapl85,Sred85,Sik86}.

\subsubsection{Electromagnetic coupling}

The axion coupling to two photons is 
\begin{equation}
{\cal L}_{a\gamma\gamma} = - g_\gamma {\alpha\over \pi} 
{1 \over f_a} a(x) \vec{E}(x)\cdot\vec{B}(x)
\label{agamgam}
\end{equation}
where $\alpha$ is the fine structure constant and
\begin{equation}
g_\gamma = {1 \over 2}\left({N_e \over N} 
- {5 \over 3} - {m_d - m_u \over m_d + m_u}\right)~~\ .
\label{ggam}
\end{equation}
$N$ and $N_e$ are respectively the color anomaly and 
electromagnetic anomaly of the PQ charge. They are 
given by
\begin{equation}
N \delta^{ab} = Tr(Q_{\rm PQ} Q^a Q^b)~~~,~~~
N_e = Tr(Q_{\rm PQ} Q_e Q_e)
\label{anom}
\end{equation}
where the trace symbol indicates a sum over all left-handed Weyl 
fermions in the model, $Q_{\rm PQ}$ is PQ charge, the $Q^a$ ($a$ 
= 1,2, ..., 8) are the color charges, and $Q_e$ is electric charge.  
In the original PQWW model and in the DFSZ invisible axion model, 
$N = 6$, $N_e = 16$, and therefore $g_\gamma \simeq$ 0.36.  In 
the KSVZ invisible axion model, $N=1$, $N_e = 0$, and therefore 
$g_\gamma \simeq$ - 0.97.  In any grand unified model, $N/N_e = 
\sin^2 \theta_W^0$ where $\theta_W^0$ is the value of the 
electroweak angle at the grand unification scale.  A favored 
value is $\sin^2 \theta_W^0 = 3/8$ since this is consistent with 
the measured value of $\sin^2 \theta_W$ at the electroweak scale 
\cite{Geor74}.  For $N_e/N = 8/3$ 
\begin{equation}
g_\gamma = {m_u \over m_u + m_d} \simeq 0.36~~,
\label{GUT}
\end{equation}
the same as in the PQWW and DFSZ models because these models are 
grand unifiable with $\sin^2 \theta_W^0 = 3/8$.  Because the axion 
mixes with the neutral pion, Eq.~(\ref{ggam}) has contributions 
both from the PQ charges of quarks and leptons and from the two 
photon coupling of the neutral pion.   As a result $g_\gamma$ 
can only vanish if there is a cancellation between unrelated 
contributions.  The electromagnetic coupling is relevant to 
many approaches to invisible axion detection.

\subsubsection{Coupling to nucleons and electrons}

The coupling of the axion to a Dirac fermion $f(x)$ has the 
general form
\begin{equation}
{\cal L}_{a\bar{f}f} = {1 \over f_a}
\left[- {g_f \over 2} \partial_\mu a(x)~\bar{f}(x) \gamma^\mu \gamma^5 f(x)
~~+~~ \theta_f m_f a(x) \bar{f}(x) f(x)\right]~~~\ ,
\label{CPv2}
\end{equation}
where the $g_f$ are model dependent numbers that are generically 
of order one, whereas the $\theta_f$ are generically of order 
$10^{-17}$ assuming that SM weak interactions are the only 
source of CP violation.   The $\theta_f$ would vanish if CP were 
conserved.  The known CP violation of the weak interactions induces, 
through loop diagrams, small values for $\theta_{\rm QCD}$ and for 
the $\theta_f$ that are generically of order $10^{-17}$ 
\cite{Elli79,Geor86b}.  In the non-relativistic limit, 
Eq.~(\ref{CPv2}) implies the interaction energy
\begin{equation}
H_{a\bar{f}f} = {1 \over f_a}
\left[{g_f \over 2} \left(\vec{\sigma}\cdot\vec{\nabla}a(\vec{x},t) +
{\vec{p}\cdot\vec{\sigma} \over m_f} \partial_t a(\vec{x},t)\right)
- \theta_f m_f a(\vec{x},t)\right]
\label{intfa2}
\end{equation}
where $\vec{x}$, $\vec{p}$, $m_f$ and ${1 \over 2} \vec{\sigma}$ 
are respectively the position, momentum, mass and spin of the 
fermion.  For axion searches, the most relevant fermions are the 
proton, the neutron and the electron.

For nucleons ($f = p, n$), the coefficients $g_f$ that appear
in Eqs.~(\ref{CPv2}) and (\ref{intfa2}) are given by
\begin{equation}
g_{p \atop n} = - {1 \over 2} [ \pm g_{A3} 
({m_d - m_u \over m_d + m_u} - {g_u - g_d \over N})
+ g_{A0}(1 - {g_u + g_d \over N})]~~~\ ,
\label{axnuc2}
\end{equation}
where $g_{A3}$ = 1.25 is the isotriplet axial vector coupling. 
The isosinglet axial vector coupling $g_{A0}$ has not been 
measured directly.   It is estimated in ref. \cite{Adle75}
to be 0.74 using the quark model, and 0.65 using the MIT 
bag model.  The $g_u$ and $g_d$ coefficients are related
to the PQ charges of the up and down quarks in a way that 
depends on whether the PQ field spontaneously breaks
SM gauge symmetries in addition to $U_{\rm PQ}(1)$.  In 
the KSVZ model, $g_u = g_d = 0$. In the PQWW and DSVZ 
models, 
\begin{equation}
g_u = {v_2^2 \over 3(v_1^2 + v_2^2)}~~,~~
g_d = {v_1^2 \over 3(v_1^2 + v_2^2)}~~~\ ,
\label{gugd}
\end{equation}
where $v_1 (v_2)$ is the vacuum expectation value of the Higgs 
field that gives mass to the up (down) quarks.  Because the axion 
mixes with the neutral pion, $g_p$ and $g_n$ receive contributions 
from the pion-nucleon coupling as well as from the PQ charges of 
the up and down quarks.  Each may vanish only if there is a 
fortuitous cancellation between unrelated contributions.  

The coupling to the electron ($f = e$) is more model dependent 
than the others.  In the PQWW and DFSZ models, $g_e = g_d$ given 
in Eq.~(\ref{gugd}).   In the KSVZ model, $g_e = 0$ at tree level.  
However, a one loop correction yields a contribution of order 
$g_e \sim 10^{-3}$ \cite{Sred85}. 

\subsubsection{Stellar evolution constraints}

Stellar evolution arguments constrain the axion couplings.
The two photon coupling causes axions to be produced in 
stellar cores by the Primakoff process, the conversion of 
a photon to an axion in the Coulomb field of a nucleus
($\gamma + N \rightarrow N + a$). The lifetime of horizontal 
branch stars in globular clusters implies the constraint 
\cite{Raff08}
\begin{equation}
g_{a\gamma\gamma} \equiv g_\gamma {\alpha \over \pi} {1 \over f_a}
< 10^{-10}~{\rm GeV}^{-1}~~\ .
\label{HB} 
\end{equation}
The coupling to electrons causes stars to emit axions through 
the Compton-like process $\gamma + e^-\rightarrow e^- + a$ and 
through axion bremstrahlung 
$e^- + (Z,A) \rightarrow (Z,A) + e^- + a$.  The resulting energy 
losses excessively delay the onset of helium burning in globular
cluster stars unless \cite{Raff95,Cate96}
\begin{equation}
g_{a\bar{e}e} \equiv {g_e \over f_a} 
< 5 \cdot 10^{-10}~{\rm GeV}^{-1}~~\ . 
\label{Hefl}
\end{equation}
The increase in the cooling rate of white dwarfs resulting from 
these processes produces a similar bound \cite{Raff86,Blin94}. The 
coupling 
to  nucleons causes axions to be radiated by the collapsed stellar
core produced in a supernova explosion.  The requirement that the 
observed neutrino pulse from SN1987a not be quenched by axion
emission implies \cite{Elli87,Raff88a,Turn88,Raff08}.
\begin{equation}
f_a > 4 \cdot 10^8~{\rm GeV}
\label{SN87a}
\end{equation}
or $m_a < 1.6 \cdot 10^{-2}$ eV.

Very light axions ($6 \cdot 10^{-13} < m_a < 2 \cdot 10^{-11}$ eV) 
are constrained by stellar mass black hole superradiance, as 
discussed in refs. \cite{Arva10,Arva11,Arva15}.   

Updates on the bounds of Eqs.~(\ref{HB}) and (\ref{Hefl}) can be found 
in refs. \cite{Ayal14,Viau13}.

\subsubsection{Solar axion flux}

The solar axion flux on Earth was calculated by Raffelt \cite{Raff08}:
\begin{equation}
{d \Phi_a \over d E}~=~
{6.0 \cdot 10^{10} \over {\rm cm}^2~{\rm sec}~{\rm keV}}~
\left({g_{a\gamma\gamma} \over 10^{-10}~{\rm GeV^{-1}}}\right)^2~
\left({E \over {\rm keV}}\right)^{2.481}
\exp\left(- {E \over 1.205~{\rm keV}}\right)~~~\ .
\label{solaxfl2}
\end{equation}
The integrated flux is
\begin{equation}
\Phi_a = {3.75 \cdot 10^{11} \over {\rm cm}^2~{\rm sec}}~
\left({g_{a\gamma\gamma} \over 10^{-10}~{\rm GeV^{-1}}}\right)^2~~~\ .
\label{intfl2}
\end{equation}
The energy   
spectrum in Eq.~(\ref{solaxfl2}) is nearly isothermal with temperature
that of the solar core, approximately 1.3 keV.  Eq.~(\ref{solaxfl2}) 
includes only solar axions produced by the Primakoff process.  There 
may be additional axions from processes involving the electron coupling
\cite{Redo13}.  Also, axions with specific energies are emitted in 
nuclear deexcitations in the solar core\cite{Avig18a}.

\subsubsection{Cold axion cosmological energy density}

The present cosmological energy density in cold axions, as a fraction 
of the critical energy density, may be written \cite{Siki08} 
\begin{equation}
\Omega_a \equiv \rho_a {8 \pi G \over 3 H_0^2}= 0.3~X~
\left({f_a \over 10^{12}~{\rm GeV}}\right)^{7 \over 6}
\label{axcden}
\end{equation}
where $X$ is a poorly known fudge factor reflecting cosmological
uncertainties.  According to the discussion in ref. \cite{Siki08},
$X$ is of order two if the axion field does not get homogenized 
by inflation and the string decay contribution is of the same 
order of magnitude as that from vacuum realignment.  If the 
string decay contribution dominates, $X$ may be as large as 
ten. If inflation homogenizes the axion field, $X$ is of order 
${1 \over 2} \theta_{\rm in}^2$ where $\theta_{\rm in}$ is the 
initial misalignment angle.  Lattice QCD simulations may help
remove uncertainties associated with the dependence of the axion 
mass on temperature.  For a discussion and list of references 
see ref. \cite{Dine17}

\subsubsection{Galactic halo models}

When discussing axion dark matter detection, we will 
consider two contrasting proposals for the local density 
and velocity distribution of dark matter axions.  Proposal 
A assumes that galactic halos are in thermal equilibrium.  
By fitting the isothermal model to the Milky Way rotation 
curve, one finds \cite{Turn86} 
\begin{equation}
\rho_{\rm dm} \simeq 300~{\rm MeV/cm}^3
\label{dmA}
\end{equation}
for the local dark matter density.  The velocity 
distribution is a Maxwell-Boltzmann with dispersion 
$\sqrt{\langle \vec{v}\cdot\vec{v} \rangle} \simeq$ 
270 km/s at any location in the halo.

Proposal B is based on the observation that dark matter
particles accreting onto a galactic halo do not, as a 
result of their gravitational interactions, thermalize 
over the age of the universe \cite{Sik92}.  A galactic 
halo is then a set of overlapping cold flows with sharp 
features, called ``caustics", in the physical density. 
The caustic ring model \cite{Duff08} is a particular 
realization motivated by observation.  According to 
the model, we on Earth are located close to a caustic.
As a result our local dark matter velocity distribution 
is dominated by the flows that form this caustic.  Most 
prominent among these is the `Big Flow' \cite{Sik03}.  
It has velocity vector \cite{Duff08,Chak20} 
\begin{equation}
\vec{v}_{\rm BF} \simeq 
[509~ \hat{\phi} - 104 ~ \hat{r}
+ 6~\hat{z}]~{\rm km/s}
\label{BFv}
\end{equation}
in a non-rotating galactic reference frame.  $\hat{\phi}$ 
is the unit vector in the direction of galactic rotation,
$\hat{r}$ in the direction away from the galactic center, and 
$\hat{z}$ in the direction of the north galactic pole. The 
Big Flow has velocity dispersion less than 71 m/s \cite{Bani16}.
The uncertainty in the speed (520 km/s) of the Big Flow is of 
order 9\%.   It is due mainly to the uncertainty in the galactic 
rotation velocity.   The uncertainty in its direction is of 
order $1^\circ$.  The density of the Big Flow on Earth 
depends sharply on our distance to a cusp in the nearby 
caustic and is poorly constrained for this reason. According
to ref. \cite{Chak20}, it is at least 6 GeV/cm$^3$.

\section{Axion to photon conversion in a magnetic field}

This section discusses the conversion of axions to photons in a 
static magnetic field in the absence of cavity or reflecting walls 
for the photons \cite{Sik83b,Sik85,Ans85,Maia86,VanB87,Raff88b,VanB89}.  
We allow the presence of a homogeneous and static dielectric constant 
$\epsilon$ and magnetic susceptibility $\mu$.

\subsection{Axion electrodynamics}

Consider the action density for the electromagnetic and axion fields:
\begin{eqnarray}
{\cal L}_{{\rm e.m.}+a} &=& {1 \over 2}(\epsilon \vec{E}\cdot\vec{E}
- {1 \over \mu} \vec{B}\cdot\vec{B}) - \rho_{\rm el} \Phi + 
\vec{j}_{\rm el}\cdot\vec{A}\nonumber\\
&+& {1 \over 2}\left((\partial_t a)^2 - (\vec{\nabla} a)^2\right)
- {1 \over 2} m_a^2 a^2 - g a \vec{E}\cdot\vec{B}
\label{emaLag}
\end{eqnarray}
where $\vec{E} = - \vec \nabla \Phi - \partial_t \vec A$~,~ 
$\vec{B} = \vec \nabla \times \vec A$~, and 
$g \equiv g_{a\gamma\gamma} = g_\gamma {\alpha \over \pi} {1 \over f_a}$.  
$\rho_{\rm el}$ and $\vec{j}_{\rm el}$ are the charge and current densities due to 
ordinary charged particles.  Eq.~(\ref{emaLag}) implies the modified 
Maxwell's equations \cite{Sik84,Sik83b}
\begin{eqnarray}
\vec \nabla\cdot (\epsilon \vec E - g a \vec B) &=& 
\rho_{\rm el}\nonumber\\
\vec\nabla\times({1 \over \mu} \vec B + g a \vec E) - 
\partial_t (\epsilon \vec E - g a \vec B) &=& 
\vec{j}_{\rm el}\nonumber\\
\vec\nabla\times \vec E + \partial_t \vec B &=& 0\nonumber\\
\vec\nabla\cdot\vec B &=& 0~~~\ ,
\label{Maxwa}
\end{eqnarray}
and
\begin{equation}
\partial_t^2 a - \nabla^2 a + m_a^2 a = - g \vec{E}\cdot\vec{B}~~~~\ .
\label{eoma}
\end{equation}
The set of equations (\ref{Maxwa}) and (\ref{eoma}) is referred 
to as ``axion electrodynamics".

The first two Eqs.~(\ref{Maxwa}) may be rewritten
\begin{eqnarray}
\vec \nabla\cdot (\epsilon \vec E) &=&
g~\vec B \cdot\vec\nabla a + \rho_{\rm el}\nonumber\\
\vec\nabla\times({1 \over \mu} \vec B) -
\partial_t (\epsilon \vec E) &=& g~(\vec E\times \vec\nabla a -
\vec B \partial_t a) + \vec{j}_{\rm el}~~~\ ,
\label{inhMaxwa}
\end{eqnarray}
showing that in background magnetic $\vec{B}_0(\vec{x}, t)$ and 
electric $\vec{E}_0(\vec{x}, t)$ fields the axion is a source of 
electric charge and current density 
\begin{equation}
\rho_a = g~\vec B_0\cdot \vec\nabla a~~~~,~~~~
\vec j_a = g (\vec E_0 \times \vec\nabla a -
\vec B_0\partial_t a)~~~~\ .
\label{axc}
\end{equation}
In covariant form, $j_a^\mu = - g \tilde{F}^{\mu\nu} \partial_\nu a$.
The axion induced electric current is separately conserved:
$\partial_\mu j_a^\mu \equiv 0$.  

$j_a^\mu$ is a source of electromagnetic waves, implying the conversion
of energy from the axion to the electromagnetic field.  For practical 
reasons, it is magnetic rather than electric fields that are used to 
cause the conversion.  Hence, for simplicity, we set $\vec{E}_0 = 0$ 
below.  We will assume furthermore that $\vec{B}_0$ is static and, 
henceforth in this section, that $\epsilon$ and $\mu$ are constant 
in space and time.

Let us set $\rho_{\rm el} = \vec{j}_{\rm el} = 0$ and consider an 
axion plane wave
\begin{equation}
a(\vec x,t) = Re (A~e^{i(\vec k_a\cdot \vec x - \omega t)})~~~\ ,
\label{aFour}
\end{equation}
where $\omega = \sqrt{m_a^2 + \vec{k}\cdot\vec{k}}$.  We choose the 
gauge~$\epsilon \mu \partial_t \Phi + \vec\nabla\cdot\vec A = 0$.
The inhomogeneous Maxwell's equations are then
\begin{eqnarray}
(- \nabla^2 + \epsilon \mu \partial_t^2) \Phi &=&
{1 \over \epsilon} \rho_a \nonumber\\
(- \nabla^2 + \epsilon \mu \partial_t^2) \vec A 
&=& \mu \vec{j}_a~~~~\ .
\label{inhM}
\end{eqnarray}
Provided the first equation is satisfied at an initial time, 
it is satisfied at all times as a consequence of the second 
equation.  The second equation is solved by
$\vec{A}(\vec{x}, t) = Re (\vec{A}(\vec{x}) e^{-i \omega t})$
provided
\begin{equation}
(- \nabla^2 - \epsilon \mu \omega^2) \vec A(\vec{x})
= \mu \vec{j}_a(\vec{x})
\label{eqAx}
\end{equation}
where
\begin{equation}
\vec{j}_a(\vec{x}) = i g \omega A \vec{B}_0(\vec{x}) 
e^{i \vec{k}_a \cdot \vec{x}}~~~~\ .
\label{jax}
\end{equation}
The solution of interest, involving the retarded Green's  
function, is
\begin{equation}
\vec A (\vec x) = {\mu \over 4 \pi}
\int_V d^3 x^{\,\prime}~ 
{e^{ik\vert\vec x - \vec x^{\,\prime}\vert} \over 
\vert\vec x - \vec x^{\,\prime}\vert}~ 
\vec j_a (\vec x^{\,\prime})
\label{retsol}
\end{equation}
with $k = \sqrt{\epsilon \mu} \omega$.  $V$ is the volume 
of the region over which the magnetic field extends.  Let 
$\vec{x} = r \hat{n}$ and $r \rightarrow \infty$.  In that limit
\begin{equation}
\vec{A}(\vec{x}) = \mu~{e^{ikr} \over 4\pi r}~\vec j_a (\vec k)
+ 0({1 \over r^2})
\label{vecpot}
\end{equation}
where $\vec k = k \hat{n}~$ and
\begin{equation}
\vec j_a (\vec k) = 
\int_V d^3 x~e^{- i \vec k \cdot \vec x}~\vec j_a (\vec x)
= i \omega~g~A\int_V d^3 x~
e^{i(\vec k_a - \vec k) \cdot \vec x} \vec B_0(\vec x)~~~\ .
\label{jFour}
\end{equation}
The electromagnetic power radiated per unit solid angle in 
direction $\hat{n}$ is 
\begin{equation}
{dP \over d\Omega} = \lim_{r \to \infty} 
\langle \hat{n} \cdot(\vec E \times \vec H)\rangle r^2
=  {\mu k \omega \over 32\pi^2} 
\vert \hat{n} \times \vec j_a (\vec k)\vert^2~~~~\ .
\label{convPow}
\end{equation}
The $\langle ... \rangle$ brackets indicate that a time average 
is being taken.  

We derived Eq.~(\ref{convPow}) by a classical field theory
calculation but the actual world is quantum-mechanical.  Whereas
the conversion of axion field energy to electromagnetic field energy 
happens continuously in the classical description, in reality it 
happens one quantum at a time.  Because the magnetic field is static, 
the energy of each photon produced is exactly the energy of the axion 
that disappeared.  Eq.~(\ref{convPow}) gives the time averaged power 
for the quantum process of axion to photon conversion.

\subsection{Conversion cross-section}

Dividing by the magnitude of the incident axion 
energy flux
\begin{equation} 
\vec {\cal P}_a = \langle - \dot a \vec \nabla a\rangle = 
{1\over 2} \vert A\vert^2 \omega \vec k_a~~~~\ ,
\label{incafl}
\end{equation}
we obtain the differential cross-section \cite{Sik83b}:
\begin{equation}
{d \sigma \over d\Omega} (a \rightarrow \gamma) = 
{1 \over \vert \vec {\cal P}_a\vert} {dP \over d\Omega}
= g^2 {\mu k \omega \over 16\pi^2\beta_a}  
\bigg| \int_V d^3 x~e^{i(\vec k_a -\vec k) \cdot \vec x}~ 
\hat{n} \times \vec B_0 (\vec x)\bigg|^2
\label{convxs}
\end{equation}
where $\beta_a = \vert \vec k_a \vert/\omega$ is the speed of 
the incident axions.  We may rewrite the RHS of Eq.~(\ref{convxs}) 
as a sum over final state photon polarizations, $\hat{e}_1(\hat{n})$ 
and $\hat{e}_2(\hat{n})$, using the completeness relation
\begin{equation}
\delta_{ij} = n_i~n_j~+~e_{1i}~e_{1j}~+~
e_{2i}~e_{2j}~~~~\ .
\label{complete}
\end{equation}
In that form
\begin{equation}
{d \sigma \over d\Omega} (a \rightarrow \gamma) =
g^2 {\mu k \omega \over 16\pi^2\beta_a}
\sum_{\lambda = 1,2}
\bigg| \int_V d^3 x~e^{i(\vec k_a -\vec k) \cdot \vec x}~
\hat{e}_\lambda(\hat{n}) \cdot \vec B_0 (\vec x)\bigg|^2~~~\ .
\label{convxs2}
\end{equation}
Because the axion and photon have equal energy 
but satisfy different dispersion relations, their momenta differ 
in general.  The momentum transfer $\vec q \equiv \vec k - \vec k_a$ 
is provided by the inhomogeneity of the magnetic field.  The conversion 
cross-section is proportional to the power in the Fourier 
component of $\vec B_0 (\vec x)$ with wavevector $\vec q$. An 
analogous calculation, starting with Eq.~(\ref{eoma}), yields
the differential cross-section for the inverse process, the 
conversion in a static magnetic field of a photon with 4-momentum 
$(p_\gamma^\mu) = (\omega, \vec{k})$ to an axion with 
4-momentum $(p_a^\mu) = (E_a, \vec{k_a})$:
\begin{equation}
{d \sigma \over d \Omega}(\gamma \rightarrow a) = g^2 
{\omega k_a \over 16 \pi^2} \sqrt{\mu \over \epsilon}~  
\bigg|\int_V d^3 x~e^{i(\vec{k} - \vec{k}_a) \cdot \vec{x}}~
\hat{t} \cdot \vec{B}_0(\vec{x})\bigg|^2~~~~\ ,
\label{invxs}
\end{equation} 
where $\hat{t}$ is the polarization vector of the initial 
photon.  

\subsection{Colinear conversion}

Consider the particular case where the magnetic field is smooth 
on a length scale $\lambda_B$  much larger than $k_a^{-1}$ and  
$k^{-1}$.   The conversion process is co-linear then since 
$\vert \vec q \vert = \vert \vec k - \vec k_a \vert \sim 
\lambda_B^{-1} << k_a,~k$.  Let $z$ be the position coordinate 
along the path of the axion and photon.  The conversion probability 
depends only on the magnetic field along the path.  To calculate 
it, we may take $\vec{B}_0$ to be independent of the coordinates 
orthogonal to $z$ over a cross-sectional area $S$.  Since 
$\hat{n} = \hat{z}$ in this case 
\begin{equation}
\bigg| \int_V d^3 x~e^{- i \vec q \cdot \vec x} 
~\hat{n} \times \vec B_0 (\vec x)\bigg|^2
= (2\pi)^2 \delta^2 (\vec q_\perp) S~ 
\bigg| \int_0^L dz~e^{-iqz} \vec B_{0\perp} (z)\bigg|^2
\label{colinc}
\end{equation}
where $L$ is the depth over which the magnetic field extends, 
the subscript $\perp$ indicates the component perpendicular
to the direction of propagation, and 
\begin{equation}
q =  k - k_a =
\sqrt{\epsilon\mu} \omega - \sqrt{\omega^2-m_a^2}~~~~~\ .
\label{momtran}
\end{equation}
The conversion probability is 
\begin{equation}
p(a \rightarrow \gamma) = 
{1 \over S} \int d\Omega_{\vec{k}}~{d \sigma \over d \Omega_{\vec{k}}}
= {g^2 \over 4 \beta_a} \sqrt{\mu \over \epsilon}~
\bigg| \int_0^L dz~e^{-iqz}~\vec B_{0\perp} (z)\bigg|^2~~~~\ .
\label{conprob}
\end{equation}
The produced photon is linearly polarized in the direction of 
$\vec{B}_{0\perp}(z)$ in case $\vec{B}_{0\perp}(z)$ has everywhere
the same direction.  Similarly, from Eq.~(\ref{invxs}) we find the 
conversion probability of a photon to an axion 
\begin{equation}
p(\gamma \rightarrow a) =
{g^2 \over 4 \beta_a} \sqrt{\mu \over \epsilon}~
\bigg| \int_0^L dz~e^{+iqz}~\hat{t} \cdot \vec B_0 (z)\bigg|^2~~~~\ .
\label{revprob}
\end{equation}
For a given polarization state of the photon, $p(a \rightarrow \gamma)$
and $p(\gamma \rightarrow a)$ are equal, as required by the principle 
of detailed balance.  

For $\vec B_0 = \hat t B_0 \cos({2 \pi \over d} z)$ with
$\hat{t}\cdot\hat{z} = 0$, we have
\begin{equation} 
p = {g^2 B_0^2 \over 4 \beta_a} \sqrt{\mu \over \epsilon} 
\Bigg|{\sin(q+{2 \pi \over d}){L \over 2} \over q+ {2 \pi \over d}} 
+ e^{2i\pi {L \over d}}~ 
{\sin (q- {2 \pi \over d}){L \over 2}\over q - {2 \pi \over d}} \Biggr|^2 
~~~~\ . 
\label{perprob} 
\end{equation} 
The conversion is resonant when $q = \pm {2 \pi \over d}$, with 
probability 
\begin{equation} 
p = {g^2 B_0^2 L^2 \over 16 \beta_a} \sqrt{\mu \over \epsilon} 
\label{reson} 
\end{equation} 
assuming $d << L$. 

If the magnetic field is homogeneous
\begin{equation}
p = {g^2 B_0^2 \over \beta_a} \sqrt{\mu \over \epsilon}~
\sin^2 ({q L \over 2}) {1\over q^2}~~\ .
\label{classicp}
\end{equation}
The axion and photon oscillate into each other, with oscillation length 
$\ell_{\rm osc} = {\pi \over q}$.  After a 
distance $\ell_{\rm osc}$, a fraction 
${g^2 B_0^2 \over \beta_a q^2} \sqrt{\mu \over \epsilon}$ of the axions 
has converted to photons; after a distance $2\ell_{\rm osc}$, those 
photons have converted back to axions, and so forth.  This is similar 
to neutrino flavor oscillations.  In fact, the conversion probability 
can be derived \cite{Maia86, Raff88b} using this analogy; see Section 
9.5.  

In a homogeneous magnetic field, the conversion is resonant when $q << 1/L$.  
In that case
\begin{equation}
p = {g^2 \over 4 \beta_a} \sqrt{\mu \over \epsilon} B_0^2 L^2~~~~~\ .
\label{sim1p}
\end{equation}
The $L^2$ behaviour of the conversion probability in
Eqs.~(\ref{reson}) and (\ref{sim1p}), characteristic of 
resonant conversion, persists only as long as coherence 
between the axion and photon excitations is  maintained.  
Various effects may limit this coherence, e.g. the 
absorption or scattering of the photon out of the path 
of the axion. If coherence persists up to a distance 
$\ell < L$, $L^2$ should be replaced by $L \ell$.

To convert Eq.~(\ref{sim1p}) into practical units, we
note that the energy stored in a volume $V$ permeated 
by a magnetic field $B_0$ is
\begin{equation}
E = {1 \over 2} V B_0^2
\label{storen}
\end{equation}
in the Heaviside-Lorentz units used here, whereas in Gaussian units
\begin{equation}
E = {1 \over 8 \pi}~{\rm erg}~\left({V \over {\rm cm}^3}\right)~
\left({B_0 \over {\rm Gauss}}\right)^2~~~\ .
\label{EGauss}
\end{equation}
The implied conversion factor is:
\begin{equation}
{\rm Gauss} = \sqrt{{\rm erg} \over 4 \pi ~{\rm cm}^3} =
1.9535~10^{-2}~{\rm eV}^2~~~~\ .
\label{Gauss}
\end{equation}
Eq.~(\ref{sim1p}) becomes then:
\begin{equation} 
p =  1.71 \cdot 10^{-17} \Biggl({g_\gamma \over 0.36}\Biggr)^2 
\Biggl({10^7~{\rm GeV} \over f_a}\Biggr)^2 
\Biggl({B_0 \over 10~{\rm T}}\Biggr)^2 
\Biggl({L \over 10~{\rm m}}\Biggr)^2~
{1 \over \beta_a} \sqrt{\mu \over \epsilon}~~~~\ . 
\label{pracp} 
\end{equation}
When $\omega >> m_a$, 
\begin{equation}
q \simeq (\sqrt{\epsilon\mu} - 1)\omega + {m_a^2 \over 2 \omega}~~~\ .
\label{relax}
\end{equation} 
The resonance condition ($q L < 1$) can be satisfied even for 
large $L$ by using a dielectric medium with a plasma-like 
dispersion law \cite{VanB89}:
\begin{equation} 
\epsilon(\omega) = 1 - {\omega_{\rm pl}^2 \over \omega^2}~~~~\ .
\label{plasm} 
\end{equation} 
For $\mu = 1$, resonance is obtained when $\omega_{\rm pl} = m_a$.

Ref. \cite{Flam18} proposes to replace axion-photon conversion in a 
magnetic field by axion-photon conversion through resonant forward 
scattering on atoms or molecules.

\subsection{Applications}

Axion to photon conversion in a magnetic field was originally proposed 
as a method to detect dark matter axions and axions emitted by the Sun 
\cite{Sik83b}.  These applications will be discussed in Sections 5.1 and 
6.0.1 respectively.  Other applications are "shining light through walls" 
and the conversion of axions to photons in astrophysical magnetic fields.

In a ``shining light through walls" experiment, photons are converted 
to axions in a magnetic field on one side of a wall and the axions 
converted back to photons in a magnetic field on the other side of 
that wall \cite{VanB87}.  The sensitivity of the experiment can be 
improved by introducing matched Fabry-P\'erot cavities in the two 
conversion regions, producing a resonance \cite{Hoog91,Fuku96,Sik07}.
Shining light through walls with resonant axion-photon reconversion 
is discussed in Section 8. 

Axion-photon conversion can occur in astrophysical magnetic fields, 
and may have implications for observation.  Axions can readily convert 
to photons, and vice-versa, in the magnetospheres of neutron stars 
\cite{Morr86,Huan18,Hook18}.  With $B_0 = 10^{13}$ Gauss and $L = 10$ km, 
the conversion probability is of order one for $f_a$ up to 10$^{10}$ GeV 
provided $q \simeq 0$.  The latter condition is satisfied if the 
axions are sufficiently energetic. For example, if the axion energy 
is 1 keV and the axion mass $10^{-4}$ eV, the oscillation length 
${2 \pi \omega \over m_a^2}$ = 126 km.  The neutron star may 
therefore convert axions produced in its core or axions emitted 
by a companion star.  It may also convert dark matter axions 
in regions of its magnetosphere where the resonance condition 
is satisfied because the plasma frequency is near the axion mass.

The magnetic fields in galaxies and galaxy clusters are very 
weak, of order $10^{-6}$ Gauss, but extend over enormous distances.  
With $B_0 = 10^{-6}$ Gauss and $L$ = 1 Mpc the conversion probablity 
is of order one for $g$ larger than $10^{-12}$ GeV$^{-1}$ provided 
$q L < 1$.  The latter condition cannot easily be satisfied by  
QCD axions since they are massive, but may be satisifed by light 
ALPs.  Conceivable  phenomena involving the conversion of 
Nambu-Goldstone bosons/ALPs into photons, or vice-versa, in 
large scale astrophysical magnetic fields include the production 
of high energy gamma-rays \cite{Sik88}, distortions of the cosmic 
microwave background spectrum \cite{Hara92}, and alterations in 
the apparent luminosity of faraway sources \cite{Csa02}.

Refs.~\cite{Broc96,Grif96,Paye15} place a limit on light ALPs
from the non-observation of gamma ray photons from the direction 
of SN1987a coincident with that supernova's neutrino signal.  
ALPs are emitted by the Primakoff process in the supernova core
and convert to photons in the magnetic field of the Milky Way.  
A recent published limit is $g < 5.3 \cdot 10^{-12}$ 
GeV$^{-1}$ for $m < 4.4  \cdot 10^{-10}$ eV \cite{Paye15}.

It has been proposed that the apparently excessive transparency 
of the universe to high energy gamma rays is due to the existence 
of ALPs \cite{DeAn07,DeAn08,Sanc09,Horn12}.  High energy gamma 
rays above approximately 100 GeV are absorbed over cosmological 
distances because they produce $e^+e^-$ pairs by colliding with 
extragalactic background photons.  Observations show the universe 
to be more transparent than expected.  The proposed explanation 
is that the high energy photons convert to ALPs in astrophysical 
magnetic fields and that the ALPs, after traveling unimpeded over 
great distances, convert back to high energy photons by the inverse 
process.

Ref. \cite{Conl17} provides a guide to the literature of 
axion-photon conversion in astrophysical magnetic fields and
places an upper limit $g \lesssim 2 \cdot 10^{-12}$ GeV$^{-1}$ 
on ALPs of mass $m \lesssim 10^{-12}$ eV from the non-observation 
of spectral modulations of X-rays from chosen active galactic nuclei, 
caused by the conversion of the X-rays to ALPs in the magnetic fields 
of foreground galaxy clusters.  

\section{The cavity haloscope}

The dark halo of our Milky Way galaxy has density of order $10^{-24}$ 
gr/cm$^3$ in the solar neighborhood.  The halo particles have velocities 
$v$ of order $10^{-3}~c$.  If the dark matter is axions, we are 
surrounded by a pseudo-scalar field oscillating with angular 
frequency:
\begin{equation}
\omega_a = E_a = m_a + {1\over 2} m_a v^2 
=  m_a \Bigl( 1 + {\cal O}(10^{-6})\Bigr)~~~~\ .
\label{halen}
\end{equation}
In an externally applied magnetic field $\vec B_0$, the axion 
electromagnetic interaction (\ref{agamgam}) becomes
\begin{equation}
{\cal L}_{a\gamma\gamma} =
- g_\gamma {\alpha \over \pi} {1 \over f_a} a~
\vec{E} \cdot \vec{B}_0~~~~~\ .
\label{emint} 
\end{equation}
It allows the conversion of axions to photons, and vice-versa, 
as was discussed in the previous section.  In the case of dark 
matter axions, assuming their mass is in the $10^{-6}$ to 
$10^{-4}$ eV range, it is useful to have the conversion process 
occur inside an electromagnetic cavity \cite{Sik83b,Sik85}. The 
cavity captures the photons produced and enhances the conversion 
process through resonance when one of the cavity modes equals the 
angular frequency of the axion signal.  

Cavity searches for galactic halo axions have been carried 
out at Brookhaven National Laboratory \cite{DePa87,Wuen89}, 
the University of Florida \cite{Hagm90b,Hagmth}, Kyoto 
University \cite{Mats91,Yamam99}, Lawrence Livermore 
National Laboratory \cite{Hagm98,Aszt01,Aszt02,Aszt04,
Duff05,Duff06,Aszt10}, the University of Washington 
\cite{Aszt11,Hosk11,Hosk16,Du18,Bout18,Brai19}, Yale 
University \cite{Brub16,Brub17,Zhon18}, the Universty 
of Western Australia \cite{McAl17a}, the INFN National 
Laboratory in Legnaro, Italy \cite{Ales19a} and the Center
for Axion and Precision Physics (CAPP) in Daejeon, Korea 
\cite{Lee20}.  New cavity detectors are under construction 
at CAPP \cite{Petr17,Seme19}, and at CERN \cite{Melc20}.  
A large cavity detector is proposed at the INFN National 
Laboratory in Frascati \cite{Ales19b}.  A summary of limits 
from axion dark matter searches using the cavity technique 
is shown in Fig. 1.

\subsection{The signal}

Axion to photon conversion occurs in large externally imposed electric 
$\vec E_0$ and/or magnetic $\vec B_0$ fields because the axion induced 
electric charge and current densities, Eqs.~(\ref{axc}), are sources of 
electromagnetic waves.  For non-relativistic axions, the
\begin{equation}
\vec j_a = - g~\vec B_0 \partial_t a
\label{ja}
\end{equation}
term in the current density is most relevant since
$\mid \partial_t a \mid >> \mid \vec\nabla a\mid$. 

Consider an electromagnetic cavity, of volume $V$, inside of which 
exists a large static magnetic field $\vec B_0 (\vec x)$, dielectric 
constant $\epsilon (\vec x)$ and magnetic permeability $\mu(\vec{x})$.  
We choose $\Phi = 0$ gauge and expand the vector potential into 
cavity eigenmodes:
\begin{equation}
\vec A (\vec x, t) = 
\sum_\alpha \vec e_\alpha (\vec x) \psi_\alpha (t)~~~\ .
\label{modex}
\end{equation}
In the limit of vanishing skin depth, the normalized mode functions 
$\vec e_\alpha (\vec x)$ satisfy:
\begin{eqnarray}
\vec\nabla\cdot (\epsilon \vec e_\alpha) &=& 0\nonumber\\
\vec\nabla\times({1 \over \mu}\vec\nabla\times\vec e_\alpha) - 
\epsilon \omega_\alpha^2 \vec e_\alpha &=& 0\nonumber\\
\hat{n}\times\vec e_\alpha\mid_S &=& 0\nonumber\\
\int_V d^3x~\epsilon(\vec{x})~
\vec e_\alpha(\vec{x})\cdot\vec e_\beta(\vec{x}) &=&
\delta_{\alpha\beta}~~~\ ,
\label{modprop}
\end{eqnarray}
where $S$ is the surface of the cavity volume and $\hat{n}$ the unit 
normal to the surface.  The $\omega_\alpha$ are the eigenfrequencies.  
In the absence of axions, the amplitudes $\psi_\alpha (t)$ satisfy
\begin{equation}
\Biggl( {d^2\over dt^2} + \gamma_\alpha {d\over dt} 
+ \omega_\alpha^2\Biggr) \psi_\alpha(t) = 0
\label{daho}
\end{equation}
where the term proportional to $\gamma_\alpha$ describes energy 
dissipation.  $Q_\alpha = {\omega_\alpha \over \gamma_\alpha}$ is 
the quality factor of the cavity in its $\alpha$-eigenmode.

We write the axion field as
\begin{equation}
a(\vec x,t) = Re (A~e^{-i \omega_a t})~~~~\ . 
\label{axfld}
\end{equation}
Its $\vec x$-dependence is ignored because the cavity size 
is generally of order ${1 \over m_a}$ whereas the de Broglie 
wavelength of halo axions is of order ${10^3 \over m_a}$.
Eq.~(\ref{axfld}) implies the local axion energy density:
\begin{equation}
\rho_a = {1 \over 2}\left((\partial_t a)^2 + 
(\vec{\nabla} a)^2 + m_a^2~a^2\right) =
{1 \over 2} m_a^2 |A|^2~~~~\ .
\label{axenden}
\end{equation}
In the presence of axions, the $\psi_\alpha(t)$ satisfy the 
equation of motion 
\begin{equation}
\Biggl( {d^2\over dt^2} + \gamma_\alpha {d\over dt} 
+ \omega_\alpha^2\Biggr) \psi_\alpha(t) = 
- g~\int_V d^3 x~\vec B_0 (\vec x)\cdot \vec e_\alpha (\vec x)~
Re ( -i \omega_a A e^{- i \omega_a t})~~~~\ ,
\label{drdaho}
\end{equation}
obtained by substituting Eqs.~(\ref{modex}) and (\ref{axfld}) into 
Eqs.~(\ref{inhMaxwa}), setting $\rho_{\rm el} = \vec{j}_{\rm el} 
= 0$, and using Eqs.~(\ref{modprop}).  The term describing 
energy dissipation was added by hand.  Up to transients, the 
solution of Eq.~(\ref{drdaho}) is 
\begin{equation}
\psi_\alpha (t) = g~\omega_a~
\Bigl(\int_V d^3x~\vec B_0\cdot\vec e_\alpha\Bigr) 
~Re \Bigl({i A~e^{-i \omega_a t} \over 
\omega_\alpha^2-\omega_a^2-i\gamma_\alpha \omega_a}\Bigr) ~~~~\ .
\label{emsol}
\end{equation}
The time-averaged power from axion conversion into the $\alpha$-mode 
of the cavity is therefore
\begin{eqnarray}
P_\alpha &=&
\gamma_\alpha \int_V d^3 x 
\left({1 \over 2} \epsilon \vec E_\alpha \cdot \vec E_\alpha + 
{1 \over 2\mu} \vec B_\alpha \cdot \vec B_\alpha \right) =
{\gamma_\alpha \over 2}[({d \psi_\alpha \over dt})^2 +
\omega_\alpha^2 \psi_\alpha^2]\nonumber\\
&=& {\gamma_\alpha \over 4}~
{g^2 \omega_a^2 (\omega_a^2 + \omega_\alpha^2)
\over (\omega_\alpha^2- \omega_a^2)^2 + \gamma_\alpha^2 \omega_a^2}
~|A|^2 	\left(\int_V d^3x \vec B_0(\vec x)\cdot 
\vec e_\alpha (\vec x)\right)^2~~~~\ .
\label{kpow}
\end{eqnarray}
The ratio of the energy of galactic halo axions to their energy 
spread is usually called the ``quality factor'' $Q_a$ of the 
axion signal.  Eq.~(\ref{halen}) indicates that $Q_a$ is of 
order $10^6$.  If $Q_a >> Q_\alpha$ and the axion signal falls 
at the center of the cavity bandwidth ($\omega_\alpha = \omega_a$), 
Eq.~(\ref{kpow}) implies \cite{Sik83b,Sik85,Krau85}
\begin{equation}
P_\alpha = 
g^2 \rho_a B_0^2 V C_\alpha {1\over m_a} Q_\alpha
\label{sigpow}
\end{equation}
where $B_0$ is a nominal magnetic field inside the cavity and
\begin{equation}
C_\alpha \equiv {1 \over B_0^2 V} 
\Biggl( \int_V d^3 x~\vec B_0 (\vec x) \cdot 
\vec e_\alpha (\vec x)\Biggr)^2
= {\Biggl( \int_V d^3 x \vec B_0 (\vec x) \cdot 
\vec E_\alpha (\vec x)\Biggr)^2 
\over B_0^2 V~\int_V d^3 x~\epsilon (\vec x) 
\vec E_\alpha (\vec x)\cdot \vec E_\alpha (\vec x)}~~~\ .
\label{formf}
\end{equation}
$C_\alpha$ expresses the coupling strength of mode $\alpha$ to 
galactic halo axions, and is called its ``form factor".

The conversion factor between mass and frequency is ($\hbar = c =1$) 
\begin{equation}
10^{-5}~{\rm eV} = 2 \pi (2.418~{\rm GHz})~~~~~\ .
\label{conv}
\end{equation}
The GHz region is good hunting ground since $10^{-5}$ eV is 
a likely mass for axion dark matter. It is also convenient 
since an electromagnetic cavity whose fundamental mode has 
GHz frequency has size of order GHz$^{-1}$ = 30 cm.  Expressed 
in practical units, Eq.~(\ref{sigpow}) is
\begin{eqnarray}
P_\alpha &=&
1.34 \cdot 10^{-26}~{\rm Watt} \Biggl({g_\gamma \over 0.36}\Biggr)^2
\Biggl({\rho_a \over {1\over 2}~10^{-24}~{\rm gr/cm^3}}\Biggr)
\cdot\nonumber\\
&~&~~~~~~~~\cdot \Biggl({B_0 \over 8~{\rm Tesla}}\Biggr)^2
\Biggl({V \over {\rm m}^3}\Biggr)~C_\alpha~
\Biggl({m_a \over 2\pi~{\rm GHz}}\Biggr) Q_\alpha~
\left({6 \cdot 10^{15}~{\rm eV}^2 \over f_a~m_a}\right)^2~~~\ .
\label{sigpowp}
\end{eqnarray}
The last factor in Eq.~(\ref{sigpowp}) is approximately one in view 
of Eq.~(\ref{mass}).  We include it here so that the numerical 
prefactor in Eq.~(\ref{sigpowp}) may be written with precision 
unmarred by the uncertainty in the relationship between $m_a$ 
and $f_a$.

Because the axion mass is unknown, the cavity should be tunable.
In all experiments so far, tunability is achieved by inserting 
movable metal and/or dielectric posts inside the cavity.
For the sake of definiteness, consider a cylindrical cavity 
in which exists a longitudinal homogeneous magnetic field
$\vec B_0 = B_0 \hat z$~ and a $z$-independent dielectric 
constant $\epsilon(x,y)$.  By “cylindrical” cavity we mean one 
that is invariant under translations in the $\hat{z}$-direction, 
except for the endcaps \cite{Jack99}.  The cross-sectional shape 
is arbitrary.  Only the transverse magnetic (TM) modes of a 
cylindrical cavity couple to the axion field.  Indeed the 
transverse electric (TE) and transverse electromagnetic (TEM) 
modes have vanishing form factor since their electric fields 
are perpendicular to $\vec{B}_0$.  TM modes are labeled by 
three integers $\alpha = (l,n,p)$ with 
$\hat{z} \cdot \vec e_{lnp} \propto \cos {p \pi z \over L}$ and
$p = 0,1,2, ...$, where $L$ is the length of the cavity.  Only 
the TM$_{ln0}$ have non-zero form factor.  For TM$_{ln0}$,
\begin{eqnarray}
\vec E_{ln0} (\vec x) &=& \hat z~\phi_{ln}(x,y)\\
\Biggl( {\partial^2 \over \partial x^2} + 
{\partial^2 \over \partial y^2} &+& \epsilon(x,y) 
\omega_{ln0}^2\Biggr) \phi_{ln} = 0\\
\phi_{ln} \mid_S &=& 0~~~~\ .
\label{longhom}
\end{eqnarray}
Here we assumed that the magnetic permeability $\mu = 1$. 

For a circular cross-section of radius R and $\epsilon = \mu = 1$, 
\begin{equation}
\phi_{ln} \propto J_l\left(x_{ln}{\rho \over R}\right) 
e^{il\theta}~~,~~\omega_{ln0} = {x_{ln} \over R}~~,~~
C_{ln0} = {4 \over (x_{on})^2} \delta_{l0}
\label{cyl}
\end{equation}
where $(\rho, \theta)$ are axial coordinates and $x_{ln}$ is the 
$n^{th}$ zero of the Bessel function $J_l (x)$.
In particular, $C_{010} = 0.69$.  

For a rectangular cross-section 
\begin{eqnarray}
\phi_{ln} &\propto& \sin \left({l \pi x \over L_x}\right) 
\sin \left({n \pi y \over L_y}\right)\nonumber\\
C_{ln0} &=& {64\over \pi^4 l^2 n^2}~~~~~~~~~~~~ 
{\rm for}~l~{\rm and}~n~{\rm odd}
\nonumber\\
&=&~~0~~~~~~~~~~~~~~~~~{\rm otherwise}~~~~\ ,
\label{rect}
\end{eqnarray}
where $L_x$ and $L_y$ are the transverse sizes.

Eqs.~(\ref{cyl}) and (\ref{rect}) show that, when $\vec{B}_0$ 
is homogeneous, the lowest TM mode has the strongest coupling.
Indeed the electric field profiles $\phi_{ln} (x,y)$ of the 
higher TM modes have nodes, so that the contributions to the 
form factor from different regions of the cavity tend to cancel 
each other out.

\subsection{Signal to noise and search rate}

The microwave power from axion conversion is coupled out through a
small hole in the cavity walls and brought to the front end of a 
microwave receiver.  The quality factor $Q$ of the cavity may be 
written 
\begin{equation}
{1\over Q} = {1 \over Q_{\rm h}} + {1\over Q_{\rm w}}~~~\ .
\label{qual}
\end{equation}
In Eq.~(\ref{qual}) and henceforth we are suppressing 
the label $\alpha$ that indicates the mode dependence.
$\gamma_{\rm h} = {\omega \over Q_{\rm h}}$ is the contribution 
to $\gamma$ from emission through the hole and $\gamma_{\rm w} = 
{\omega \over Q_{\rm w}}$ the contribution from absorption by the 
cavity walls.  The maximum power that can be brought to the microwave 
receiver is $P_{\rm d} = {Q \over Q_{\rm h}} P$ where $P$ is given 
by Eq.~(\ref{sigpowp}).  

Because the cavity volume is permeated by a strong magnetic 
field, the cavity walls are ordinarily made of normal metal, 
although superconducting material can be used for the side 
walls \cite{VanB15}\cite{Ales19a,Ahn19}. At low temperatures 
$(T \lesssim$ few K) and frequencies $f$ in the GHz range, a 
cavity made of high purity copper has $Q_{\rm w} \sim 2 \cdot 10^5$.  
In that case, the cavity bandwidth $B_{\rm c} \equiv f/Q$ is larger 
by a factor 10 or so than the bandwidth $B_a \equiv f/Q_a$ of 
the axion signal.

The axion signal is searched for by tuning the cavity to successive
frequencies, separated by a cavity bandwidth $B_{\rm c}$ or less, 
and by integrating for an amount of time $t$ at each tune. To 
proceed at a reasonably fast rate, e.g. to cover a factor 2 in 
frequency in one year, the amount of time $t$ spent at each tune 
is of order ${1 \over 3} {{\rm year} \over Q} \sim$ 100 seconds.  
${1 \over 3}$ is an assumed duty factor.  During each time interval 
$t$, the power leaving the cavity is amplified by a receiver, 
shifted down in frequency by mixing with one or more local 
oscillators, digitized and spectrum analyzed.  The signal can 
be analyzed with different resolutions.  For example, the Axion 
Dark Matter Experiment (ADMX) \cite{Aszt01} has a 125 Hz medium 
resolution channel, hereafter called MedRes, obtained by co-adding 
many ($\sim 10^4$) short spectra taken during the measurement 
integration time $t$, and a high resolution channel, hereafter 
called HiRes, with 0.01 Hz resolution, the highest possible 
when $t = 100$ sec.  Any resolution less than $\delta f = 1/t$ 
can be obtained by averaging the highest resolution spectrum. 

When an axion signal is found, the energy spectrum of halo 
axions will immediately become known in great detail.  So it 
is interesting to try and anticipate what that spectrum will 
look like.

As do all other cold effectively collisionless dark matter 
candidates, axions lie on a thin continuous 3-dimensional 
hypersurface in phase space.  This hypersurface wraps and 
folds but does not break.  This fact implies that, at any 
location and any time, dark matter axions form a discrete 
set of flows, each with a well defined density and velocity 
vector \cite{Sik92,Nat05}.   Predictions for the velocity 
vectors and densities of the dark matter flows at our location 
in the Milky Way halo have been made \cite{Sik95,Sik97,Duff08}.  
Discrete flows are also produced when satellites, such as the 
Sagittarius dwarf galaxy \cite{Newb02,Maje03}, are tidally 
disrupted by the gravitational field of the Milky Way.  
Discrete flows are called ``streams" in this context. Each 
flow or stream at our location produces a narrow peak in 
the cavity detector, since the axions in the flow or stream 
have well defined kinetic energy in the laboratory frame.  
The peaks have a daily frequency modulation due to the 
Earth's rotation and an annual frequency modulation due 
to the Earth's orbital motion \cite{Ling04}.  During 100 
seconds of data taking, the frequency of a peak at 1 GHz 
shifts at most by $10^{-2}$ Hz due to the Earth's rotation, 
and stays therefore within the $10^{-2}$ Hz highest possible 
resolution bandwidth.  Because of each peak's diurnal and 
annual modulations it is possible to measure the velocity 
vector of the associated flow or stream.  Searching for 
narrow peaks increases the sensitivity of the cavity 
experiment provided a sufficiently large fraction of 
the local halo density is in one or more cold flows 
\cite{Duff05,Duff06}.  

The output of the receiver chain is mostly noise, thermal noise 
from the cavity plus electronic noise from the receiver.  If the 
axion signal frequency $\omega_a$ falls within the cavity bandwidth
$B_c$, the output spectrum has extra power within the axion signal 
bandwidth $B_a$.  The ratio $s \over n$ of the signal to a $1~\sigma$ 
fluctuation in the noise within a $B_a$ bandwidth is given by 
Dicke's radiometer equation:
\begin{equation}
{s \over n} = {P_{\rm d} \over T_{\rm n}} \sqrt{{t\over B_a}}
\label{radio}
\end{equation}
where $T_{\rm n}$ is the total noise temperature.  Each candidate 
peak is checked by taking more data.   If the peak is a statistical 
fluctuation in the noise, it averages away.  If a peak does not 
average away, it is a signal of something but most likely not an 
axion signal. The non-statistical peaks found so far have all been 
the result of leakage of microwave power into the cavity from the 
environment of the experiment. Such spurious signals are referred 
to as ``environmental peaks".  It is straightforward to distinguish 
an axion signal from an environmental peak by exploiting the following
properties:  1) an axion signal does not depend on the degree of 
microwave isolation of the cavity, 2) it cannot be picked up by 
a simple antenna outside the apparatus, 3) its dependence on the 
central frequency $\omega_\alpha$ of the cavity mode is a Lorentzian 
[see Eq.~(\ref{kpow})], and 4) it is proportional to $B_0^2$.

In a search, every $p \sigma$ candidate peak is checked to see whether 
or not it is due to galactic halo axions.  $p$ should be chosen neither 
too high nor too low.  If too high, the search loses sensitivity.  If too 
low, an excessive amount of time is wasted investigating fluctuations in 
the noise.  In the ADMX MedRes channel, the noise is Gaussian-distributed 
because each spectrum is the sum of many independent spectra.  There is 
therefore a 2.3\% chance that the background fluctuates downward by $2\sigma$ 
or more in each $B_a$-wide bin.  Hence, to put a 97.7\% confidence level limit 
on the product $g_\gamma^2 \rho_a$, the $s/n$ ratio must be $p+2$ and every 
candidate peak larger than $p \sigma$ ruled out as an axion signal.  Through 
Eq. (\ref{radio}), this determines the minimum measurement integration time 
$t$ per cavity bandwidth $B_{\rm c}$  and hence the maximum rate at which the 
search may proceed in frequency space:
\begin{equation}
{df \over dt} \simeq {B_{\rm c}\over t} = 
{Q_a \over Q} {1\over (s/n)^2} 
\left({P_{\rm d} \over T_{\rm n}}\right)^2
= {1\over (s/n)^2} \left({P_0\over T_n}\right)^2 Q_a 
{Q^3 \over Q_{\rm h}^2}
\label{srate}
\end{equation}
where we used $P_{\rm d} = {Q \over Q_{\rm h}} P$ and defined 
$P_0 \equiv P/Q$.  We may choose $Q_{\rm h}/Q_{\rm w}$ to maximize 
the search rate.  One readily finds that the optimum occurs at 
$Q = {1 \over 3} Q_{\rm w}$, in which case
\begin{eqnarray}
{d \ln f \over dt} &\simeq& {1 \over f} 
{1\over (s/n)^2} \left({P_0\over T_n}\right)^2 Q_a
\left({2 \over 3}\right)^2 Q\nonumber\\
&\simeq& {27 \over {\rm year}} 
\left({4 \over s/n}\right)^2 \left({V \over {\rm m}^3}\right)^2 
\left({B_0 \over 8~{\rm T}}\right)^4 C^2\cdot\nonumber\\
&\cdot& \left({g_\gamma\over 0.36}\right)^4 
\left({\rho_a \over {1 \over 2}~10^{-24}~{\rm gr/cm}^3}\right)^2 
\left({1~{\rm K} \over T_{\rm n}}\right)^2 
\left({f \over {\rm GHz}}\right) 
\left({Q \over {1 \over 3} \cdot 10^5}\right) 
\left({Q_a \over 10^6}\right)\ ~~\ .
\label{lograte}
\end{eqnarray}
At GHz frequencies, electronic noise temperatures of order 2 K are 
achieved by using cooled Heterostructure Field-Effect Transistors 
(HFET) as microwave amplifiers \cite{Brad99}. The cavity is then 
cooled to liquid He temperatures so that the thermal noise qualitatively 
matches the electronic noise.  This was the approach of the earliest 
experiments \cite{DePa87,Wuen89,Hagm90b}.  The experiment at Kyoto 
University \cite{Mats91,Ogaw96} explored the use of a beam of Rydberg 
atoms to detect the microwave photons from axion conversion.  The more 
recent experiments \cite{Aszt10,Brub16} use Superconducting Quantum 
Interference Devices (SQUIDs) \cite{Muck98,Muck03} or Josephson 
Parametric Amplifiers (JPAs) \cite{AlKe17}.  These devices approach 
the so-called 
`quantum limit' defined by a noise temperature equal to the angular 
frequency in units where $k_B = \hbar = 1$:
\begin{equation}
T_n = \omega = 48~{\rm mK}\left({f \over {\rm GHz}}\right)~~\ .
\label{TnOm}
\end{equation}
To reduce thermal noise accordingly, the cavity is cooled to 
temperatures in the 100 mK range by a dilution refrigerator.
The sensitivity of microwave photon detection for axion 
haloscopes may be boosted further by `vacuum squeezing' 
\cite{Maln19} or single photon counting \cite{Lamo13,Kuzm18}
techniques.

We may use Eq.~(\ref{lograte}) to estimate the search rate in the 
ADMX HiRes channel as well.  When searching for peaks of width less 
than 0.01 Hz, the HiRes channel is sensitive to cold axion flows 
with quality factor $Q_a \gtrsim 10^{11}$ at GHz frequencies. A 
large increase in $Q_a$ is the main motivation for the HiRes channel.  
However, it is partially offset by decreases in other parameters.  
The relevant $\rho_a$ is the density of the largest flow that 
produces a peak of width less than 0.01 Hz.  If $\delta v$ is the 
velocity dispersion of a flow of cold axions, its energy dispersion 
is $\delta E \sim m_a v \delta v$ where $v \sim 10^{-3}$ is the flow 
velocity.  Hence $Q_a > 10^{11}$ requires $\delta v <$ 3 m/s. An
additional consideration is that the noise is exponentially 
distributed in the HiRes channel\cite{Duff05,Duff06} whereas 
it is Gaussian distributed in the MidRes channel.  Because 
each HiRes spectrum has on the order of $Q_c/Q_a \sim 10^6$ bins, 
the threshold for a peak to be admitted as a candidate signal has 
to be set very high. The signal to noise ratio for a practical 
HiRes search was found to be of order 20 \cite{Duff05,Duff06}.

Ref. \cite{Chau19} studies the sensitivity of a cavity
haloscope that searches for a signal both inside and
outside the cavity's resonant bandwidth and optimizes
the frequency-integrated sensitivity of such a search.

\subsection{Cavity design}

After many years of improvement, the cavity technique has 
reached sufficient sensitivity to detect dark matter axions 
even with the weaker DFSZ value of the electromagnetic 
coupling \cite{Du18,Bout18}.  The next challenge is to 
extend the technique to the widest possible axion mass 
range.

A large superconducting solenoid is the type of magnet that has 
been most commonly used for the experiment, although dipole, 
wiggler and toroidal magnets have specific advantages and are 
being considered as well \cite{Bake12,Mice15,Melc18}.  At first, 
the bore of a solenoidal magnet is filled with a single cylindrical
cavity.  Its resonant frequency may be tuned upwards approximately 
50\% by moving a metal post transversely from the side of the 
cavity to its center, and 30\% downwards by similarly moving 
a dielectric rod \cite{Hagm90a}.  Provided longitudinal symmetry 
is maintained (the rods must extend from endcap to endcap and 
remain parallel to the cavity walls), the form factor $C$ stays 
of order one over the tuning range.  If longitudinal symmetry 
is broken, the mode may become localized in a small part of 
the cavity.  The form factor is then severely degraded.

To reach higher frequencies, one may fill up the volume available 
inside a magnet bore with many identical cavities and power-combine 
their outputs \cite{Hagm90a,Hagmth}.  A two-port Wilkinson power 
combiner produces the output voltage 
${1 \over \sqrt{2}} (a + b~e^{-i \varphi})~e^{-i \omega t}$ 
when the input voltages are 
$a~e^{-i \omega t}$ and $b~e^{-i \varphi~- i \omega t}$.  
The power combiner adds the axion signals from the cavities 
provided that they are equal in magnitude and in phase.  Thus
one may power-combine the outputs of identical cavities provided 
that the largest distance between the cavities is less than the 
de Broglie wavelength ($\sim 10^{+3}/ m_a$) of galactic halo 
axions, that the cavities are in tune, and that the phase-shifts 
between the individual cavities and the power combiner are identical.
Since the noise in the different cavities is uncorrelated in phase,
the noise temperature at the output of the power combiner is the 
average of the noise temperatures at its input ports.  Properly 
built multi-cavity arrays have effective form factors of order one 
and allow, at the cost of engineering complexity, the upward extension 
of the frequency range over which a galactic halo axion search can be 
carried out with a given magnet.  

Alternatively, one may reach higher frequencies by dividing a cavity 
into cells separated by metal vanes \cite{Ster16,Jeon18a,Jeon18b}. 
Such multi-cell cavities must be carefully designed to avoid mode 
crowding and mode localization.  Ref. \cite{Kim19} presents a 
design achieving a large form factor for the TM$_{030}$ mode 
of a cylindrical cavity by inserting dielectric vanes.  Another 
proposal is to introduce materials that produce a plasma frequency 
for the electromagnetic field inside the cavity \cite{Laws19}.

The frequency range of cavity haloscopes can also be extended 
upward by controlling the spatial variation of the magnetic 
field inside the cavity, or by introducing dielectric plates 
to control the mode structure.   These two approaches 
are discussed in Section 5.1.  

Refs. \cite{Sik10,Berl19,Lase19} propose to search for 
axion dark matter in an electromagnetic cavity which is 
driven with input power instead of being permeated by a 
magnetic field.  The relevant process is $a + \gamma \rightarrow
\gamma^\prime$ where $\gamma$ is a microwave photon in the 
mode that is driven by input power and $\gamma^\prime$ is 
a microwave photon, in another mode of the cavity, to be 
detected as signal.  This approach can be pursued using 
an optical cavity as well \cite{Meli09}.

Ref. \cite{Gory19} proposes to search for dark matter axions 
by detecting the phase noise induced by the oscillating axion 
field in driven cavity modes separated in frequency by the 
axion mass.  An experiment of this type is reported on 
in ref. \cite{Thom19}.

\section{Other approaches to axion dark matter detection}

The cavity technique works well for axion masses between 
perhaps $10^{-7}$ eV and a few times $10^{-5}$ eV but not, 
at any rate, for all masses that dark matter axions may 
plausibly have.  So there is good motivation to look for 
alternatives.  Over the years, different approaches have 
been proposed which collectively address the whole QCD 
axion mass range, from $10^{-2}$ eV to $10^{-12}$ eV.  
They are the topic of this Section.  Several methods
were anticipated in ref. \cite{Voro95} and rediscovered
later.

\subsection{Wire arrays and dielectric plates}

The conversion of axions to photons in a magnetic field 
can be enhanced by controlling the spatial variation of 
the magnetic field or by introducing dielectric plates to 
modify the mode structure of the electromagnetic field.  
In such schemes, it is likely useful to introduce a 
cavity as well. 

\subsubsection{Wire arrays}

The differential cross-section for axion to photon conversion 
in a static magnetic field is given in Eq.~(\ref{convxs2}).  
Multiplying by the axion flux $\beta_a n_a$ and integrating 
over solid angles yields the conversion rate
\begin{equation}
R = {g^2 n_a \over 16 \pi^2 \epsilon}
\int d^3 k ~\delta({k \over \sqrt{\epsilon\mu}} - \omega)
\sum_{\lambda = 1,2}
\bigg| \int_V d^3 x~e^{i(\vec k_a -\vec k) \cdot \vec x}~
\hat{e}_\lambda(\hat{n}) \cdot \vec B_0 (\vec x)\bigg|^2~~~\ .
\label{rate}
\end{equation}
To maximize $R$ for given field strength and volume, the 
magnetic field should be made inhomogeneous on the length scale 
set by the momentum transfer $\vec{q} = \vec{k} - \vec{k}_a$.  
Since dark matter axions are non-relativistic, $k_a << k$ and 
hence $q \simeq k = \sqrt{\epsilon\mu} \omega \simeq 
\sqrt{\epsilon\mu} m_a$.  So the inhomogeneity length scale 
should be of order ${1 \over \sqrt{\epsilon \mu} m_a}$.  

In view of this, it was proposed to build an array of superconducting 
wires embedded in a dielectric medium transparent to microwave 
radiation \cite{Sik94}.  Magnetic fields are produced by passing 
electric currents through the wires.  The dielectric medium keeps 
the wires in place.  We set $\mu = 1$ here for simplicity.

A possible realization consists of wires parallel to the $y$-axis
whose intersections with the $xz$ plane form a regular lattice with
lattice constant $d$.  The wires intersect the $xz$ plane at
$(n_x d, n_z d)$ where $n_x$ and $n_z$ are integers that range
from $-{L_x \over 2d}$ to $+{L_x \over 2d}$, and $-{L_z \over 2d}$
and $+{L_z \over 2d}$ respectively. $L_x$ and $L_z$ are the dimensions
of the detector in the $\hat{x}$ and $\hat{z}$ directions.  The  
currents $I(n_x,n_z)$ in the wires are chosen to produce a particular
magnetic field profile.  For example
\begin{equation}
I(n_x,n_z) = I(n_z) = I_0 \sin(\kappa n_z d)
\label{wirecurr}
\end{equation}
produces the magnetic field
\begin{equation}
\vec{B} = - \hat{x} {I_0 \over \kappa d^2} \cos(\kappa z)
\label{wireB}
\end{equation}
in the limit $L \rightarrow \infty$ and $d \rightarrow 0$.  In
practice the magnetic field deviates from Eq.~(\ref{wireB})
because of finite size $L$ and finite lattice constant $d$  
effects. Such deviations, which can be calculated without 
much difficulty, are ignored here for simplicity.

For the sake of definiteness we assume Eq.~(\ref{wireB}) within
a rectangular volume $V = L_x L_y L_z$.  Because the photons 
produced are polarized in the direction $\hat{x}$, perpendicular 
to the wires, the effect of the wires on their propagation is 
minimized.  In using Eq.~(\ref{rate}) we are assuming that the 
photons propagate as if the wires were absent.  For the magnitude 
squared of the space integral in Eq.~(\ref{rate}) we have
\begin{eqnarray}
\bigg| \int_V d^3x~
e^{- i \vec{q}\cdot\vec{x}} B_0 \cos(\kappa z) \bigg|^2 
&=& (2 \pi)^2 \delta_{L_x}(q_x) \delta_{L_y}(q_y) L_x L_y B_0^2
\cdot\nonumber\\
&\cdot& \left({\sin((q_z + \kappa)L_z/2) \over q_z + \kappa} +
{\sin((q_z - \kappa)L_z/2) \over q_z - \kappa}\right)^2
\label{mssi}
\end{eqnarray}
where $\delta_L(q)$ is a Dirac delta-function spread over a width of 
order $1/L$. Resonant conversion is obtained for $q_z = \pm \kappa$. 
Since $\vec{k} \simeq \vec{q}$, the photons are emitted in the $\pm 
\hat{z}$ direction and can therefore be focused by mirrors onto one 
or two microwave receivers.

The wavevector $\kappa$ of the current configuration can be changed 
to tune the detector over a range of possible axion masses.  The 
detector bandwidth is $\Delta k_z \simeq {\pi \over L_z}$ 
whereas the bandwidth of the axion signal is 
$\Delta k_{az} \simeq 2 \cdot 10^{-3} m_a$.  The conversion rate 
is obtained by inserting Eq.~(\ref{mssi}) into Eq.~(\ref{rate})
and carrying out the integral over $\vec{k}$.  Provided the axion 
signal falls entirely within the bandwidth of the detector, the 
signal power is 
\begin{eqnarray}
P &=& m_a R =
{g^2 \over 8 \sqrt{\epsilon}} V L_z B_0^2 \rho_a \nonumber\\
&=& 2 \cdot 10^{-25}~{\rm W} \left({V L_z \over {\rm m}^4}\right)
\left({B_0 \over 8~{\rm T}}\right)^2 
\left({g_\gamma \over 0.36}\right)^2
\left({m_a \over 10^{-5}~{\rm eV}}\right)^2 {1 \over \sqrt{\epsilon}}
\left({\rho_a \over {1 \over 2} \cdot 10^{-24}~{\rm g/cm}^3}\right)\ .
\label{inhmagpow}
\end{eqnarray}
The discussion of the signal to noise and search rate is similar 
to that for the cavity detector in Section 4.2, and need not be 
repeated here.

The above design is convenient for signal calculation but not 
so convenient for construction and operation.  In practice one 
wishes to minimize the number of connections between wires.  A 
possible way to do this is to deform the above rectangular array 
into a cylinder so that all the wires at given $n_z$ combine to 
form a spiral.  The spiral could be a NbTi strip etched by 
photolithographic techniques onto a low loss insulating sheet.  
The sheets would then be stacked to form the body of the 
detector.  

Comparing Eqs.~(\ref{sigpow}) and (\ref{inhmagpow}), the 
expression for the conversion power of a wire array is seen 
to be similar to that of a cavity haloscope but with the 
product $CQ$ of the cavity form and quality factors replaced 
by $L_z m_a/8 \sqrt{\epsilon}$.  If the axions have velocity 
dispersion $\delta v \sim 10^{-3}$, the requirement $\Delta k_z 
\gtrsim \Delta k_a$ implies $L_z m_a \lesssim 1,600$.  When 
searching for low velocity dispersion flows, such as the 
Big Flow of Eq.~(\ref{BFv}), $L_z m_a$ can be made much
larger.  However the detector must in that case be kept 
aligned with respect to a particular flow.

It is generally advantageous to place the wire array inside
an electromagnetic cavity \cite{Rybk15}.  A small wire array 
was built at the University of Washington and placed in an 
open Fabry-Perot resonator, in an experiment called ORPHEUS 
\cite{Rybk15}.  A schematic drawing of such a setup is shown 
in Fig. 2.  The detector is now in effect a cavity haloscope 
and the considerations of Section 4 apply to it.  If the 
electric field for the Fabry-Perot mode is 
\begin{equation}
\vec{E}_\omega = \hat{x} E_\omega \cos(\kappa z)
\label{FPE}
\end{equation}
within the volume of the wire array, and the magnetic field 
is as in Eq.~(\ref{wireB}), the conversion power is given by  
Eq.~(\ref{sigpow}) with $V$ being the volume of the wire 
array and $C = 0.5~F$, where $F$ is the fraction of the 
distance between the mirrors that is occupied by the wire 
array.  The detector is tuned by changing the distance 
between the mirrors.  In the ORPHEUS detector, the 
distances between the wire planes were changed 
proportionately.

\subsubsection{Dielectric plates}

Instead of making the magnetic field inhomogeneous on the 
$q^{-1} \sim m_a^{-1}$ length scale, one may instead have 
the dielectric constant vary on that length scale 
\cite{Morr84,Cald17,McAl18,Ioan17,Mill17,Bary18}.  MADMAX  
\cite{Brun19} is a proposed experiment using dielectric 
plates, although in a different manner from the setup
described below.  MADMAX evolved from an earlier broadband 
axion dark matter detection scheme, called the dish antenna 
\cite{Horn12b}.

Here we consider a stack of parallel plates of thickness $d$ 
and dielectric constant $\epsilon$ placed in a Fabry-Perot 
resonator, as shown schematically in Fig.~\ref{dielpl}.  
The distance $D$ between the plates is chosen to be the 
half-wavelength ${\pi \over m_a}$ in vacuum of the photons 
produced by axion conversion, whereas the plate thickness 
$d$ is chosen to be of order the half-wavelength 
${\pi \over \sqrt{\epsilon} m_a}$ of those photons in the 
dielectric material.  The intended electric field profile 
of the electromagnetic mode in the region occupied by the 
dielectric plates is 
\begin{eqnarray}
\vec{E}_\omega &=& \hat{x} E_\omega \sin(\omega (z - z_j))  
~~~~~~~~~~{\rm for}~~ 0 \leq z - z_j \leq D \nonumber\\
&=& \hat{x} {E_\omega \over \sqrt{\epsilon}} 
\sin(\sqrt{\epsilon}\omega (z - z_j))
~~~~~~{\rm for} - d \leq z - z_j \leq 0
\label{dpE}
\end{eqnarray}
where the $z_j$ are the positions of the right faces of 
the plates; see Fig.~\ref{dielpl}.  With this electric field 
profile and a unifrom magnetic field $\vec{B}_0 = B_0 \hat{x}$,
the conversion power is given by Eq.~(\ref{sigpow}) with $V$ 
being the volume of the stack of plates (including the spaces 
between plates) and 
\begin{equation}
C = {8 \over \pi^2}~
{(\epsilon - 1)^2 \over \epsilon (\sqrt{\epsilon} + 1)^2}~F~\ ,
\label{dpC}
\end{equation}
where $F$ is the fraction of the distance between the mirrors 
that is occupied by the stack of plates. Some materials (e.g. 
Al$_2$O$_3$) have high dielectric constant ($\epsilon \sim 10$) 
but low dielectric losses ($\tan \delta_e \sim 10^{-4}$). The 
mirrors, if placed outside the magnetic field region, can be 
made of superconducting material so that their contribution 
to dissipative losses is small.

\subsection{Magnetic resonance}

Ignoring the small CP violating term shown explicitly in 
Eq.~(\ref{intfa2}), the interaction energy of the axion 
with a non-relativistic electron is 
\begin{equation}
H_{a\bar{e}e} = {g_e \over 2 f_a}
(\vec{\nabla} a \cdot \vec{\sigma} +
\partial_t a~{\vec{p} \cdot \vec{\sigma} \over m_e})
\label{nonrelf}
\end{equation}
where $\vec{p}$ is the electron momentum, $m_e$ its mass
and $\vec{S} = {1 \over 2} \vec{\sigma}$ its spin.  The 
first term in Eq.~(\ref{nonrelf}) is similar to the coupling 
of a magnetic field to electron spin. The effective magnetic 
field associated with a gradient in the axion field is
\begin{equation}
\vec{B}_{\rm eff} = 
- {g_e \over \gamma_e f_a} \vec{\nabla} a
\label{Beff1}
\end{equation}
where $\gamma_e$ is the electron gyromagnetic ratio.  The 
axion has analogous interactions (\ref{intfa2}) with quarks.  
We therefore expect an interaction energy of the axion field 
with nuclear spin 
$\vec{I}$ 
\begin{equation}
H_{a\bar{N}N} = {g_N \over f_a} \vec{\nabla}a  \cdot \vec{I}
\label{nucc}
\end{equation}
where the $g_N$ are dimensionless couplings of order one that
are determined by nuclear physics in terms of $g_p$ and $g_n$
~\cite{Stad15}.  Eqs.~(\ref{nonrelf}) and (\ref{nucc}) suggest 
that one may search for dark matter axions using magnetic resonance 
techniques.  Refs. \cite{Barb89,Barb17} proposed to detect the 
power from axion to magnon conversion in a medium containing a 
high density of aligned electron spins.  Refs. \cite{Grah13,Budk14} 
proposed to detect the transverse magnetization induced by the 
axion field onto a sample of aligned nuclear spins.  

Let us briefly recall basic aspects of magnetic resonance 
\cite{Kitt68}.  A macroscopic sample of particles with spin 
$\vec{I}$ and magnetic moment
\begin{equation}
\vec{\mu} = \gamma \vec{I}
\label{magmom}
\end{equation}
is polarized in a static magnetic field $\vec{B_0} = B_0 \hat{z}$,
or by some other means, resulting in a magnetization $M_0 \hat{z}$.  
We use $\vec{I}$ to represent electron spin ${1 \over 2}\vec{\sigma}$ 
or nuclear spin, whichever applies.  In addition to $\vec{B}_0$, a 
weak transverse time-dependent magnetic field
$\vec{B}_\perp = \hat{x} B_x(t) + \hat{y} B_y(t)$ is applied.  The
transverse components of the magnetization satisfy the Bloch equations 
\begin{eqnarray}
{d M_x \over dt} &=& \gamma (\vec{M} \times \vec{B})_x - {1 \over t_2} M_x
= - \gamma M_0 B_y + \gamma B_0 M_y - {1 \over t_2} M_x \nonumber\\
{d M_y \over dt} &=& \gamma (\vec{M} \times \vec{B})_y - {1 \over t_2} M_y
= + \gamma M_0 B_x - \gamma B_0 M_x - {1 \over t_2} M_y
\label{Bloch}
\end{eqnarray}  
where $t_2$ is the transverse relaxation time.  When $\vec{B}_\perp = 0$, 
an initial transverse magnetization precesses about the $z$-axis with 
angular frequency $\omega_0 = - \gamma B_0$ and decays in a time $t_2$.
For the sake of definiteness, we assume that the $\hat{z}$-axis is 
chosen so that $\omega_0 > 0$.   If the transverse field has the 
form
\begin{equation}
\vec{B}_\perp(t) = 
B_\perp (\hat{x} \cos \omega t + \hat{y} \sin \omega t)~~\ ,
\label{oscmag}
\end{equation}
the sample acquires in steady state the transverse magnetization
\begin{equation}
\vec{M}_\perp = M_\perp [\hat{x} \cos(\omega t - \phi)
+ \hat{y} \sin(\omega t - \phi)]
\label{tranmag}
\end{equation}
with
\begin{equation}
\tan \phi =  {1 \over (\omega_0 - \omega) t_2}
\label{tanphi}  
\end{equation}
and
\begin{equation}
M_\perp = {\gamma M_0 t_2 B_\perp \over 
\sqrt{1 + (\omega_0 - \omega)^2 t_2^2}}~~\ .
\label{tranmagn}
\end{equation}
On resonance ($\omega = \omega_0$) the transverse magnetization has 
its maximum magnitude $\gamma M_0 t_2 B_\perp$ and its phase is $\pi/2$ 
behind that of $\vec{B}_\perp$. 

We now consider a magnetized sample bathed in a flow of axions 
described by the field
\begin{equation}
a(\vec{x}, t) = A \sin(\vec{k}\cdot\vec{x} - \omega t)~~~\ ,
\label{trax}
\end{equation}
with $\omega = \sqrt{m_a^2 + \vec{k}\cdot\vec{k}}
\simeq m_a + {k^2 \over 2 m_a}$.  The energy density 
of such a flow is 
\begin{equation}
\rho_a = {1 \over 2} \omega^2 A^2~~\ .
\label{axden}
\end{equation}
Comparing (\ref{nucc}) with the interaction $H_B = - \gamma 
\vec{I}\cdot\vec{B}$ of a magnetic field with spin, the axion 
field (\ref{trax}) is seen to produce an effective tranverse 
magnetic field
\begin{equation}
\vec{B}_{\perp, {\rm eff}} = - {1 \over \gamma} {g_N \over f_a}
A \vec{k}_\perp \cos(\omega t)
\label{efftrH}
\end{equation}
where $\vec{k}_\perp = \vec{k} - \hat{z} (\hat{z}\cdot\vec{k})$.
In contrast to Eq.~(\ref{oscmag}), it drives the transverse 
magnetization in only one spatial direction.  Also, the field 
due to dark matter axions in the Milky Way halo does not have 
the infinite coherence time implied by Eq.~(\ref{oscmag}) or 
(\ref{trax}).  The direction and time-dependence of  
$\vec{B}_{\perp, {\rm eff}}$ depends on the model of the 
galactic halo. Two contrasting proposals were mentioned in 
Section 2.  In the isothermal model, the energy dispersion 
$\delta \omega \simeq 10^{-6} m_a$, and hence the coherence 
time $t_c = 1/ \delta \omega \simeq 0.16~{\rm sec}({\rm MHz}/\nu_a)$ 
where $\nu_a$  is the frequency associated with the axion mass: 
$m_a = 2 \pi \nu_a$.  In the caustic ring model, the local dark 
matter density is dominated by a single flow, the Big Flow, with 
velocity dispersion $\delta v \lesssim$ 70 m/s.  Its energy 
dispersion $\delta \omega = m_a v \delta v \lesssim 2 \cdot 10^{-10} m_a$ 
and hence its coherence time $t_c \gtrsim 700$ sec (MHz/$\nu_a$).

\subsubsection{Nuclear magnetic resonance (NMR)}

When the transverse magnetic field is in only one spatial 
direction, say $\vec{B}_\perp (t) = B_\perp \cos(\omega t) \hat{x}$, 
the Bloch equations are solved by 
\begin{equation}
\vec{M}_a(t) = 
M_{\perp a} [\hat{x} \cos(\omega t - \phi) 
+ \hat{y} \sin(\omega t - \phi)]
+ {\cal O}({1 \over \omega + \omega_0})
\label{tranmag2}
\end{equation}
with $\phi$ given in Eqs.~(\ref{tanphi}) and 
\begin{equation}
M_{\perp a} = {1 \over 2} 
{\gamma M_0 t_2 B_\perp \over
\sqrt{1 + (\omega_0 - \omega)^2 t_2^2}}~~\ .
\label{tranma}
\end{equation}  
The terms of order $(\omega + \omega_0)^{-1}$ in Eq.~(\ref{tranmag2})
are nonresonant and can be ignored.  The effect of frequency dispersion 
in the axion field is included by replacing $t_2$ with $\min(t_2,t_c)$. 
We have then on resonance $(\omega_0 = m_a)$ 
\begin{eqnarray}
M_{\perp a} &=&  
{g_N \over f_a} v_\perp \sqrt{\rho_a \over 2} M_0 \min(t_2,t_c)
\nonumber\\
&=& 0.5 \cdot 10^{-14}~M_0~
g_N \left({m_a \over 10^{-8}~{\rm eV}}\right)
\left({v_\perp \over 10^{-3}}\right)
\left({\rho_a \over {\rm GeV/cm}^3}\right)^{1 \over 2}
{\min(t_2, t_c) \over {\rm sec}}~~\ .
\label{nmrsig2}
\end{eqnarray}
The transverse magnetization may be detected by a SQUID 
magnetometer.  The present sensitivity of such devices 
is of order $10^{-16}~{\rm T}/\sqrt{\rm Hz}$.  The 
CASPEr-Wind experiment \cite{Budk14,Garc17} searches 
for axion dark matter using this technique.

Refs.~\cite{Grah13,Budk14,Garc17} propose a second approach 
to axion dark matter detection using NMR techniques, called 
CASPEr-Electric.  In it a static electric field $\vec{E}_0$ 
is applied in a direction transverse to $\vec{M}_0$.  The 
electric field interacts with the oscillating electric dipole 
moment induced onto the nucleus by the local axion field
\begin{equation}
\vec{d}_e(t) = D_N \bar{\theta}(t)\vec{I} = 
D_N {a(t) \over f_a}\vec{I}
\label{eldip}
\end{equation}  
where $D_N \sim 3 \cdot 10^{-16}~e$ cm; see Eq.~(\ref{dnest}).
The relevant interaction is thus
\begin{equation}
H_{a\bar{N}N}^\prime = - \vec{d}_e \cdot\vec{E}_N
= - D_N {a(t) \over f_a} \vec{I}\cdot\vec{E}_N
\label{inprim}  
\end{equation}
where $\vec{E}_N$ is the electric field at the location of 
the nucleus. In an atom, a static externally applied electric 
field $\vec{E}_0$ is screened at the location of the nucleus
by the electron cloud, implying $\vec{E}_N = 0$.  Indeed, 
if $\vec{E}_N \neq 0$, the nucleus moves till $\vec{E}_N = 0$.
However, because of finite nuclear size effects, $\vec{E}_N$ does 
not vanish entirely but is suppressed by a factor $\epsilon_S$, 
called the Schiff factor, of order $10^{-2}$ for a large nucleus 
\cite{Grah13,Budk14}.  Relative to 
$H_{a\bar{N}N}$ the strength of the interaction 
$H_{a\bar{N}N}^\prime$ is then 
\begin{equation}
{\epsilon_S D_N E_0 \over g_N v_\perp m_a} \sim
10^4~\left({\epsilon_s \over 10^{-2}}\right)
\left({E_0 \over 3\cdot10^8~{\rm V/cm}}\right)
\left({10^{-10}~{\rm eV} \over m_a}\right)   
\label{relstren}
\end{equation}
suggesting that this approach is attractive for small axion masses.

\subsubsection{Axion to magnon conversion}

When the axion field excites transverse magnetization, axions are 
converted to magnons \cite{Barb89,Barb17,Chig20}.  Whereas an 
amplitude measurement, such as CASPEr is more sensitive at low 
frequencies, a power measurement is more sensitive at hign frequencies.  
The power from axion to magnon conversion on resonance ($\omega_0 = m_a$) 
is 
\begin{eqnarray}
P_{m} &=& 
- \vec{M}_{\perp a} \cdot {d \vec{B}_{\perp,{\rm eff}} \over dt} V 
= m_a M_{\perp a} B_{\perp,{\rm eff}} V \nonumber\\
&=& \left({g_e \over f_a}\right)^2 m_a \rho_a (v_\perp)^2 
{M_0 V \over \gamma}\min(t_2, t_c)
\nonumber\\
&=& 7.9 \cdot 10^{-20}~{\rm W}~g_e^2 \left({m_a \over 10^{-4}~{\rm 
eV}}\right)^3
\left({n_s V \over 10^{24}}\right)  
\left({v_\perp \over 10^{-3}}\right)^2 
\left({\rho_a \over {\rm GeV/cm}^3}\right)
{\min(t_2, t_c) \over {\rm sec}}
\label{axmag}
\end{eqnarray}
in a volume $V$ of aligned electron spins with density $n_s$.  
The QUAX experiment at the INFN Laboratory in Legnaro, Italy, 
searches for axion dark matter using a magnetized sample placed 
in an electromagnetic cavity \cite{Barb17,Cres18,Cres20}.   The 
electron spins are coupled to a cavity resonant mode, tuned to 
the frequency $\omega_0$, so that the magnons convert to microwave 
photons.  The electromagnetic power is coupled out and detected by 
a microwave receiver. The cavity is cooled to temperatures of order 
100 mK to suppress thermal noise.  The approach is discussed also 
in \cite{Flow19} with results from an initial experiment.

\subsection{LC circuit}

For axion masses below $10^{-7}$ eV, the size of the cavity 
detector is of order 10 m or larger.  For such small masses 
it may be advantageous to replace the cavity by a LC circuit 
\cite{Sik14a,Kahn16,McAl16,Silv17,Chu18,Cris18,Chau18,
Ouel19a,Ouel19b,Cris20}
\footnote{\label{unpLC} Unpublished work on the LC circuit axion 
dark matter detector was done in the early 2000's by P. Sikivie,
N. Sullivan and D.B. Tanner, and independently by B. Cabrera and
S. Thomas. The work of Cabrera and Thomas was presented in a talk, 
http://www.physics.rutgers.edu/~scthomas/talks/Axion-LC-Florida.pdf,
at the Axions 2010 Conference in Gainesville, Florida, January 15-17,
2010.}.

Eqs.~(\ref{inhMaxwa}) tell us that in an externally applied 
magnetic field $\vec{B}_0$ dark matter axions produce an electric 
current density $\vec{j}_a = - g \vec{B}_0 \partial_t a$.  Assuming 
the magnetic field is static, $\vec{j}_a$ oscillates with frequency
$\omega = m_a (1 + {1 \over 2} \vec{v}\cdot\vec{v})$ where $\vec{v}$ 
is the axion velocity.  If the spatial extent of the externally applied 
magnetic field is much less than $\omega^{-1}$, the Maxwell displacement 
current $\partial_t(\epsilon \vec{E})$ can be neglected in the second 
equation (\ref{inhMaxwa}).  The magnetic field $\vec{B}_a$ produced by 
$\vec{j}_a$ satisfies then $\vec{\nabla}\times\vec{B}_a = \vec{j}_a$.  
We set $\mu = 1$ for simplicity.  One may amplify $\vec{B}_a$ using an 
LC circuit and detect the amplified field with a SQUID magnetometer.

Fig. \ref{LCsch} shows a schematic drawing in case the magnet
producing $\vec{B}_0$ is a solenoid.  The field $\vec{B}_a$ has
flux $\Phi_a$ through a loop of superconducting wire. Because the 
wire is superconducting the total magnetic flux through the loop 
circuit is constant.  In the limit where the capacitance $C$ of 
the circuit is infinite (the capacitance is removed) the current 
in the wire is $I = - \Phi_a/L$ where $L$ is the inductance of 
the circuit.  The magnetic field seen by the magnetometer is 
\begin{equation}
B_d \simeq {N_d \over 2 r_d} I = - {N_d \over 2 r_d~L} \Phi_a     
\label{detf}
\end{equation}
where $N_d$ is the number of turns and $r_d$ the radius of the small  
coil facing the magnetometer.  Ignoring for the moment the mutual 
inductances of the LC circuit with neighboring circuits in its 
environment, $L$ is a sum
\begin{equation}
L \simeq L_m + L_c + L_d
\label{Lsum}
\end{equation}
of contributions $L_m$ from the large pickup loop inside the externally
applied magnetic field, $L_d$ from the small coil facing the magnetometer,
and $L_c$ from the co-axial cable in between. We have
\begin{equation}
L_d = r_d N_d^2 c_d
\label{dind}
\end{equation}
with
\begin{equation}
c_d \simeq \ln\left({8 r_d \over a_d}\right) - 2
\label{CD}
\end{equation}
where $a_d$ is the radius of the wire in the small coil.  The mutual 
inductances of the LC circuit with neighboring circuits can be measured 
in any actual setup and taken into account when optimizing the circuit 
and estimating the detector's sensitivity.

For finite $C$, the LC circuit resonates at frequency
$\omega = 1/\sqrt{LC}$.
When $\omega$ equals the axion rest mass, the magnitude of the current in
the wire is multiplied by the quality factor $Q$ of the circuit and hence
\begin{equation}
B_d \simeq {Q N_d \Phi_a \over 2 L r_d} ~~~\ .
\label{fbd}
\end{equation}
Let us consider the case where the externally applied magnetic field is 
homogeneous, $\vec{B}_0 = B_0 \hat{z}$, as is approximately true inside 
a long solenoid.  In such a region
\begin{equation}
\vec{B}_a = - {1 \over 2} g (\partial_t a) B_0 \rho \hat{\phi}
\label{Ba}
\end{equation}
where ($z$, $\rho$, $\phi$) are cylindrical coordinates. For the pickup 
loop depicted in Fig. \ref{LCsch}, a rectangle whose sides $l_m$ and 
$r_m$ are approximately the length and radius of the magnet bore, the 
flux of $\vec{B}_a$ through the pickup loop is
\begin{equation}
\Phi_a = - V_m g B_0 \partial_t a 
\label{genflux}
\end{equation}
with $V_m = {1 \over 4} l_m r_m^2$.  Assuming $l_m >> r_m$, 
the self-inductance of the pickup loop is 
$L_m \simeq {1 \over \pi} l_m \ln\left({r_m \over a_m}\right)$
where $a_m$ is the radius of the wire.

The time derivative of the axion field is given in terms of the axion
density by $\partial_t a = \sqrt{2 \rho_a} \cos(\omega t)$. Hence, 
combining Eqs.~(\ref{fbd}) and (\ref{genflux}), we obtain the 
magnitude of the magnetic field seen by the magnetometer:
\begin{eqnarray}
B_d &\simeq& {N_d Q \over 2 r_d L} V_m g \sqrt{2 \rho_a} B_0
= 1.25 \cdot 10^{-15}~{\rm T}
\left({\rho_a \over {\rm GeV/cm}^3}\right)^{1 \over 2}
\left({Q \over 10^4}\right)\cdot\nonumber\\
&\cdot&
\left({g \over 10^{-17}~{\rm GeV}^{-1}}\right) N_d
\left({{\rm cm} \over r_d}\right)\left({V_m \over {\rm m}^3}\right)
\left({\mu{\rm H} \over L}\right) \left({B_0 \over 10~{\rm T}}\right)\ .
\label{signal}
\end{eqnarray}
In comparison, the sensitivity of today's best magnetometers
is $\delta B = B_n \sqrt{\Delta\nu \over {\rm Hz}}$ with $B_n$ 
of order $10^{-16}~{\rm T}$.  The detector bandwidth is $\nu/Q$.  
If a factor 2 in frequency is to be covered per year, and the 
duty factor is 30\%, the amount of time $t$ spent at each tune 
of the LC circuit is of order $10^7~{\rm s}/Q$.  The signal to 
noise ratio depends on the signal coherence time $t_c$ and 
hence on the axion velocity distribution.  The coherence 
times of two contrasting galactic halo models were given 
in the previous subsection.  The magnetometer is sensitive 
to magnetic fields of magnitude 
$\delta B = B_n~({\rm t~Hz})^{-{1 \over 2}}$ when $t < t_c$ and
$\delta B = B_n~({\rm Hz})^{-{1 \over 2}}~(t_c~t)^{-{1 \over 4}}$
when $t > t_c$.

An important source of noise, in addition to the flux noise in
the magnetometer, is the thermal (Johnson-Nyquist) noise in the
LC circuit.  It causes voltage fluctuations
$~\delta V_T = \sqrt{4 k_B T~R~\Delta \nu}~$ \cite{Nyq28} and hence
current fluctuations
\begin{eqnarray}
\delta I_T &=& {\delta V_T \over R} =
\sqrt{4 k_B T Q \Delta\nu \over L \omega}\nonumber\\
&=& 2.96 \cdot 10^{-13} {\rm A}
\sqrt{\left({{\rm MHz} \over \nu}\right)
\left({\mu{\rm H} \over L}\right)\left({Q \over 10^4}\right)
\left({T \over {\rm mK}}\right)\left({\Delta \nu \over {\rm mHz}}\right)} 
\label{John}
\end{eqnarray}
where we used the relation $R = {L \omega \over Q}$ between the
resistance and quality factor of a LC circuit.  Eq.~(\ref{John})
should be compared with the current due to the signal
\begin{eqnarray}
I &=& {Q \over L} V_m g B_0 \partial_t a=
1.99 \cdot 10^{-11} {\rm A} \left({Q \over 10^4}\right)
\left({V_m \over {\rm m}^3}\right)
\left({\mu{\rm H} \over L}\right) \cdot\nonumber\\
&\cdot& 
\left({g \over 10^{-17} {\rm GeV}^{-1}}\right)
\sqrt{\rho_a \over {\rm GeV}/{\rm cm}^3}\left({B_0 \over 10~{\rm T}}\right)
\label{sig}
\end{eqnarray}
and with the fluctuations in the measured current due to the noise in the
magnetometer
\begin{equation}
\delta I_B \simeq {2 r_d \over N_d} \delta B = 5.03 \cdot 10^{-14} {\rm A}
{1 \over N_d} \left({r_d \over {\rm cm}}\right)
\left({B_n \over 10^{-16}~{\rm T}}\right)
\sqrt{\Delta \nu \over {\rm mHz}}\ .
\label{mag}
\end{equation}
Another possible source of noise is jumps in the $\vec{B}_0$
field, caused by small sudden displacements in the positions
of the wires in the magnet windings. 

The LC circuit detector appears well suited to axion dark matter 
detection in the $10^{-7}$ to $10^{-9}$ eV range.  The ABRACADABRA
experiment at MIT \cite{Kahn16,Ouel19a,Ouel19b} and ADMX SLIC 
experiment at the University of Florida \cite{Cris18,Cris20} 
have published results. An experiment is also under construction 
at Stanford (DM Radio) \cite{Silv17,Chau18}.

A reentrant cavity is an electromagnetic cavity with properties 
similar to those of an LC circuit.  Ref. \cite{McAl17} describes
such a cavity and computes its form factor as a function of 
frequency.

\subsection{Atomic transitions}

The interaction of an axion with a non-relativistic electron, 
Eq.~(\ref{nonrelf}) and the interaction of an axion with nuclear 
spin, Eq.~(\ref{nucc}), allow atomic transitions in which an axion 
is emitted or absorbed.  The transitions are resonant between atomic 
states that differ in energy by an amount equal to the axion mass.  
Such energy differences can be conveniently tuned using the Zeeman 
and Stark effects.  One approach to axion dark matter detection is 
to cool a kilogram-sized sample to milli-Kelvin temperatures and 
count axion induced atomic or molecular transitions using laser 
techniques \cite{Sik14b,Sant15,Brag17,Avig18b}.

Eq.~(\ref{nucc}) and the first term on the RHS of Eq.~(\ref{nonrelf}) 
are similar to the coupling of the magnetic field to spin.  Those 
interactions may cause magnetic dipole (M1) transitions in atoms
and molecules.  The second term in Eq.~(\ref{nonrelf}) allows 
$\Delta j = 0$, $\Delta l = 1$, parity changing transitions.  As 
usual, $l$ is the quantum number giving the magnitude of orbital 
angular momentum, and $j$ that of total angular momentum.  We will 
not consider that last interaction further because, starting from 
the ground state ($l=0$), it causes atomic transitions only if the 
energy absorbed is of order eV, much larger than the axion mass.  
Molecular transitions in the eV range are discussed as a technique 
for dark matter detection in ref.~\cite{Arva18}.

The ground state of most atoms is accompanied by several
other states related to it by hyperfine splitting, i.e. 
by flipping the spin of one or more valence electrons
or by changing the $z$-component $I_z$ of the nuclear 
spin.  The transition rate by axion absorption from an 
atomic ground state $|0\rangle$ to an excited state 
$|i\rangle$ is
\begin{equation}
R_i = {1 \over 2 m_a f_a^2} \min(t,t_1,t_c)
~\int d^3p~~{d^3 n \over dp^3}(\vec{p})~
|\langle i|(g_e \vec{S} + g_N \vec{I})\cdot\vec{p}|0\rangle|^2
\label{arate}
\end{equation}
on resonance.  Here $\vec{S}$ is electron spin, $t$ is the 
measurement integration time,  $t_1$ the lifetime of the 
excited state, and $t_c$ the coherence time of the signal.
The latter is related to the energy dispersion of dark matter 
axions, $t_c \sim 1/\delta E$, as was discussed already.  The 
resonance condition is $m_a = E_i - E_0$ where $E_i$ and $E_0$ 
are the energies of the two states.  The detector bandwidth is 
$B = 1/\min(t,t_1)$.  ${d^3 n \over dp^3}(\vec{p})$ is the local 
axion momentum distribution.  The local axion energy density is
\begin{equation}
\rho_a \simeq m_a \int d^3p~{d^3 n \over dp^3}(\vec{p})~~\ .
\label{axden2}
\end{equation}
Let us define $g_i$ by
\begin{equation}
g_i^2 \overline{v^2} m_a \rho_a \equiv
\int d^3p~{d^3 n \over dp^3}(\vec{p})~
|\langle i|(g_e \vec{S} + g_N \vec{I})\cdot\vec{p}|0\rangle|^2~~\ ,
\label{gi}
\end{equation}
where $\overline{v^2} \sim 10^{-6}$ is the average velocity squared
of dark matter axions.  $g_i$ is a number of order one giving the 
coupling strength of the target atom.  $g_i$ depends on the atomic 
transition used, the direction of polarization of the atom, and 
the momentum distribution of the axions.  It varies with time 
of day and of year since the momentum distribution changes on 
those time scales due to the motion of the Earth. 

For a mole of target atoms, the transition rate on
resonance is
\begin{eqnarray}
N_A R_i &=& g_i^2 N_A~\overline{v^2}~
{\rho_a \over 2 f_a^2} \min(t,t_1,t_c)
\nonumber\\
&=&~{535 \over {\rm sec}}
\left({\rho_a \over {\rm GeV}/{\rm cm}^3}\right)
\left({10^{11}~{\rm GeV} \over f_a}\right)^2 g_i^2~\left({\overline{v^2} \over 
10^{-6}}\right) {\min(t,t_1,t_c) \over {\rm sec}}
\label{mrate}
\end{eqnarray}
where $N_A$ is Avogadro's number.  There is an (almost) equal transition
rate for the inverse process, $|i\rangle \rightarrow |0\rangle$ with emission 
of an axion.  It is proposed to allow axion absorptions only by cooling the
target to a temperature $T$ such that there are no atoms in the excited
state.  The requirement $N_A e^{- m_a \over T} < 0.1$ implies 
\begin{equation}
T =  12~{\rm mK} \left({10^{11}~{\rm GeV} \over f_a}\right)~~\ .
\label{temp}
\end{equation}
The transitions are detected by shining a tunable laser on the   
target.  The laser's frequency is set so that it causes transitions
from state $|i\rangle$ to a highly excited state (with energy of order 
eV above the ground state) but does not cause such transitions from the
ground state or any other low-lying state.  When the atom de-excites,
the associated photon is counted.  The efficiency of this technique
for counting atomic transitions is between 50 and 100\%.

Consider a sweep in which the frequency is shifted by the bandwidth $B$ 
after each measurement integration time $t$.  The expected number of events 
per tune and per mole on resonance is $t N_A R_i$.  If $B_a < B$, events 
occur only during one tune, whereas events occur during $B_a/B$ successive 
tunes if $B_a > B$.  Thus the total number of events per mole during a 
sweep through the axion frequency $\nu_a = m_a/2\pi$ is
\begin{equation}
{\#{\rm events} \over {\rm mole}} =
t N_A R_i {\min(t,t_1) \over \min(t,t_1,t_c)}~~\ .
\label{evnum} 
\end{equation}
To proceed at a reasonably fast pace, the search should cover
a frequency range of order $\nu_a$ per year.  Assuming a 30\%
duty cycle, one needs a search rate
\begin{equation}
{B \over t} = {1 \over t \min(t,t_1)} =
{\nu_a \over 0.3~{\rm year}} = {1.5~{\rm kHz} \over {\rm sec}}
\left({10^{11}~{\rm GeV} \over f_a}\right)~~\ .
\label{sr}
\end{equation}
The expected number of events per sweep through the axion
frequency is then
\begin{equation}
{\#{\rm events} \over {\rm mole}} = 0.35~g_i^2~
\left({\overline{v^2} \over 10^{-6}}\right)
\left({\rho_a \over {\rm GeV}/{\rm cm}^3}\right)
\left({10^{11}~{\rm GeV} \over f_a}\right)~\ .
\label{evnum2}
\end{equation}
Note that when the search rate is fixed, as in Eq.~(\ref{sr}),
the number of events per sweep through the axion frequency is
independent of $t$, $t_1$ and $t_c$.

A suitable target material may be found among the numerous 
salts of transition group ions that have been studied extensively 
using electron paramagnetic resonance techniques \cite{Abra70}.  
C. Braggio et al. \cite{Brag17} carried out a pilot experiment 
on a a small crystal of YLiF$_4$ doped with Er$^{3+}$ target 
ions at concentrations of 0.01 and 1\%.  They studied the 
heating of the sample by the laser and found that it did not 
produce an unmanageable background in the case studied.

\subsection{Axion echo}

Electromagnetic radiation of angular frequency equal to half the
axion mass ($\omega = m_a/2$) stimulates the decay of axions to 
two photons and produces an echo, i.e. faint electromagnetic
radiation traveling in the opposite direction.  Hence one may 
search for axion dark matter by sending to space a powerful beam 
of microwave radiation and listening for its echo \cite{Arza19}.  
Stimulated axion decay is described below, first in the rest frame 
of a perfectly cold axion fluid, followed by the case where the 
observer is moving with respect to a perfectly cold axion fluid, 
and finally the case where the axion fluid has velocity dispersion.

\vskip 0.5cm

{\it Perfectly cold axion fluid at rest}

\vskip 0.2cm

In the rest frame of a perfectly cold axion fluid of density 
$\rho$ the axion field is 
\begin{equation}
a(t) = A \sin(m_a t)
\label{axfl}
\end{equation}
with $\rho = {1 \over 2} m_a^2 A^2$.  In radiation gauge 
($\vec{\nabla} \cdot \vec{A} = 0$), the second equation 
(\ref{inhMaxwa}) becomes
\begin{equation}
(\partial_t^2 - \nabla^2) \vec{A} = 
- g (\vec{\nabla} \times \vec{A}) \partial_t a~~\ . 
\label{inM}
\end{equation}
We set $\epsilon = \mu = 0$ for simplicity.  Let 
the vector potential of the outgoing radiation be
\begin{equation}
\vec{A}_0(\vec{x}, t) = Re \int d^3k~\vec{A}_0(\vec{k}) 
e^{i(\vec{k}\cdot\vec{x} - \omega t)}
\label{A0}
\end{equation}
where $\omega = |\vec{k}|$.  In the presence of the axion 
fluid, $\vec{A}_0$ is itself a source of electromagnetic 
radiation $\vec{A}_1(\vec{x}, t)$:
\begin{equation}
(\partial_t^2 - \nabla^2) \vec{A}_1 =
- g (\vec{\nabla} \times \vec{A}_0) \partial_t a
+ {\cal O}(g^2)~~\ .
\label{A1}
\end{equation}
We have therefore
\begin{equation}
\vec{A}_1 (\vec{x}, t) = Re \int d^3k~ 
\vec{A}_1(\vec{k}, t) e^{i \vec{k}\cdot\vec{x}}
\label{A1e}
\end{equation}
with 
\begin{equation}
(\partial_t^2 + \omega^2) \vec{A}_1(\vec{k}, t) 
= - g A m_a \cos(m_a t)~ i \vec{k} \times \vec{A}_0(\vec{k}) 
~e^{- i \omega t}~~\ .
\label{A1kt}
\end{equation}
The frequencies appearing on the RHS of Eq.~(\ref{A1kt})  
are $\omega \pm m_a$.  Resonance occurs when $\omega - m_a = 
- \omega$, i.e. when $\omega = m_a/2$.  

Let us write
\begin{equation}
\vec{A}_1(\vec{k}, t) = \vec{\cal A}_1(\vec{k}, t) e^{i \omega t}
~~\ .
\label{cA1}
\end{equation}
In terms of $\vec{\cal A}_1$, Eq.~(\ref{A1kt}) is
\begin{equation}
\partial_t \vec{\cal A}_1 (\vec{k}, t) = - {1 \over 4 \omega} 
g A m_a ~\vec{k} \times \vec{A}_0(\vec{k})~ 
e^{i (m_a - 2 \omega)t}
\label{cA1kt}
\end{equation}
when $\partial_t^2 \vec{\cal A}_1$ is neglected versus
$\omega \partial_t \vec{\cal A}_1$ and only the resonance 
producing term is kept on the RHS.  Solving Eq.~(\ref{cA1kt}) 
with $\vec{\cal A}_1(\vec{k}, 0) = 0$ yields
\begin{equation}
\vec{\cal A}_1(\vec{k}, t) = - {1 \over 4} g A m_a 
~\hat{k} \times \vec{A}_0(\vec{k})~ 
e^{i \epsilon t}~ {\sin(\epsilon t) \over \epsilon}
\label{A1sol}
\end{equation}
where $\hat{k} = {\vec{k} \over \omega}$ 
and  $\epsilon = m_a/2 - \omega$.  For large $t$, 
\begin{equation}
\left({\sin(\epsilon t) \over \epsilon}\right)^2 \rightarrow
\pi t \delta(\epsilon) ~~\ .
\label{fermi}
\end{equation}
Hence, if we write the power in the outgoing $\vec{A}_0$ wave as 
\begin{equation}
P_0 = \int d \omega {d P_0 \over d \omega}(\omega)
\label{outpow}
\end{equation}
the power in the $\vec{A}_1$ wave is \cite{Arza19}
\begin{equation}
P_1 = {1 \over 16} g^2 A^2 m_a^2 t \int d\omega 
{d P_0 \over d \omega}(\omega) \pi \delta(m_a/2 - \omega)
= {1 \over 16} g^2 \rho {d P_0 \over d \nu} t 
\label{epow}
\end{equation}
where ${d P_0 \over d \nu}$ is the spectral density of
the outgoing power at frequency $\nu = {\omega \over 2 \pi}
= {m_a \over 4 \pi}$.

Only outgoing power of frequency $\nu = {m_a \over 4 \pi}$ 
stimulates axion decay and produces an echo.  If the outgoing 
wave is stationary, with angular frequency $\omega = m_a/2$ 
and linear polarization $\vec{e}$:
\begin{equation}
\vec{A}_0(\vec{x}, t) = Re \left[e^{-i {m_a \over 2} t}~\vec{e}~
\int_{|\vec{k}| = m_a/2} d^2k~A_0(\vec{k})~
e^{i \vec{k}\cdot\vec{x}} \right]~~\ , 
\label{rA0}
\end{equation}
the echo wave is 
\begin{equation}
\vec{A}_1(\vec{x}, t) = + {1 \over 4} g A m_a t~ 
Re \left[e^{i {m_a \over 2} t}~\vec{e}\times~
\int_{|\vec{k}| = m_a/2} d^2k~\hat{k}~A_0(\vec{k})~
e^{i \vec{k}\cdot\vec{x}}\right]~~\ .
\label{rA1}
\end{equation}
The echo wave is linearly polarized at $90^\circ$ relative to 
the outgoing wave and traces it exactly backwards in time since 
it has the same spatial Fourier transform but the opposite
frequency. If the outgoing beam is emitted as a parallel 
beam of finite cross-section, it will spread as a result 
of its transverse wavevector components.  The echo wave 
retraces the outgoing wave backward in time, returning to 
the location of emission of the outgoing wave with the 
latter's original transverse size.  If the outgoing power
$P_0$ is turned on for a time $t$ and then turned off, the
echo power $P_1$ given by Eq.~(\ref{epow}) lasts forever 
in the future under the assumption that the perfectly cold 
axion fluid has infinite spatial extent.  In the rest frame 
of a perfectly cold axion fluid it does not matter in which 
direction $P_0$ is emitted.  A finite amount of energy emitted 
at angular frequency $\omega = m_a/2$ in any direction produces 
an everlasting faint echo.

\vskip 0.5cm

{\it Perfectly cold axion fluid in motion}

\vskip 0.2cm

Next let us consider the case where the perfectly cold axion
fluid is moving with velocity $\vec{v}$ with respect to the 
source of outgoing power.  Nothing changes in the above 
discussion except that each increment of outgoing energy 
$d E_0 = P_0 dt$ is emitted from a different location in 
the cold axion fluid rest frame.  The incremental echo power, 
given by the RHS of Eq.~(\ref{epow}) with $t$ replaced by $dt$
\begin{equation}
dP_1 = {1 \over 16} g^2 \rho {d P_0 \over d \nu}~dt~~\ ,
\label{depow}
\end{equation}
returns forever to the location in the axion fluid rest
frame from which the increment $dE_0$ of outgoing energy 
was emitted.  In the rest frame of the outgoing power source, 
the echo from outgoing power emitted a time $t_e$ ago arrives 
displaced from the point of emission of the outgoing power by 
$\vec{d} = \vec{v}_\perp t_e$ where $\vec{v}_\perp$ is the 
component of $\vec{v}$ perpendicular to the direction $\hat{k}$ 
of emission.  Fig. \ref{echo} illustrates the relative 
locations of the outgoing power and echo power in case the 
outgoing power is turned on for a while and then turned off.   
The echo moves away from the place of emission of the outgoing 
power with velocity $\vec{v}_\perp$.  To detect as much echo 
power as possible at or near the place of emission of the 
outgoing power, the observer wants $\vec{v}_\perp$ as small 
as possible, i.e. $\hat{k}$ in the same direction as $\vec{v}$ 
or the opposite direction.  In the frame of its source, the 
angular frequency at which the outgoing power stimulates 
axion decay is
\begin{equation}
\omega_0 = {m_a \over 2} (1 + \vec{v}\cdot\hat{k}) 
+ {\cal O}(v^2)
\label{om0}
\end{equation}
whereas 
\begin{equation}
\omega_1 = {m_a \over 2} (1 - \vec{v}\cdot\hat{k})
+ {\cal O}(v^2)~~\ .
\label{om1}
\end{equation}
is the angular frequency of the echo.

An attractive target for the echo method is the 
Big Flow, the locally prominent cold dark matter flow
in the caustic ring model of the Milky Way halo.  Its 
velocity vector $\vec{v}_{\rm BF}$ in a non-rotating 
galactic reference frame is given in Eq.~(\ref{BFv}).  
In a reference frame attached to the surface of the 
Earth its velocity is
\begin{equation}
\vec{v}(t) = \vec{v}_{\rm BF} - \vec{v}_{\rm LSR}
-\vec{v}_\odot - \vec{v}_\otimes(t)
\label{BFvE}
\end{equation}
where $\vec{v}_{\rm LSR}$ is the velocity of the Local
Standard of Rest (LSR) in that same reference frame,
$\vec{v}_\odot$ is the velocity of the Sun with respect
to the LSR, and $\vec{v}_\otimes$ is the velocity of the
observer with respect to the Sun as a result of the orbital
and rotational motions of the Earth.  We are particularly 
interested in the extent to which the uncertainties in the 
several terms on the RHS of Eq.~(\ref{BFvE}) affect our 
ability to minimize $\vec{v}_\perp$.  $\vec{v}_\otimes(t)$ 
is known with great precision.  The components of 
$\vec{v}_\odot$ are known with a precision of order two 
or three km/s.  $\vec{v}_{\rm LSR}$ is in the direction 
of galactic rotation by definition.  Its magnitude (often 
quoted to be 220 km/s) is known within an uncertainty of 
order 20 km/s.  The magnitude of $\vec{v}_{\rm BF}$ scales 
with the magnitude of $\vec{v}_{\rm LSR}$ and has been 
estimated to be approximately 520 km/s \cite{Duff08}.
The direction of $\vec{v}_{\rm BF}$ is mostly in the
direction of galactic rotation [see Eq.~(\ref{BFv})]
and is known with a precision of order $1^\circ$ 
\cite{Chak20}.  So we expect that it is possible to 
reduce $|\vec{v}_\perp|$ to approximately 5 km/s, the 
nominal value we use below.

Consider a dish (e.g. a radiotelescope) of radius $R$ 
collecting echo power near the location of the outgoing 
power source.   Because the echo from outgoing power 
emitted a time $t_e$ ago is displaced by $\vec{d} = 
\vec{v}_\perp t_e$, the amount of echo power collected 
by the dish is
\begin{equation}
P_c = {1 \over 16} g^2 \rho {d P_0 \over d \nu}
C {R \over |\vec{v}_\perp|}
\label{echo2}
\end{equation}
where $C$ is a number of order one which depends
on the configuration of the source relative to
the receiver dish:
\begin{equation}
C = {|\vec{v}_\perp| \over 2 R P_0} \int dt
\int_{S_0}  d^2x ~I_0(\vec{x})
\Theta_c(\vec{x} + \vec{v}_\perp t)~~\ .
\label{C}
\end{equation}
Here $S_0$ is the surface from which the outgoing
power is emitted, $I_0(\vec{x})$ is the outgoing
power per unit surface, and $\Theta_c(\vec{x})$ is
a function that equals one if $\vec{x}$ belongs
to the receiver dish area and zero otherwise.
For example, $C = 0.5$ if the outgoing power
is emitted from the center of the receiver dish,
whereas $C = 0.424$ if the outgoing power is
emitted uniformly from the area of the receiver
dish.  However neither of these configurations
is likely to be optimal.  It is probably better
to place several source dishes around the receiver
dish. $C$ can be straightforwardly calculated for
each configuration.

Let us assume that a pulse of outgoing power $P_0$, 
with frequency $\nu_0$ and uniform spectral density 
${dP_0 \over d\nu} = {P_0 \over \Delta \nu_0}$
over bandwidth $\Delta \nu_0$, is emitted during 
a time $t_m$.  Provided that
\begin{equation}
t_m \gtrsim {R \over 2 |\vec{v}_\perp|}
= 0.5 \times 10^{-2}~{\rm sec}~{R \over 50~{\rm m}}~
{5~{\rm km/s} \over |\vec{v}_\perp|}
\label{cond}
\end{equation}
echo power
\begin{eqnarray}
P_c &=& 2.33 \times 10^{-31} P_0
\left({10~{\rm kHz} \over \Delta \nu_0}\right)
\left({g_\gamma \over 0.36}\right)^2
\left({10^{12}~{\rm GeV} \over f_a}\right)^2 \cdot\nonumber\\
&\cdot& \left({\rho \over {\rm GeV/cm}^3}\right)
\left({C \over 0.30}\right)
\left({R \over 50~{\rm m}}\right)
\left({5~{\rm km/s} \over |\vec{v}_\perp|}\right)~~\ ,
\label{PC}
\end{eqnarray}
is received over the same time interval $t_m$. Since the
magnitude of the velocity of the Big Flow relative to us
$v \simeq$ 520 km/s - 220 km/s = 300 km/s, the frequency
of the echo power is red- or blue-shifted from $\nu_0$ by
$\Delta \nu \simeq 2 \times 10^{-3} \nu_0$.  The echo power 
has bandwidth $B = 2 \delta v~\nu < 5 \times 10^{-7} \nu$
since the velocity dispersion of the Big Flow is less than
70 m/s \cite{Bani16}.  The frequency range of interest 
is approximately 30 MHz to 30 GHz because the Earth's
atmosphere is mostly transparent at those frequencies.
It corresponds to the mass range $2.5 \times 10^{-7} <
m_a < 2.5 \times 10^{-4}$ eV, which happens to be prime
hunting ground for QCD axions.

The cosmic microwave background and radio emission by
astrophysical sources are irreducible sources of noise.
In addition there is instrumental noise.  The total noise
temperature depends on frequency, on the location of the
telescope and on the direction of observation.  As an example
we may consider the system noise temperature of the Green
Bank Telescope 
\footnote{\label{GBT} 
https://science.nrao.edu/facilities/gbt/proposing/GBTpg.pdf}:
approximately 20 K from 1 GHz
to 8 GHz, approximately linearly rising from 20 K at
8 GHz to 40 K at 30 GHz, and exponentially rising
towards low frequencies from 20 K at 1 GHz to 100 K
at 300 MHz.  The rise at low frequencies is due
to Galactic emission and is strongly direction
dependent.  100 K at 300 MHz is a typical value.
The rise at high frequencies is due to atomic and
molecular transitions in the atmosphere.  It can
be mitigated by placing the telescope at a high
elevation.  We use below a nominal system noise
temperature of 20 K at all frequencies for the
purpose of stating the results of our sensitivity
calculations.

The signal to noise ratio with which the echo power 
is detected when $\omega_0$ falls within the angular 
frequency range of the emitted power is given by 
Dicke's radiometer equation (\ref{radio}). Combining 
Eqs.~(\ref{PC}) and (\ref{radio}) and setting 
$B = 5 \times 10^{-7} \nu$, the total outgoing
energy per logarithmic frequency interval necessary to
detect the axion echo with a given signal to noise ratio
is found to be:
\begin{eqnarray}
{d E_0 \over d \ln \nu}\Bigg|_{\rm BF} &=& 7.2~{\rm MW year}
\left({s/n \over 5}\right)
\left({10~{\rm GHz} \over \nu}\right)^{1 \over 2}
\left({0.36 \over g_\gamma}\right)^2
\left({T_n \over 20~{\rm K}}\right)
\left({{\rm GeV/cm}^3 \over \rho}\right)
\cdot\nonumber\\
&\cdot&
\left({0.30 \over C}\right)
\left({t_m \over 10^{-2}~{\rm sec}}\right)^{1 \over 2}
\left({50~{\rm m} \over R}\right)
\left({|v_\perp| \over 5~{\rm km/s}}\right)\ .
\label{outpow2}
\end{eqnarray}
We used Eq.~(\ref{mass}) and $m_a = 4 \pi \nu$.  

\vskip 0.5cm

{\it General axion fluid}

\vskip 0.2cm

In the most general case, the axion fluid is moving with 
respect to the observer and has velocity dispersion.  Its 
density can be expressed as an integral over cold flows
\begin{equation}
\rho = \int d^3v~{d^3 \rho \over dv^3}(\vec{v})~~\ .
\label{veldis}
\end{equation}
Everything said in the previous subsection holds true for 
each infinitesimal cold flow increment.  The echo frequency 
has a spread $\delta \omega_- = {m \over 2} \delta v_\parallel$ 
where $\delta v_\parallel$ is the spread of axion velocities in   
the $\hat{k}$ direction.  The echo of power emitted a time
$t_e$ ago is spread over a transverse size $\delta \vec{d}
= \delta \vec{v}_\perp t_e$ where $\delta \vec{v}_\perp$
is the spread of axion velocities perpendicular to $\hat{k}$.

Let us consider the isothermal model of the Milky Way in 
particular.  According to it, the dark matter has density 
300 MeV/cm$^3$ on Earth.  Its velocity distribution is 
Gaussian and isotropic in a non-rotating galactic reference
frame with dispersion $\sqrt{\vec{v}\cdot\vec{v}}$ = 270 km/s 
$\equiv \sqrt{3} \sigma$.  In the LSR, the axion fluid moves 
with speed 220 km/s in the direction opposite to that of 
galactic rotation.  Assuming the direction $\hat{k}$ of 
the outgoing power is parallel (anti-parallel) to the 
direction of galactic rotation the echo power is blue 
(red)-shifted in frequency by a fractional amount whose 
average is
$\langle {\Delta \nu \over \nu} \rangle \simeq$ 440 km/s
= $1.47 \times 10^{-3}$ and whose rms deviation is
${\delta \nu \over \nu} = 2 \sigma  \simeq 1.04 \times 10^{-3}$.
The echo from outgoing energy that was emitted a time $t_e$
ago is spread in space over a fuzzy circular region whose
radius is Gaussian-distributed with average value $\sigma t_e$.
Eq.~(\ref{echo2}) holds with
${1 \over |\vec{v}_\perp|}$ replaced by
\begin{equation}
\langle {1 \over |\vec{v}_\perp|} \rangle
= \sqrt{\pi \over 2} {1 \over \sigma} =
{1 \over 124~{\rm km/s}}~~\ .
\label{sub}
\end{equation}
In view of Eq.~(\ref{cond}) we now require
$t_m > 2 \times 10^{-4} {\rm sec} {R \over 50~{\rm m}}$.
Using Eq.~(\ref{radio}) with $B = 4 \sigma \nu
= 2.1 \times 10^{-3} \nu$ and setting $\rho$ = 
0.3 GeV/cm$^3$, we find
\begin{eqnarray}
{d E_0 \over d \ln \nu}\Bigg|_{\rm iso} = 5.3~{\rm GWyear}
\left({s/n \over 5}\right)
\left({10~{\rm GHz} \over \nu}\right)^{1 \over 2}
\left({0.36 \over g_\gamma}\right)^2
\cdot\nonumber\\
\cdot
\left({T_n \over 20~{\rm K}}\right)
\left({0.30 \over C}\right)
\left({t_m \over 2 \cdot 10^{-4}~{\rm sec}}\right)^{1 \over 2}
\left({50~{\rm m} \over R}\right)\ .
\label{outpowi}
\end{eqnarray}
In case of a known cold flow, the echo method appears an 
attractive approach because it uses relatively old technology 
and is applicable over a wide range of axion masses.

\section{Solar axion detection}

The solar axion flux presents an attractive search opportunity.  It
has been pursued using a number of methods.  Eqs.~(\ref{solaxfl2})
and (\ref{intfl2}) provide an estimate of the flux on Earth of 
axions produced in the Sun by the Primakoff process.

\subsubsection{The axion helioscope}

Solar axions can be searched for by converting them to photons 
in a magnetic field \cite{Sik83b,Sik85,VanB87}; see Section 3.  
Multiplying the axion flux by the conversion probability 
Eq.~(\ref{pracp}) yields the photon flux
\begin{equation}
\Phi_\gamma =
{0.79 \over {\rm cm}^2~{\rm day}}
\left({g_{a\gamma\gamma} \over 10^{-10}~{\rm GeV^{-1}}}\right)^4
\left({L \over 10~{\rm m}}\right)^2
\left({B_0 \over 10~{\rm T}}\right)^2
\label{solaxr}
\end{equation}
for the case $q L << 1$, $\epsilon = \mu  = 1$.  The photons 
produced are x-rays with approximately 4 keV average energy.  
They point back to the solar core.  If a signal is found, it 
becomes possible to see directly into the solar interior.  The 
inverse of the momentum transfer in the axion-photon conversion 
process is
\begin{equation}
{1\over q} \simeq {2 E \over m_a^2}
= 15.8~{\rm m} \left({10^{-2}~{\rm eV} \over m_a}\right)^2
\left({E \over 4~{\rm keV}}\right)~~~~~\ .
\label{oscl}
\end{equation}
The resonance condition $q L < 1$ is satisfied in vacuum for axion 
masses up to approximately $1.5 \cdot 10^{-2}$ eV if, for example, 
$L = 10$ m.  One may extend this range by introducing gas under 
pressure in the conversion region \cite{VanB87}, giving the photon 
an effective mass equal to the plasma frequency $\omega_{\rm pl}$.  
Alternatively one may make the magnetic field periodic with 
wavelength $d = {2 \pi \over q}$.

Axion helioscope experiments have been carried out at Brookhaven 
National Laboratory \cite{Laza92}, at the University of Tokyo 
\cite{Mori98,Inou02,Inou08}, and by the CAST collaboration at 
CERN \cite{Ziou05,Andr07,Arik09,Anas17a}.  New experiments have 
been proposed by the IAXO \cite{Arme19,Dafn19} and TASTE \cite{Anas17b}
collaborations.  The Tokyo helioscope magnet (2.3 m long, 4 T 
field) was mounted on a platform which allowed the Sun to be 
tracked at all times.  The CAST magnet tracks the Sun for only 
part of the day, approximately 1.5 hour during sunrise and 1.5 
hour during sunset, but is longer (9.3 m) and more powerful (9 T).  
Both experiments have introduced He gas in the conversion region to 
extend the search range upwards in axion mass \cite{Inou08,Arik14,
Arik15}.  The limits obtained by the CAST and Tokyo experiments are 
shown in Fig. 6, in rough outline.  For details, see the original 
publications.  Additional limits on axion couplings were obtained 
by the CAST Collaboration \cite{Andr09,Andr10} from a search 
for mono-energetic axions emitted in nuclear M1 transitions in 
the Sun. Axion emission rates in the M1 transitions of $^{57}$Fe, 
$^{55}$Mn and $^{23}$Na were calculated in ref. \cite{Haxt91}.

It has been proposed to search for solar axions converting to 
x-rays in the magnetic field of the Earth \cite{Davo06} or that 
of the Sun \cite{Hong19} using an x-ray detector placed in orbit 
around the Earth or Sun.

\subsubsection{Axioelectric and M\"ossbauer effects}

Solar axions may be searched for using the axioelectric effect,
which is the same as the photoelectric effect but with an axion 
instead of a photon.  It uses the coupling of the axion to the 
electron.  Through this coupling, axions are produced in the Sun by 
axion bremstrahlung, Compton-like scattering and axion recombination. 
Theoretical discussions of the axioelectric effect are given in 
refs.~\cite{Zhit79,Dimo86,Dere10}.  Results from experimental 
searches are reported in refs.~\cite{Avig87,Ljub04,Avig09,Bell12,
Arme13,Abe13,Cuor13,Derb13,Apri14,Liu17,Aker17,Fu17,Wang19}. The 
bounds on the electron coupling obtained by these searches are of 
order 
\begin{equation}
g_{a\bar{e}e} = g_e {m_e \over f_a} \lesssim 10^{-11}~~\ .
\label{ellim}
\end{equation}
The most severe limit reported is $3.5\cdot10^{-12}$ (90\% CL)
\cite{Aker17}.   

Solar axions may also be searched for using the M\"ossbauer effect 
\cite{deRu89,Mori95}.  Nearly monochromatic axions are emitted in 
nuclear transitions in the Sun, e.g. 14.4 keV axions in transitions 
between the first excited and ground states of $^{57}$Fe.  Such 
axions may be searched for by resonant absorption on the same 
nucleus on Earth.  The nucleus emits an x-ray when it de-excites.  
The process uses the coupling of axions to nucleons. An experimental 
search of this type is reported in ref. \cite{Krcm01}.

\subsubsection{Primakoff effect}

Solar axions may be searched for by converting them to photons in the 
Coulomb field of nuclei, i.e. the Primakoff effect.  The cross-section
for Primakoff conversion of an axion to a photon in the Coulomb field 
of a nucleus can be obtained by repeating the steps of Section 3.2
but keeping the first term in 
$\vec{j}_a = g(\vec{E}_0 \times \vec{\nabla}a - \vec{B}_0 \partial_t a)$
instead of the second term, and setting $E_0(\vec{x}) = Ze 
\vec{x}/4\pi|\vec{x}|^3$. For $\epsilon = \mu = 1$, this yields
\begin{equation}
{d \sigma \over d \Omega} =
{g^2 Z^2 \alpha \over 4 \pi \beta_a} {\omega^2 \over q^4}
(q^2 - (\hat{n}\cdot\vec{q})^2)
\label{Primxs}
\end{equation}
where $\vec{q} = \vec{k} - \vec{k}_a$, 
$\vec{k}_a = \omega \vec{\beta}_a$ is the initial
axion momentum and $\vec{k} = \omega \hat{n}$ the final photon momentum.
The conversion rate on nuclei forming a crystal lattice is resonantly
enhanced when the Bragg scattering condition is satisfied \cite{Pasc94}.
As the Earth spins, the varying orientation of the detector with respect
to direction of the Sun produces a distinctive temporal pattern of the 
counting rate, which helps to distinguish signal from background.  Searches
of this type have been carried out by the SOLAX collaboration \cite{Avig98}
in Sierra Grande, Argentina, the DAMA collaboration \cite{Bern01} in the
Gran Sasso Laboratory, Italy, the COSME collaboration \cite{Mora02} in the
Canfranc Laboratory, Spanish Pyrenees, and the CDMS collaboration \cite{Ahme09}
in the Soudan Underground Laboratory in Minnesota.  The limits obtained are 
shown in Fig. 6.

\section{Dichroism and birefrigence in a magnetic field}

The existence of an axion field causes the vacuum to be birefrigent 
and dichroic in the presence of a magnetic field \cite{Maia86,Raff88b}.  
The purpose of this section is to derive these properties.  A medium 
is called ``birefringent" if it has different indices of refraction 
for the two states of linear polarization of light.  It is ``dichroic"
if it has different absorption coefficients for those two states.  In
general, light traveling in the $z$-direction has
amplitude
\begin{equation}
\vec{A} = Re[({\cal A}_x \hat{x} + {\cal A}_y \hat{y}) e^{-i \omega t}]
\label{ampli}
\end{equation}
where ${\cal A}_x$ and ${\cal A}_y$ are complex numbers.  Eq.~(\ref{ampli}) 
implies that, as a function of time $t$, the vector $\vec{A}$ describes an 
ellipse in the $xy$-plane.  By definition, the ellipticity of the light is 
the ratio of the minor to major axes of that ellipse.  If the light is 
linearly polarized, ${\cal A}_x$ and ${\cal A}_y$ have the same phase, 
and we may write $~~{\cal A}_x = {\cal A} \cos\alpha~,~{\cal A}_y = 
{\cal A} \sin\alpha~~$ where $\alpha$ is the angle of the plane of 
polarization with the $x$-axis.  The ellipticity of such light is zero.  
When light that is initially linearly polarized travels through a 
birefringent material, the relative phase between ${\cal A}_x$ and 
${\cal A}_y$ changes and the light acquires ellipticity.  When light 
that is initially linearly polarized travels through a dichroic material, 
the plane of polarization rotates towards the direction with least 
absorption.

Let us recall that, even if there is no axion, the vacuum is birefringent 
in the presence of a magnetic field as a consequence of the box diagram 
of quantum electrodynamics.  Euler and Heisenberg showed that it implies 
the effective interaction \cite{Heis36,Itzy80}:
\begin{equation}
{\cal L}_{\rm EH} = {2 \alpha^2 \over 45 m_e^4}
\left[(E^2 - B^2)^2 + 7 (\vec{E} \cdot \vec{B})^2\right]~~~\ ,
\label{EHlag}   
\end{equation}
as a consequence of which light traveling through a magnetic field 
$\vec{B}_0$ has different indices of refraction depending on whether 
it is polarized parallel or perpendicular to $\vec{B}_0$.  

Let us assume that an axion exists and consider light traveling
in the $\hat{z}$ direction in a constant magnetic field 
$\vec{B} = B_0 \hat{x}$, and initially linearly polarized 
($\vec{A}_{\rm in}$) at an angle $\alpha$ relative to the 
direction of the magnetic field.  See Fig.~7.  In view of 
Section 3, we expect the $x$-component of light to become 
depleted by $\gamma \rightarrow a$ conversion whereas the 
$y$-component propagates as usual. Specifically, after a 
distance $L$, the magnitude of the $x$ component of light is 
reduced $~~|{\cal A}_x| \rightarrow (1 - {1 \over 2} p(L))|{\cal A}_x|$, 
where $p(L)$ is the conversion probability given in Eq.~(\ref{classicp}),
whereas $|{\cal A}_y|$ is unchanged.  The plane of polarization rotates 
therefore by the angle
\begin{equation}
\delta\alpha(L) = {1 \over 4}~p(L)~\sin(2\alpha)~~~\ .
\label{dichr}   
\end{equation}
In addition the light acquires ellipticity because the relative phase 
between ${\cal A}_x$ and ${\cal A}_y$ changes \cite{Maia86,Raff88b}, i.e.
\begin{equation}
{\cal A}_y \rightarrow {\cal A}_y~~~~,~~~~
{\cal A}_x \rightarrow [1 - {1 \over 2} p(L) + i \phi(L)] {\cal A}_x~~~\ .
\label{changes}
\end{equation}
In terms of $\phi(L)$, the acquired ellipticity is 
\begin{equation}
e(L) = {1 \over 2} |\phi(L)| \sin(2 \alpha)~~~\ .
\label{ellipt}
\end{equation}
Thus, if an axion exists, there is dichroism and a new source of 
birefringence in the presence of a magnetic field.  We now derive 
the expression for the phase shift $\phi(L)$ \cite{Maia86}.  As a 
byproduct, Eq~(\ref{classicp}) for the conversion probability $p(L)$ 
will be rederived \cite{Maia86,Raff88b} as well.

Consider a region of homogeneous dielectric constant $\epsilon$ and   
homogeneous static external magnetic field $\vec B_0$.  Axion-photon
dynamics in such a region is described by the modified Maxwell's
equations
\begin{eqnarray}
\epsilon \vec{\nabla} \cdot \vec{E} &=& g \vec{B}_0 \cdot \vec{\nabla} a
\nonumber\\
\vec{\nabla} \times \vec{B} - \epsilon \partial_t \vec{E} &=&
- g \vec{B}_0 \partial_t a
\label{eom4}
\end{eqnarray}
and the equation of motion for the axion field 
\begin{equation}
(\partial_t^2 - \vec{\nabla}^2 + m_a^2) a = -g \vec B_0 \cdot \vec E~~~~~\ .
\label{reMax}
\end{equation}
We choose the gauge $\Phi = 0$.  Any solution of
Eqs.~(\ref{eom4},\ref{reMax}) is a linear superposition of plane waves:
\begin{equation}
\vec {\cal A} (\vec x,t) = 
\vec {\cal A}~e^{i(\vec k\cdot\vec x - wt)}~~~~,~~~~~
{\bf a} (\vec x,t) = {\bf a}~e^{i(\vec k\cdot \vec x - wt)}~~~~~\ ,
\label{collans}
\end{equation}
where the relation between $\omega$ and $\vec{k}$ depends on the 
direction of polarization.  Let $\vec k = k \hat{z}$ and
\begin{equation}
\vec B_0 = B_{0z}~\hat{z} + B_{0x}~\hat{x}~~~~,~~~~
\vec {\cal A} = {\cal A}_z~\hat{z} + {\cal A}_x~\hat{x} 
+ {\cal A}_y~\hat{y}~~~~\ .
\label{parperp}
\end{equation}
Eqs.~(\ref{eom4},\ref{reMax}) are equivalent to:
\begin{eqnarray}
i \omega \epsilon {\cal A}_z &=& + g B_{0z} {\bf a} \nonumber\\
(k^2 - \epsilon \omega^2){\cal A}_y &=& 0
\label{2eq}
\end{eqnarray}
and
\begin{equation}
k^2 \Biggl({{\cal A}_x \atop {\bf a}}\Biggr) =
\Biggl( \matrix{\epsilon \omega^2 &  + i g B_{0x} \omega \cr
- i g B_{0x} \omega &
\omega^2 -m_a^2 - {1 \over \epsilon} g^2 B_{0z}^2} \Biggr)
\Biggl({{\cal A}_x \atop {\bf a}} \Biggr)~~~~\ .
\label{mateq}
\end{equation} 
Thus the $A_y$ mode propagates in the usual fashion, whereas
the $A_x$ mode oscillates into the axion field and vice-versa.
From Eq.~(\ref{mateq}), one finds the $k^2$-eigenvalues for 
given $\omega$
\begin{equation}
k_\pm^2 = {1 \over 2} \left[(\epsilon + 1) \omega^2
- m_a^2 - {(g B_{0z})^2 \over \epsilon}
\pm \sqrt{\Bigl( (\epsilon - 1) \omega^2 + m_a^2 +
{(gB_{0z})^2 \over \epsilon}\Bigr)^2 + 4 g^2 B_{0x}^2 \omega^2}~\right]~~~~\ .
\label{keig}
\end{equation}
The corresponding eigenmodes are proportional to:
\begin{equation}
\Biggl( {{\cal A}_x \atop {\bf a}}\Biggr)_\pm \equiv
\Biggl( {- i g B_{0x} \omega \atop \epsilon \omega^2 - k_\pm^2 }\Biggr)
~~~~\ .
\label{eigmod}
\end{equation}  
The general solution is therefore:
\begin{eqnarray}
{\cal A}_y &=& ({\cal A}_{y+}~e^{ikz} +
{\cal A}_{y -}~e^{-ikz}) e^{-i \omega t} \nonumber\\   
\Biggl({{\cal A}_x \atop {\bf a}}\Biggr) &=&
\Biggl[ \Biggl({\cal A}_{++} e^{ik_+ z} + {\cal A}_{+-} e^{-ik_+ z}\Biggr)
\Biggl({{\cal A}_x \atop {\bf a}}\Biggr)_+ \nonumber\\
&+& \Biggl({\cal A}_{-+} e^{ik_- z} + {\cal A}_{--} e^{-ik_- z}\Biggr)
\Biggl({{\cal A}_x \atop {\bf a}}\Biggr)_- \Biggr] e^{-i \omega t}~~~~\ .
\label{gensol}
\end{eqnarray}
where ${\cal A}_{y\pm},~{\cal A}_{+\pm}$ and ${\cal A}_{-\pm}$ are 
constants, and $k = \sqrt{\epsilon} \omega$.

We are interested in the solution describing a wave traveling
in the $+\hat{z}$ direction which is initially, at z=0, a purely
electromagnetic wave linearly polarized at an angle $\alpha$
relative to $\hat{x}$.  For that case
\begin{eqnarray}
{\cal A}_{y-} &=& {\cal A}_{+-} = {\cal A}_{--} = 0 \nonumber\\
{\cal A}_y|_{z=0} &=& {\cal A}~\sin \alpha~e^{- i \omega t}~~~,
~~~{\cal A}_x|_{z=0} = {\cal A}~\cos \alpha~e^{- i \omega t}~~~,
~~~~{\bf a}|_{z=0} = 0
\label{AAa}
\end{eqnarray}  
and therefore
\begin{eqnarray}
A_y (z,t) &=& Re[{\cal A} \sin\alpha~e^{i(k z- \omega t)}]\nonumber\\
A_x (z,t) &=& Re\left[{{\cal A} \cos\alpha \over k_+^2 - k_-^2}  
\left((\epsilon \omega^2 - k_-^2)~e^{i(k_+ z - \omega t)} -
(\epsilon \omega^2- k_+^2)~e^{i(k_-z - \omega t)}\right)\right]\nonumber\\
a (z,t) &=& Re\left[- i {\cal A} \cos\alpha~{g B_{0x} \omega \over k_+^2 - 
k_-^2} \left(e^{i(k_+ z - \omega t)} - e^{i(k_- z - \omega t)}\right)\right]
~~~~\ .
\label{moreqs}
\end{eqnarray}
We used the identity
$(\epsilon \omega^2 - k_-^2) (\epsilon \omega^2 - k_+^2) =
- g^2 B_{0x}^2 \omega^2$.  From Eqs.~(\ref{moreqs}) one
may obtain the energy fluxes in the axion field   
${\cal P}_a (z) = < - \partial_t a \partial_z a>$ and
in each polarization state of the photon field:
${\cal P}_x (z) = <- \partial_t A_x \partial_z A_x>$
and ${\cal P}_y (z) = <- \partial_t A_y \partial_z A_y>$.
The brackets indicate time averages.  The $a\rightarrow \gamma$ 
conversion probability is found to be:
\begin{equation}
p(z) = {{\cal P}_a(z) \over {\cal P}_x (0)} =
{4 g^2 B_{0x}^2 \omega^2
\sin^2 ({(k_+ - k_-) z \over 2}) \over (k_+^2 - k_-^2)^2}
\left(1 + {\cal O}({k_+ - k_- \over k})\right)~~~\ .
\label{oscprob}
\end{equation}  
Since
\begin{equation}
g B_0 = {\alpha \over \pi} {g_\gamma \over f_a} B_0 =
1.63~10^{-16}~{\rm eV} \Biggl({10^7~{\rm GeV} \over f_a}\Biggr)
\Biggl({B_0 \over 10~{\rm T}}\Biggr)
\Biggl({g_\gamma \over 0.36}\Biggr)~~~~\ ,
\label{gBo}
\end{equation}
$(g B_0)^2$ is much smaller than $\epsilon \omega^2$ and
$\omega^2 - m_a^2$ in most cases of practical interest. 
If so, Eq.~(\ref{keig}) becomes  
\begin{eqnarray}
k_+^2 &=& \epsilon \omega^2 +
{g^2 B_{0x}^2 \omega^2 \over (\epsilon -1) \omega^2 + m_a^2}
+ {\cal O}(g^4 B_0^4) \nonumber\\ 
k_-^2 &=& \omega^2 - m_a^2 - {g^2 B_{0z}^2 \over \epsilon} -
{g^2 B_{0x}^2 \omega ^2 \over (\epsilon - 1) \omega^2 + m_a^2}
+ {\cal O}(g^4 B_0^4)~~~~~\ .
\label{k+k-}
\end{eqnarray}
We have then
\begin{equation}
p(z) = {g^2 B_{0x}^2 \over\epsilon q^2}
\sin^2 ({q z \over 2})~
\left(1 + {\cal O}(g^2 B_0^2, {q \over k})\right)
~~~~~\ ,
\label{again}
\end{equation}
where
\begin{equation}
q = \sqrt{\epsilon} \omega - \sqrt{\omega^2 - m_a^2}~~~~\ .
\label{q}
\end{equation}
Eq.~(\ref{again}) agrees with Eq.~(\ref{classicp}) on 
resonace, i.e. when $q \rightarrow 0$ and hence 
$\beta_a \rightarrow \sqrt{\epsilon}$.
Rewriting the second Eq.~(\ref{moreqs}) in the form
\begin{equation}
A_x = Re[ {\cal A} \cos\alpha~(1 + \delta(z) + i \phi(z))~
e^{i(k z - \omega t)}]~~\ ,
\label{2As}
\end{equation}
we find $\delta(z) = - {1 \over 2} p(z)$ as anticipated, and 
\cite{Maia86,Raff88b}
\begin{equation}
\phi(z) = {g^2 B_{0x}^2 \over 4 \epsilon q^2}
\left(qz - \sin(qz)\right)~
\left(1 + {\cal O}(g^2 B_0^2, {q \over k})\right)~~~~\ ,
\label{Maia}
\end{equation}
for the phase shift.

Dichroism and birefringence of the vacuum in the presence of a magnetic 
field was searched for by the RBFT collaboration at Brookaven National 
Laboratory \cite{Came93} and by the PVLAS collaboration at the INFN 
National Laboratory in Legnaro, Italy \cite{Zava06,Zava08}.  The 
sensitivity to optical rotation achieved in these experiments is of 
order $10^{-8}$ radians.  The Cotton-Mouton and Voigt effects are 
important backgrounds.  The Cotton-Mouton effect is the birefringence 
of liquids in the presence of a magnetic field transverse to the 
direction of propagation.  The Voigt effect is the analogous effect 
in gases.  The signal is enhanced by the use of an optical cavity 
which allows the laser beam to be passed through the magnetic field 
many times.  Eq.~(\ref{dichr}) is replaced in that case by
\begin{equation}
\delta \alpha (L) = {1 \over 4} p(L) \sin(2 \alpha) N
\label{delaN}
\end{equation}
where $N$ is the number of passes through the magnet.  With 
$N \sim 10^5$, $B \sim$ few Tesla, and $L \sim 10$ m, the
sensitivity in $g$ is of order $(10^7~{\rm GeV})^{-1}$ when
the resonance condition $q L < 1$ is satisfied.  For laser
light ($\omega \sim$ eV) in vacuum ($\epsilon$ = 1), and
$L \sim$ 10 m, the resonance condition is satisfied when
$m_a \lesssim$ 0.3 meV.  For $m_a >$ meV, the sensitivity
to $g$ decreases as $m_a^{-2}$.  The PVLAS collaboration 
claimed a signal in ref. \cite{Zava06} but retracted it 
after additional measurements were made \cite{Zava08}.  

Refs. \cite{Zare19,Shak20} propose schemes to measure the 
birefringence of light due to virtual axion exchange in a 
magnetic field or in a high intensity laser beam.

\section{Shining light through walls}

Another approach to axion detection is $\gamma \rightarrow a$ conversion
in a magnetic field followed by $a \rightarrow \gamma$ back-conversion,
also in a magnetic field \cite{VanB87}.  This type of experiment is 
commonly referred to as ``shining light through walls".  If $P_0$ is 
the power of the laser, and $p$ and $p^\prime$ the conversion 
probabilities in the magnets on the left and right hand side of 
the wall, the power in regenerated photons is $P = p^\prime~p~P_0$.  
Formulas for the conversion probabilities can be found in Section 3.  
Shining light through walls experiments have been carried out by 
several groups \cite{Ruos92,Robi07,Pugn08,Chou08,Afan08,Ehre10,Ball15}. 
For $m_a < 3 \cdot 10^{-4}$ eV, the following limit on the axion 
coupling to two photons has been obtained \cite{Ball15} 
\begin{equation}
g_{a\gamma\gamma} < 3.5 \cdot 10^{-8}~{\rm GeV}^{-1}~~\ .
\label{laserlim}
\end{equation}
It is less severe than the limit from the CAST solar axion search 
\cite{Andr07}.  This is due largely to the high intensity of the 
solar axion flux compared to the flux produced by photon conversion 
in a laboratory magnetic field, and the fact that solar axions have 
keV energies whereas axions produced by lasers have eV energies.  On 
the other hand, shining light through walls is a purely laboratory 
experiment and the simple version described above can be improved upon.

A first improvement is to introduce an optical cavity in which the 
photons, on the axion production side of the wall, are bounced back and
forth multiple times \cite{Ruos92}.  Each photon in the cavity converts 
to axions with probability $p$ for each pass through the cavity.  If the 
reflectivity of the mirrors is $R = 1 - \eta$ and $P_0$ is the laser power, 
the power of the right-moving wave in the cavity is ${1 \over 2 \eta}~P_0$. 
The axion power through the wall is then increased by the factor 
${1 \over 2 \eta}$.  Presently available mirrors may have $\eta$
as small as $10^{-5}$.

A further improvement \cite{Hoog91,Fuku96,Sik07,Muel09} is to have
tuned Fabry-P\'erot cavities on both sides of the wall.  It is shown
below that the probability of axion to photon conversion in the 
reconversion cavity is ${2 \over \pi}~{\cal F}^\prime~p^\prime$ 
where ${\cal F}^\prime$ is the finesse of the reconversion cavity, 
and $p^\prime$ the reconversion probability in the absence of cavity.  
The finesse of a cavity is ${\cal F}= {\pi \over \eta}$ if the 
reflectivity of its mirrors is $R = 1 - \eta$. Including both 
improvements, the regenerated photon power is
\begin{equation}
P = {1 \over \eta^\prime \eta}~p^\prime~p~P_0~~~\ .
\label{regpow}
\end{equation}
Half of the power $P$ is right-moving and half is left-moving.  To 
detect all the regenerated photons, detectors should be installed 
on both sides of the regeneration cavity. We implicitly assumed in 
Eq.~(\ref{regpow}) that the loss of power from the regeneration 
cavity is enirely due to transmission through its mirrors.  In 
general there are other contributions,
$\eta^\prime =  \eta^\prime_{\rm trans} +
\eta^\prime_{abs} + \eta^\prime_{scatt}$,
where the latter two terms represent absorption and scattering 
(including diffraction) losses.  If transmission is not entirely 
dominant, the RHS of Eq. (\ref{regpow}) is multiplied by the 
factor $f  = \eta^\prime_{\rm trans}/\eta^\prime$.   An experiment 
of this type, named ALPS II, is presently under construction at 
DESY \cite{Bahr13}. A derivation \cite{Muel09} of Eq.~(\ref{regpow}) 
follows.

The modes of a Fabry-Perot cavity are described in $\Phi = 0$ gauge
by standing waves:
\begin{equation}
\vec{A} = A \hat{x} \sin(kz) \cos(\omega t)
\label{FPA}
\end{equation}
with $k = \sqrt{\epsilon} \omega$.  $\epsilon$ is the dielectric 
constant in the cavity.  We set the magnetic permeability $\mu$ = 1
for simplicity.  The cavity has mirrors which enforce $E_x = 0$ at 
$z = 0$ and $z = L$.  Thus the wavevector has quantized values:  
$k = {\pi n \over L}$ with $n = 1,~2,~3~...$.  The energy stored 
in the cavity is
\begin{equation}
E = {1 \over 4} S L A^2 \epsilon \omega^2
\label{ener}
\end{equation}
where $S$ is the transverse area of the standing wave.   
Eq.~(\ref{FPA}) assumes $k S >> L$.

When a cavity mode $\vec{A} = \hat{x} A(t) \sin(kz)$ is not 
driven, its amplitude satisfies
\begin{equation}
({d^2 \over dt^2} + \gamma {d \over dt} + \omega^2) A(t) = 0~~\ .
\label{diss}
\end{equation}
The quality factor is $Q = {\omega \over \gamma}$.  
The power in the right-moving component is
\begin{equation}
P_+ = {1 \over 8} S A^2 \sqrt{\epsilon} \omega^2~~~~\ .
\label{RmP}
\end{equation}
If the mirrors have reflectivity $R = 1 - \eta$ and there are 
no other losses, the power emitted through the two mirrors is 
$2 \eta P_+ = \gamma E$.  Hence  
\begin{equation}
Q = L {\sqrt{\epsilon} \omega \over \eta} = n {\cal F}
\label{qual2}   
\end{equation}
where ${\cal F}$ is the finesse.

In the presence of a large static magnetic field $\vec{B}_0$, the 
inhomogeneous Maxwell's equations (\ref{eom4}) apply.  The axion 
beam traveling through the regeneration cavity is described by
\begin{equation}
a(z,t) = a \sin (k_a z - \omega t)~~~~\ .
\label{axbeam}
\end{equation}
It is assumed to have the same transverse area $S$ as the 
regeneration cavity mode.  In practice, $S$ varies along 
the beam path.  The calculation below assumes that the 
axion wave and the photon wave in the regeneration cavity 
follow the profile of the hypothetically unimpeded photon 
wave in the production cavity.  The power in the axion 
beam is
\begin{equation}
P_a = S <- \partial_t a ~ \partial_z a > =
{1 \over 2} S a^2 \omega k_a~~~\ .
\label{axpower}
\end{equation}  
Let $\vec{B}_0 = B_0 \hat{x}$ and $\vec{A}(z,t) = A(z,t) \hat{x}$.  
The first Eq. (\ref{eom4}) is trivially satisfied whereas the 
second becomes
\begin{equation}
(\epsilon {\partial^2 \over \partial t^2}
- {\partial^2 \over \partial z^2}) A(z,t) =
\omega g B_0 a \cos(k_a z - \omega t)~~~\ .
\label{eom6}
\end{equation}
Since $A(0,t) = A(L,t) = 0$, we may expand
\begin{equation}
A(z,t) = \sum_{n=1}^\infty A_n(t) \sin({n \pi \over L}z)~~~\ .   
\label{expand}
\end{equation}
Substituting back in Eqs.~(\ref{eom6}) one finds
\begin{equation}
({d^2 \over dt^2} + \omega_n^2) A_n(t) =
D \sin(\omega t + {q L \over 2})
\label{eom7}
\end{equation}
where $\omega_n = {n \pi \over \sqrt{\epsilon} L}$ and
\begin{equation}
D = {1 \over \epsilon} g \omega B_0 a
{2 \over Lq} \sin({qL \over 2})~~~\ .
\label{Fn}
\end{equation}
$q = k_n- k_a = {n \pi \over L} - \sqrt{\omega^2 - m_a^2}$ 
is the momentum transfer, as before.  Non-resonant terms are 
neglected.  When dissipation is included, Eq. (\ref{eom7}) 
becomes
\begin{equation}
({d^2 \over dt^2} + \gamma {d \over dt} + \omega_n^2) A_n(t)
= D \sin(\omega t + {qL \over 2})~~~~\ .
\label{eom8}
\end{equation}
Up to transients, the solution is
\begin{equation}
A_n(t) = {D \sin (\omega t - \phi) \over
\sqrt{(\omega_n^2 - \omega^2)^2 + \omega^2 \gamma^2}}
\label{inhA}
\end{equation}
with
\begin{equation}
\phi = - {qL \over 2} +
\tan^{-1}({\omega \gamma \over \omega_n^2 - \omega^2})~~~\ .
\label{phase}   
\end{equation}
When the laser cavity and the regeneration cavity are tuned 
to the same frequency ($\omega_n = \omega$)
\begin{equation}
\vec{A} = \hat{x} A \sin({n \pi \over L} z)
\sin(\omega t + {q L \over 2} - {\pi \over 2})
\label{sol}
\end{equation}
with
\begin{equation}
A = {D \over \omega \gamma} = {g B_0 a \over \epsilon \gamma}
{2 \over Lq} \sin({qL \over 2})~~~\ .
\label{ampl}
\end{equation}
The energy $E$ stored in the mode is given by Eq.~(\ref{ener}). 
Dividing the power $\gamma E$ that the cavity emits by the axion 
power $P_a$, one obtains the axion to photon conversion probability 
\begin{equation}
p_{\rm FP} =  {2 g^2 B_0^2 \over \epsilon \beta_a \omega L}
Q {1 \over q^2} \sin^2({qL \over 2})~~~~\ .
\label{FPprob}
\end{equation}
In terms of the conversion probablity $p$ in the same region without  
cavity [Eq.~(\ref{classicp}) with $\mu=1$] we have
\begin{equation}
p_{\rm FP} = {2 Q \over \sqrt{\epsilon} L \omega} p = 
{2 {\cal F} \over \pi} p ~~~~\ ,
\label{rel}
\end{equation}
as announced.

Refs. \cite{Hoog92,Casp09,Bogo19,Jani19} propose "shining light 
through walls" using microwaves instead of visible light.  Axions
are produced in one electromagnetic cavity permeated by a magnetic 
field and detected in an other.                                                     
                       
\section{Long range forces mediated by axions} 

The exchange of virtual axions produces forces between macroscopic
bodies that may manifest themselves as deviations from the $1/r^2$
gravitational law \cite{Mood84}.  It also produces effective 
interactions in atoms that modify slightly atomic spectroscopy 
\cite{Wein78}.

The general form of the interaction of the axion with a Dirac fermion
$f$ is given in Eq.~(\ref{CPv2}).  It implies the interaction energy 
given in Eq.~(\ref{intfa2}) in case the fermion is non-relativistic.  
It also implies that the fermion is a source for the axion field:
\begin{equation}
(\partial_t^2 - \nabla^2 + m_a^2) a(\vec{x},t) =
+ {g_f \over 2 f_a} \vec{\nabla}\cdot
[\chi^\dagger (\vec{x},t) \vec{\sigma} \chi(\vec{x},t)]
+ {m_f \theta_f \over f_a} \chi^\dagger(\vec{x},t) \chi(\vec{x},t)
\label{axsou}
\end{equation}
where $\chi(\vec{x},t)$ is the non-relativistic fermion's two-component 
spinor field.  Let us assume that a spin 1/2 fermion is localized at 
$\vec{x}_1$ and 
that its spin state is slowly varying on the time scale set by $m_a^{-1}$.  
Eq.~(\ref{axsou}) implies then
\begin{eqnarray}
a(\vec{x}) &=& {1 \over f_a}(m_f \theta_f +
{g_f \over 2} \vec{\sigma} \cdot \vec{\nabla})
{e^{- m_a r} \over 4 \pi r}\nonumber\\
&=& {1 \over f_a}\left(m_f \theta_f
- {g_f \over 2}(\vec{\sigma}\cdot\hat{r})(m_a + {1 \over r})\right)
{e^{- m_a r} \over 4 \pi r}
\label{axresp}
\end{eqnarray}
where $\vec{r} = r \hat{r} = \vec{x} - \vec{x}_1$ and $\vec{\sigma}$
acts on the fermion spin state. Let us assume further that a second 
fermion $f^\prime$ is localized at $\vec{x}_2$.  Eq.~(\ref{intfa2}) 
implies an interaction energy between the two fermions 
\begin{equation}
V(\vec{x}_2 - \vec{x}_1; \vec{\sigma}, \vec{\sigma}^\prime) =
{1 \over f_a}( - m_f^\prime \theta_f^\prime +
{g_f^\prime \over 2} \vec{\sigma}^\prime \cdot \vec{\nabla}_2) 
a(\vec{x}_2 - \vec{x}_1)
\label{derpoten}
\end{equation}
where primed quantities refer to the second fermion.  Substituting
Eq.~(\ref{axresp}) the interaction energy is seen to be the sum of
three terms
\begin{equation}
V(\vec{x}_2 - \vec{x}_1; \vec{\sigma}, \vec{\sigma}^\prime)     
~=~ V_{\rm mon-mon}
~+~ V_{\rm mon-dip}~+~V_{\rm dip-dip}~~~\ ,
\label{mondip}
\end{equation}
called `monopole-monopole', `monopole-dipole', and
`dipole-dipole' interactions:
\begin{eqnarray}
V_{\rm mon-mon} &=&
- {m_f \theta_f m_f^\prime \theta_f^\prime \over f_a^2}
~{e^{ - m_a r} \over 4 \pi r}\nonumber\\
V_{\rm mon-dip} &=&
- {1 \over f_a^2}
[m_f \theta_f g_f^\prime ~ \vec{r}\cdot \vec{\sigma}^\prime
- m_f^\prime \theta_f^\prime g_f ~ \vec{r} \cdot \vec{\sigma}]
~{e^{ - m_a r} \over 8 \pi r^2} (m_a + {1 \over r})
\nonumber\\
V_{\rm dip-dip} &=&
- {g_f g_f^\prime \over 4 f_a^2}~\sigma_j \sigma^\prime_k~
\bigg(\delta_{jk} [{e^{- m_a r} \over 4 \pi r^2} (m_a + {1 \over r})
+ {1 \over 3}\delta^3(\vec{r})]
\nonumber\\
&-& r_j r_k {e^{- m_a r} \over 4 \pi r^3}
(m_a^2 + 3{m_a \over r} + 3{1 \over r^2})\bigg)
\label{effpotx}
\end{eqnarray}
where $\vec{r} = \vec{x}_2 - \vec{x}_1$.  

Relative to the gravitational potential $- {G_N m_f m_f^\prime \over r}$,
the potential due to axion exchange is enhanced by the factor
$\left({M_{\rm Planck} \over f_a}\right)^2$ but has a finite range 
$m_a^{-1}$.  When the coupling $f_a^{-1} $ is large, the range $m_a^{-1}$ 
is small, and vice-versa.  The monopole-monopole force is at any rate 
small compared to the gravitational force because it is suppressed 
by two powers of $\theta_f$.  Recall that the $\theta_f$ are expected
to be of order $\theta_{\rm QCD}$ and $\theta_{\rm QCD}$ is less than 
$10^{-10}$ in view of the upper limit on the neutron electric dipole
moment.  If there is Peccei-Quinn symmetry and no other physics 
beyond the Standard Model, $\theta_{\rm QCD} \sim 10^{-17}$ due 
to CP violation in the weak interactions.  The dipole-dipole force 
is not suppressed by any powers of $\theta_f$ and therefore relatively 
large.  It has a background from ordinary magnetic forces but this 
background can be suppressed to some extent by using superconducting 
shields.  The monopole-dipole force may be the most attractive target 
as it has only one factor of $\theta_f$ and does not have such a large 
background from ordinary magnetic forces.  

The monopole-dipole force has been searched for using a variety 
of approaches including torsion balance techniques \cite{Ritt93,
Hamm07,Hoed11,Terr15}, the effect of a large unpolarized mass 
on a nearby co-magnetometer \cite{Youd96,Lee18a}, its effect 
on a paramagnetic salt sample \cite{Ni99,Cres17}, the spin 
relaxation of cold neutrons \cite{Igna09,Sere10} and $^3$He
nuclei \cite{Guig15} due to their collisions with trap walls, 
the shift in the nuclear spin precession frequency due to the 
presence of a nearby unpolarized mass \cite{Tull13}, and the 
shifts in atomic energy levels due to P and T violating 
interactions \cite{Stad18}.  These searches place lower 
bounds on $f_a/\sqrt{\theta_f}$ that reach of order $10^{15}$
GeV for ALPs that are massless or sufficiently light 
\cite{Youd96,Tull13,Cres17,Lee18a}.  The bounds do not 
reach QCD axions because the range of QCD axion mediated
forces is too short in all cases.

The dipole-dipole force has been searched for with a variety 
of approaches as well, including the shifts in nuclear Zeeman 
frequency \cite{Glen08} and nuclear precession frequency 
\cite{Vasi09} due to proximity with a polarized source,
atomic spectroscopy \cite{Ledb13,Fice17}, torsion balance
techniques \cite{Terr15}, scanning tunneling microscopy 
\cite{Luo17}, and the effect of a polarized source on a
nearby co-magnetometer \cite{Lee18b}.  The resulting lower 
bounds on $f_a$ reach of order $10^5$ GeV for ALPs that are 
massless or sufficiently light \cite{Terr15,Lee18b}.  They 
do not reach QCD axions for the same reason as above.

The ARIADNE experiment \cite{Arva14} proposes to produce an 
oscillating axion field by rotating an unpolarized non-axially 
symmetric macroscopic body at a frequency $\omega$.  The axion 
field produces an effective transverse magnetic field at the 
location of a nearby sample that is nuclear spin polarized
by a laboratory magnetic field adjusted so that its Larmor 
frequency equals $\omega$.  The oscillating axion field then 
resonantly excites transverse magnetization in the polarized body.  
The transverse magnetization is detected by a SQUID magnetometer.

\section{Photon flux from relic axion decay}

Through its electromagnetic coupling, the axion decays to two 
photons at the rate 
\begin{equation}
\Gamma_{a \rightarrow 2\gamma} = g_\gamma^2~{\alpha^2\over
64\pi^3}~{m_a^3\over f_a^2}
\simeq~{1.5 \cdot 10^{-50} \over {\rm sec}}~
\left({m_a \over 10^{-5}~{\rm eV}}\right)^5
\left({g_\gamma\over 0.36}\right)^2~~\ .  
\label{lif}
\end{equation}
One may search for the photon flux from the decay of relic axions
\cite{Bers91,Ress91,Blou01}. The signal is largest when the decay rate
is of order the Hubble rate today.  If the decay rate is much larger, 
the axions have decayed already.  If much less, too few axions are 
decaying at present.  Eq.~(\ref{lif}) indicates that the largest 
signal is for an axion mass of order 10 eV.  Such large axion masses 
are inconsistent with the constraints from laboratory searches and 
stellar evolution.  We may however consider a broader class of 
axion-like particles (ALPs). We assume the ALP is a light pseudoscalar 
similar to the axion except that its coupling to two photons 
$g_{A\gamma\gamma}$ and its mass $m_A$ are unrelated.  We use 
the letter $~A~$ to indicate the ALP field.

The two photon coupling $g_{A\gamma\gamma}$, defined by 
\begin{equation}
{\cal L}_{A\gamma\gamma} = - g_{A\gamma\gamma}~A~
\vec{E}\cdot\vec{B}~~~\ ,
\label{ALPc}
\end{equation}
is bounded from above
\begin{equation}
g_{A\gamma\gamma} < 10^{-10}~{\rm GeV}^{-1}
\label{HBlim}
\end{equation}
by requiring that ALP emission does not excessively shorten 
the lifetimes of horizontal branch stars in globular clusters 
\cite{Raff08}.  In terms of this bound and the present age of 
the universe, $t_0 = 1.38 \cdot 10^{10}$ years, the ALP decay 
rate is 
\begin{equation}
\Gamma_{A \rightarrow 2 \gamma} \simeq {1 \over t_0} 
\left({g_{A\gamma\gamma} \over 10^{-10}~{\rm GeV}^{-1}}\right)^2
\left({m_A \over 310~{\rm eV}}\right)^3
\label{ALPt}
\end{equation}
assuming no other decay modes.  ALPs are produced in the early universe 
through processes analogous to those that produce axions.  Both thermal 
and cold relic ALP populations may occur.

Let us consider an aggregate of relic ALPs of total mass $M$, e.g. 
all ALPs that are part of a galaxy cluster.  The aggregate emits 
photons with frequency $\omega = {1 \over 2} m_A$.  Its luminosity 
is $L = M \Gamma_{A \rightarrow 2\gamma}$.  In a flat static universe,
the energy flux observed at a large distance $r$ from the aggregate is
$I = {L \over 4 \pi r^2}$.  Let $S$ be the cross-sectional area of
the aggregate as seen from the direction of the observer, and therefore
$\Delta \Omega = S/r^2$ the subtended solid angle.  We  may write
\begin{equation}
M = \int_S d^2 x \int dr~\rho_A(\vec{r}) =
\int_{\Delta\Omega} r^2 d\Omega~\Sigma(\hat{n})
\label{Mint}
\end{equation}
where
\begin{equation}
\Sigma(\hat{n}) = \int dr~\rho_A(\vec{r} = r \hat{n})
\label{colum}
\end{equation}
is the column density in the direction $\hat{n}$.  The energy flux per
unit solid angle is therefore
\begin{equation}
{d I \over d \Omega}(\hat{n}) =  \Gamma_{A \rightarrow 2\gamma}
{dM \over 4 \pi r^2 d\Omega} = {1 \over 4 \pi}
\Gamma_{A \rightarrow 2\gamma}~\Sigma(\hat{n})~~\ .
\label{diffl}
\end{equation}  
Eq.~(\ref{diffl}) is valid in a flat static space-time, or on 
sufficiently small scales in a curved space-time.

We next discuss how these equations are modified in an expanding
homogeneous spatially flat universe.  The metric is 
\begin{equation}
ds^2 = - dt^2 + R(t)^2 d\vec{x} \cdot d\vec{x}
\label{flm}
\end{equation}
where $R(t)$ is the cosmological scale factor.  We assume 
that both the observer and aggregate occupy fixed positions
in comoving coordinates.  The light emitted at time $t_E$ 
and observed at time $t_0$ travels a distance
\begin{equation}
r = x R(t_0) = \int_{t_E}^{t_0} dt~{R(t_0) \over R(t)}
\label{dist}  
\end{equation}
where $x$ is the comoving coordinate distance between the
observer and the source.  Consider $N$ photons emitted over
a small time interval $\Delta t_E$ and arriving at the
observer over time interval $\Delta t_0$.  Since
\begin{equation}
x = \int_{t_E + \Delta t_E}^{t_0 + \Delta t_0} dt ~{1 \over R(t)}
\label{newx}
\end{equation}
is unchanged, we have
\begin{equation}
\Delta t_0 = \Delta t_E~{R(t_0) \over R(t_E)}~~\ .
\label{timedil}
\end{equation}
Therefore the observed photons are redshifted to the angular
frequency
\begin{equation}
\omega = {m_A \over 2} {1 \over 1 + z_E}~~~
\label{redsh}
\end{equation}
where
\begin{equation}
1 + z_E = {R(t_0) \over R(t_E)}~~~\ .
\label{red2}
\end{equation}
Eq.~(\ref{timedil}) also implies that the $N$ photons arrive
at the observer at a rate which is ${1 \over 1 + z_E}$ times
the rate at which they are emitted.  The observed energy flux
is therefore
\begin{equation}
I = {\Gamma_{A \rightarrow 2\gamma}~M \over 4 \pi r^2}
{1 \over (1 + z_E)^2}~~~\ .
\label{enfl} 
\end{equation}  
To obtain the energy flux per unit solid angle, let us imagine that
the source is spread over the sphere formed by all points that are
at comoving coordinate distance $x$ from the observer.  At the time
of emission the surface of that sphere is $4 \pi x^2 R(t_E)^2$.  Since
the actual surface of the source is $S$, its solid angle as viewed 
from the observer is
\begin{equation}
\Delta \Omega = 4 \pi {S \over 4 \pi x^2 R(t_E)^2} =
{S \over r^2} (1 + z_E)^2~~~\ .
\label{solan}
\end{equation}
Hence
\begin{equation}
{d I \over d \Omega}(\hat{n}) =
{\Gamma_{A \rightarrow 2\gamma}~dM~r^2 \over
4 \pi r^2 (1 + z_E)^2~dS(1 + z_E)^2} =
{\Gamma_{A \rightarrow 2\gamma} \over 4 \pi}~
{\Sigma(\hat{n}) \over (1 + z_E)^4}~~~\ .
\label{difl}
\end{equation}  
The wavelength of the observed light is
\begin{equation}
\lambda = {4 \pi \over m_A} (1 + z_E)
= 2.4797~\mu{\rm m}~ 
\left({{\rm eV} \over m_A}\right)(1 + z_E)~~~\ .
\label{Angst}
\end{equation} 
In general the light is not monochromatic because the ALPs are not at
rest in their aggregate.  If their velocity dispersion along the 
line of sight is $\delta v(\vec{n})$, we have
\begin{equation}
{d I \over d\Omega d\lambda}(\hat{n}) =
{\Gamma_{A \rightarrow 2\gamma}~m_A~\Sigma(\hat{n}) \over
(4 \pi)^2~\delta v(\hat{n})~(1 + z_E)^5}~~~\ .
\label{specden}
\end{equation}  
since the linewidth is $\delta \lambda = \delta v~\lambda$.

Let us assume ALPs are the dark matter and consider the ALP aggregate 
associated with a galaxy cluster of mass $M = 10^{15}~M_\odot$ and 
radius 1.3 Mpc.  Its column density is of order $\Sigma \sim 0.04$ 
gr/cm$^2$.  Its line of sight velocity dispersion is of order 
$\delta v \sim 3 \cdot 10^{-3}$.  If, for example, $m_A = 5$ eV 
and the two photon coupling saturates the bound from horizontal 
branch stars, we have
$\Gamma_{A \rightarrow 2\gamma}^{-1} = 3.3 \cdot 10^{15}$ year.  
In this case, the signal strength is
\begin{equation}
{dI \over d\Omega d\lambda}
\sim 4 \cdot 10^{-17}~{{\rm erg} \over
{\rm cm}^2 \cdot{\rm sec} \cdot {\rm \AA}~ \cdot({\rm arcsec})^2   
\cdot (1 + z_E)^5}~~~\ .
\label{phsig}
\end{equation}
This is of the same order of magnitude as the background due to 
air glow of the night sky \cite{Bers91}.  The signal to noise can 
be improved by looking at several clusters since the line from ALP 
decay will appear at different wavelengths for the different clusters, 
at ratios determined by the known cluster redshifts, whereas the 
background due to air glow is approximately constant and can be 
approximately subtracted out.  Results from searches for photons 
from relic ALP decay are reported in refs.~\cite{Bers91,Ress91,Blou01}.

\section{Optical activity of a background axion field}

The plane of polarization of light traveling through a
space-time varying axion field rotates according to the rule
\begin{equation}
\Delta \Phi = {1 \over 2} g \Delta a
\label{optic}
\end{equation}
where $\Delta a$ is the variation of the axion field along
the path traveled by the light, and $\Delta \Phi$ is the angle
by which the plane of polarization rotates in the clockwise
direction when looking in the direction of propagation.
Eq. (\ref{optic}) assumes that the wavelength of light is
short compared to the distance scale over which the axion
field varies.  A material is said to be ``optically active"
if it causes the plane of polarization to rotate as light  
travels through it.  Opical activity occurs when right and 
left circularly polarized light satisfy slightly different
dispersion laws.  Faraday rotation is a well-kown example.  
Unlike Faraday rotation, the optical activity of a background 
axion field is achromatic.  Eq.~(\ref{optic}) is derived below.

The optical activity of a background axion field was first noted   
in studies of the propagation of light through an axion domain
wall \cite{Sik84,Huan85} and in the neighborhood of an axion string
\cite{Nacu88,Harv89}.  In Ref. \cite{Carr90} it was found that
the vacuum is optically active in electrodynamics modified by 
the addition of a Chern-Simons term
\begin{equation}
{\cal L}_{\rm CS} =
- {1 \over 2} p_\alpha A_\beta(x) \tilde{F}^{\alpha\beta}(x)
\label{CS}
\end{equation}
to the action density.  $p_\alpha$ was introduced as an external 
Lorentz symmetry breaking parameter.  Upon integration by parts, 
the axion-photon-photon interaction [Eq.~({\ref{agamgam})]
is the Chern-Simons term with $p_\alpha = g \partial_\alpha a$.  
So the effect found in Ref. \cite{Carr90} is the optical activity 
of a background axion field, arrived at from a somewhat different 
point of view.   

To derive the effect, let us consider Eqs.~(\ref{Maxwa}) in the 
limit where the axion field is slowly varying on the distance 
and time scale set by the wavelength of light.  We take 
$\partial_t a$ and $\vec{\nabla} a$ to be constants, set
$\rho_{\rm el} = \vec{j}_{\rm el} = 0$, and assume $\epsilon$ and
$\mu$ to be constants as well.  We may then look for solutions in
which $\vec{E}$ and $\vec{B}$ are proportional to
$e^{i\vec{k}\cdot\vec{x} - i \omega t}$.  Eqs.~(\ref{Maxwa}) are
satisfied provided the complex amplitudes of the magnetic and electric 
fields satisfy 
\begin{eqnarray}
\vec{k}\times\vec{E} - \omega \vec{B} &=& 0 ~~~~~~~~~~{\rm and}~~~
\nonumber\\
{1 \over \mu} \vec{k}\times\vec{B} + \epsilon \omega \vec{E} &=&
- i g (\vec{E}\times\vec{\nabla} a - \vec{B} \partial_t a)~~\ .
\label{first2}
\end{eqnarray}  
These two equations may be combined to yield
\begin{equation}
(\epsilon \omega^2  - {1 \over \mu} k^2) \vec{B} =
i g (\partial_t a 
+ {\omega \over k^2} \vec{k}\cdot\vec{\nabla}a) 
~\vec{k}\times\vec{B}
+ {\cal O}(g^2)~~\ .
\label{first2p}
\end{equation}
Setting $\vec{k} = k \hat{z}$ and $\vec{B} = B_1 \hat{x} + B_2 \hat{y}$,
we have
\begin{equation}
\left(\matrix{\epsilon \omega^2 - {1 \over \mu}k^2 & i \eta k \cr
- i \eta k & \epsilon \omega^2 - {1 \over \mu}k^2}\right)
\left(\matrix{B_1 \cr B_2}\right) = 0
\label{first2q}
\end{equation}
where $\eta \equiv g (\partial_t a + {\omega \over k^2} 
\vec{k}\cdot\vec{\nabla} a)$.  The
eigenmodes are therefore the right and left circular polarization  
amplitudes $B_{\pm} = B_1 \mp i B_2$ and
\begin{equation}
\omega_{\pm} = {k \over \sqrt{\epsilon \mu}} \pm
{1 \over 2} \sqrt{\mu \over \epsilon} \eta + {\cal O}(g^2)
\label{eigenf}
\end{equation}
are the corresponding eigenfrequencies.
This implies that the plane of polarization rotates at a rate (angle
per unit time) equal to ${1 \over 2} \sqrt{\mu \over \epsilon} \eta$
clockwise when looking in the direction of propagation.  Since 
$\partial_t a + {\omega \over k^2} \vec{k}\cdot\vec{\nabla} a$ is 
the rate of change of the axion field following the motion of 
the photon, the plane of polarization rotates by an angle
\begin{equation}
\Delta \Phi = {1 \over 2} g  \sqrt{\mu \over \epsilon} \Delta a
\label{genopt}
\end{equation}
when the axion field changes by $\Delta a$.  Eq.~(\ref{optic}) is
for the particular case $\epsilon = \mu = 1$.  

The direction of polarization of light from distant galaxies and   
quasars is observed to be correlated with the direction of their
elongation on the sky \cite{Have75,Clar80}.   That correlation
disappears if there is excessive optical activity in the intervening
space.  The resulting upper limit on a constant $g \partial_0 a$ or 
$g \vec{\nabla} a$ is of order $6 \cdot 10^{-26}$ GeV \cite{Carr90}.  
The axion is massive and hence $a = 0$ on large scales.  Outside 
of domain walls the axion field has no optical activity on average. 
However a useful constraint can be placed on a massless Nambu-Goldstone 
boson $\phi$ associated with an exact global symmetry that is 
spontaneously broken by a vacuum expectation value $v$ \cite{Hara92}. 
Such a particle couples to two photons as in 
Eq.~(\ref{ALPc}) with $g_{\phi\gamma\gamma} = 
c {\alpha \over \pi v}$ where $c$ is the electromagnetic 
anomaly of the symmetry of which $\phi$ is the Nambu-Goldstone 
boson.  Provided the $\phi$ field was not homogenized during an 
inflationary epoch, its values are uncorrelated from one horizon 
to the next, implying that $\Delta \phi \sim v \pi/2$ and therefore
$\Delta \Phi \sim {c \over 2} {\alpha \over 4}$ on cosmological 
distances.  Demanding that $\Delta \Phi < 10^\circ$ to avoid 
destroying the observed correlation between the polarization and 
elongation of distant sources implies that $c \lesssim 100$.

Ref. \cite{DeRo18} proposes to search for low mass
($m_A \sim 10^{-12}$ eV) ALP dark matter by detecting 
the optical activity that the slowly oscillating ALP field 
produces in large baseline optical interferometers.  Ref. 
\cite{Obat18} discusses the use of a ring cavity for this 
purpose.

\vskip 1.0cm

{\it Acknowledgments}

I am grateful to N. Sullivan, D.B. Tanner, K. van Bibber, L. Rosenberg, 
K. Zioutas, G. Raffelt and G. Mueller for stimulating discussions and 
insights over a period of many years.  This work was supported in part 
by the U.S. Department of Energy under grant DE-SC0010296 and by the 
Heising-Simons Foundation under grant No. 2015-109.


\newpage

\appendix

\section{Units and conventions}

As in most particle physics treatises, we adopt units
in which $\hbar = c = 1$.  All dimensionful quantities  
are expressed in terms of a single unit, taken to be
the electronvolt, written as eV.  The conversion factors  
to the standard macroscopic units of energy, mass, length  
and time are
\begin{eqnarray}
{\rm eV} &=& 1.602 \cdot 10^{-12}~{\rm erg}
\nonumber\\
{\rm eV}/c^2 &=& 1.783 \cdot 10^{-33}~{\rm gr}
\nonumber\\
\hbar c/{\rm eV} &=& 1.973 \cdot 10^{-5}~{\rm cm}
\nonumber\\
\hbar /{\rm eV} &=& 6.582 \cdot 10^{-16}~{\rm sec}~~\ .
\end{eqnarray}
For describing fields we use the Heaviside-Lorentz
system of units, also of common use in particle physics.
In Heaviside-Lorentz units, the permitivity and permeability
of the vacuum $\epsilon_0 = \mu_0 = 1$.  The fine structure
constant $\alpha$ is related to the charge $e$ of the electron 
by $\alpha = {e^2 \over 4 \pi \hbar c}$.  All electric charges
and currents have values $\sqrt{4 \pi}$ times their values
in Gaussian units, i.e. the Heaviside-Lorentz unit of electric
charge is $1/\sqrt{4 \pi}$ times the Gaussian unit.  Electric 
fields and magnetic fields have values $1/\sqrt{4 \pi}$ times 
their values in Gaussian units, i.e. the Heaviside-Lorentz 
units of electric and magnetic fields are $\sqrt{4 \pi}$ 
times their Gaussian units.  The electric and magnetic 
field units are the same when $c=1$.


\newpage


\bibliography{review2}


\newpage

\begin{figure}
\includegraphics[angle=360, width=160mm]{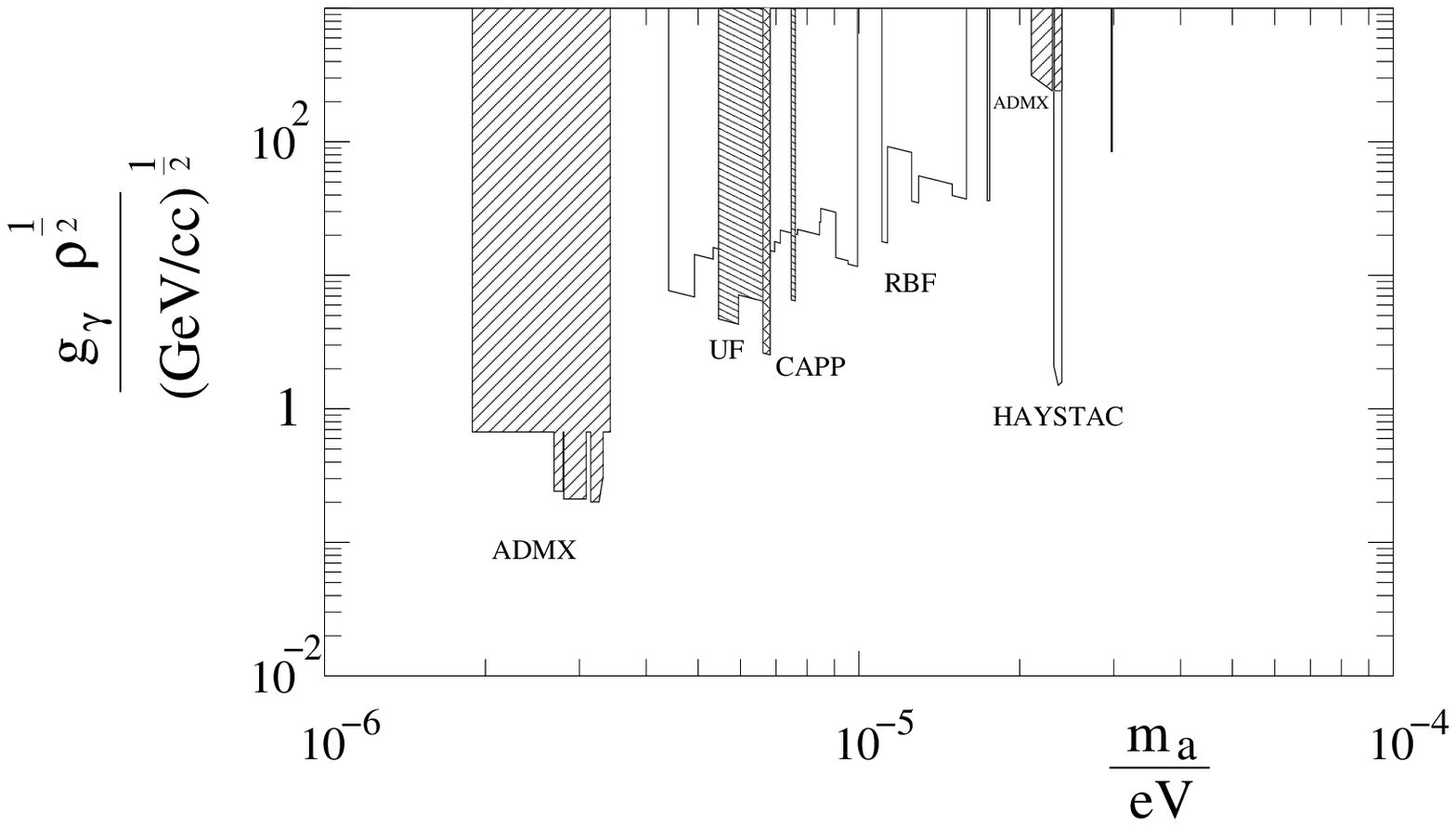}
\vspace{0.2in}
\caption{Limits on $g_\gamma \sqrt{\rho_a}$ where $g_\gamma$
is the dimensionless axion electromagnetic coupling, defined 
in Eq.~(\ref{agamgam}), and $\rho_a$ is the local axion dark 
matter density, as a function of axion mass $m_a$, obtained 
by the RBF \cite{Wuen89}, UF \cite{Hagmth}, ADMX \cite{Du18,
Bout18,Brai19}, HAYSTYAC \cite{Zhon18} and CAPP-8TB \cite{Lee20}
cavity searches.  Additional limits have been obtained by the 
ORGAN \cite{McAl17a} and QUAX$_{a \gamma}$ \cite{Ales19a} 
experiments.  The limits shown are in rough outline only.}
\label{haloaxlim}
\end{figure}

\newpage

\begin{figure}
\hspace{0.2in}
\includegraphics[angle=360, width=160mm]{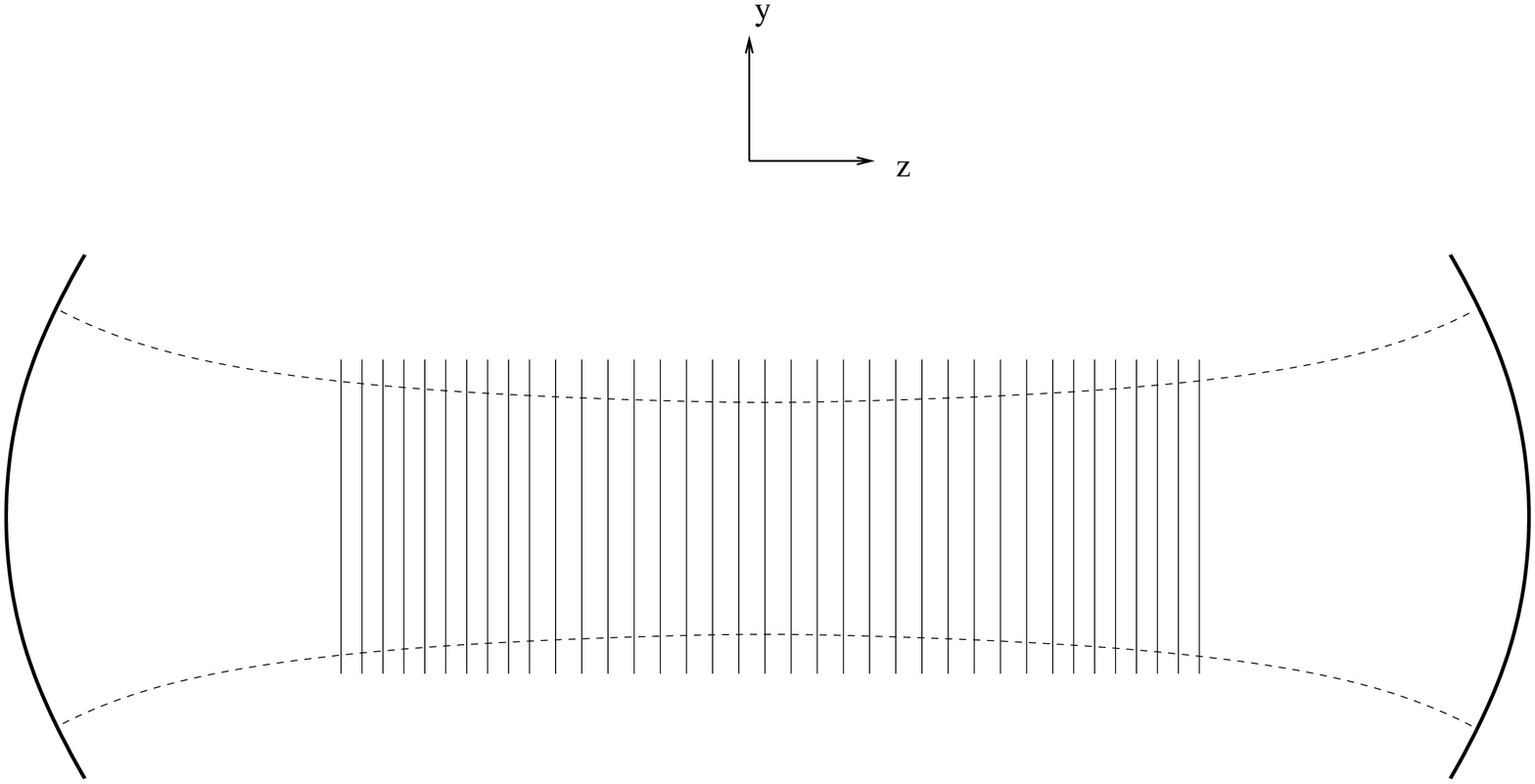}
\vspace{0.5in}
\caption{Wire array detector of dark matter axions 
discussed in Section 5.1.1.  The $\hat{y}$ and $\hat{z}$ 
directions are defined in the text.  The dashed lines 
represent the envelope of an electromagnetic mode in a 
confocal resonator.  The mode is driven by axion to photon 
conversion in the magnetic field produced by currents in 
an array of wires. The wires are represented by vertical 
lines.}
\label{warray}
\end{figure}

\clearpage

\begin{figure}
\hspace{0.2in}
\includegraphics[angle=360, width=160mm]{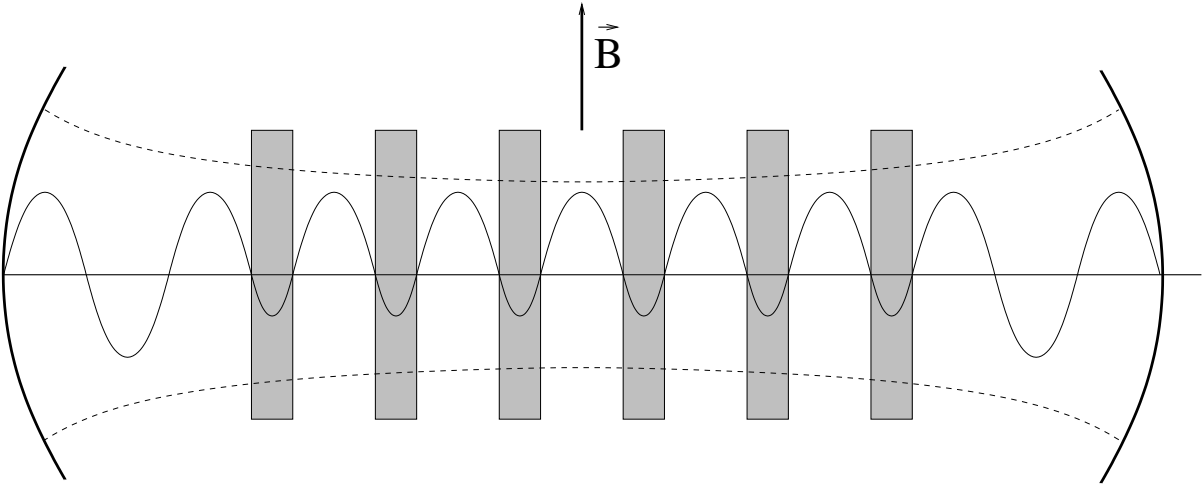}
\vspace{0.5in}
\caption{Dielectric plate detector of dark matter axions 
discussed in Section 5.1.2.  The dashed lines represent the 
envelope of an electromagnetic mode in a confocal resonator.  
The mode is driven by axion to photon conversion in a uniform 
static externally applied magnetic field $\vec{B}$.  The 
vertical shaded rectangles represent dielectric plates 
arranged in such a way as to make the overlap integral 
of the applied static magnetic field $\vec{B}$ with the 
oscillating electric field of the mode as large as possible.  
The electric field profile of the mode is shown.} 
\label{dielpl}
\end{figure}

\clearpage

\begin{figure}
\includegraphics[width=0.9\columnwidth]{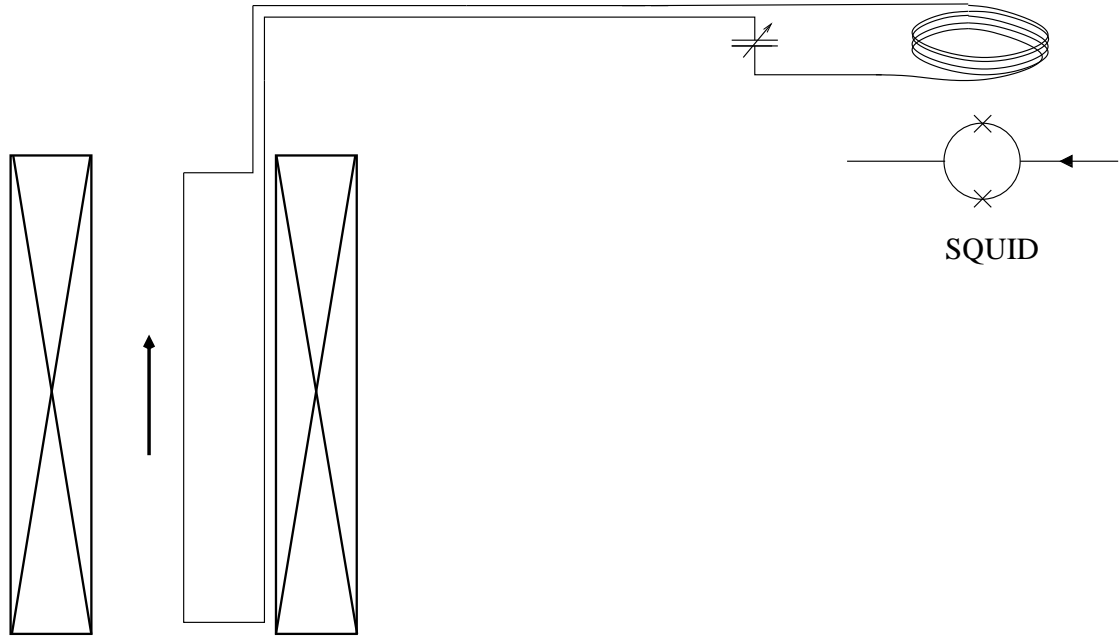}
\vspace{0.5in}
\caption{Schematic drawing of the LC circuit axion 
dark matter detector in case the magnet is a solenoid.  
The two crossed rectangles indicate cross-sections of 
the solenoid's windings.  The arrow shows the direction 
of the magnetic field that the solenoid produces. This 
figure is reproduced from ref. \cite{Sik14a}.}
\label{LCsch}
\end{figure}

\clearpage

\begin{figure}
\hspace{0.2in}
\includegraphics[angle=360, width=160mm]{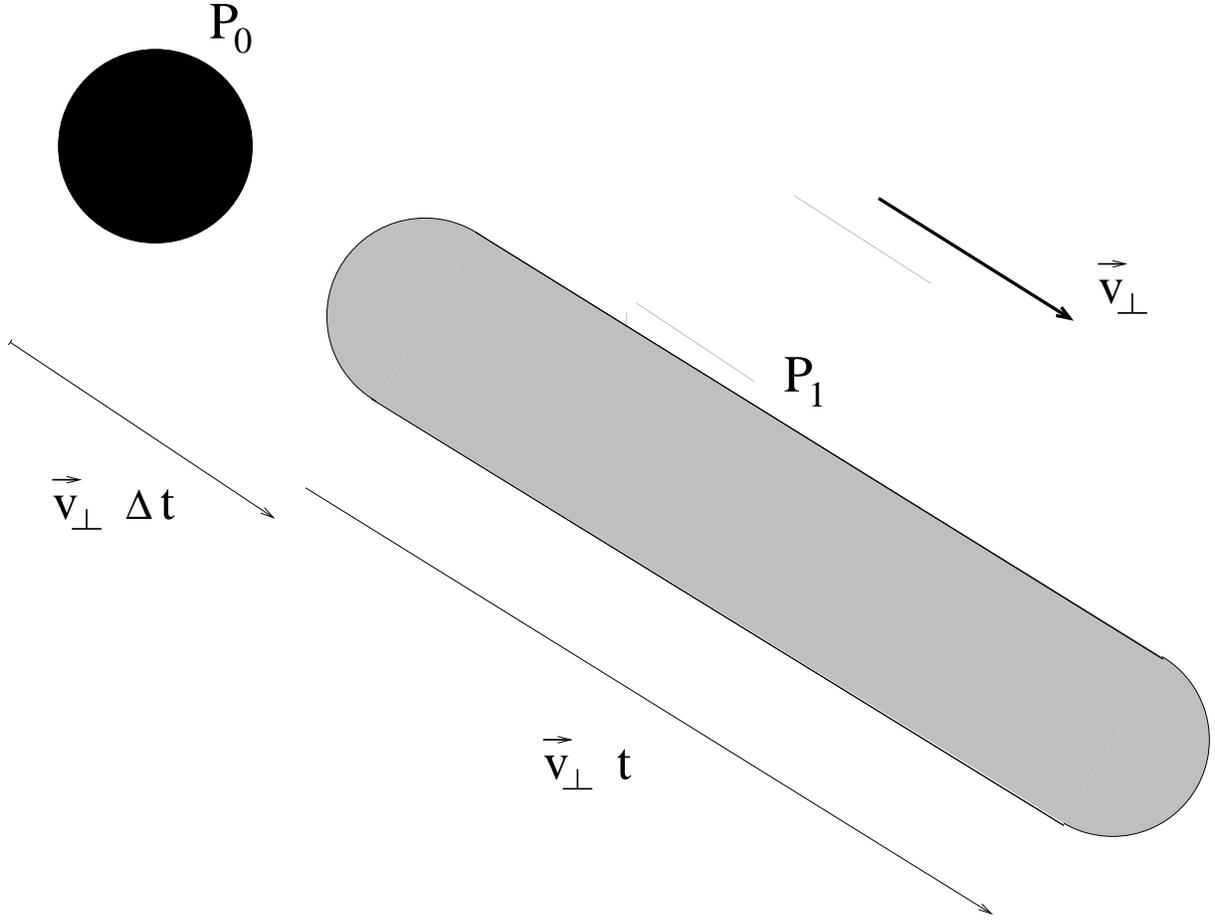}
\vspace{0.5in}
\caption{Illustration of the relative positions of the 
outgoing power and echo power in the scheme described
in Section 5.5.  The figure is drawn in a reference frame 
where the outgoing power source is at rest and where a 
perfectly cold axion fluid moves with velocity $\vec{v}$. 
If the outgoing power $P_0$ is emitted in the direction 
perpendicular to the plane of the figure from the area 
of the black circle for a time $t$ and is then turned off, 
echo power $P_1$ arrives in the grey area at a time 
$\Delta t$ after the outgoing power was turned off.  
The echo power lasts forever but moves away from the 
source of outgoing power with velocity $\vec{v}_\perp$ 
where $\vec{v}_\perp$ is the component of $\vec{v}$ 
transverse to the direction of emission of the 
outgoing power.}
\label{echo}
\end{figure}

\clearpage

\begin{figure}
\includegraphics[angle=360, width=150mm]{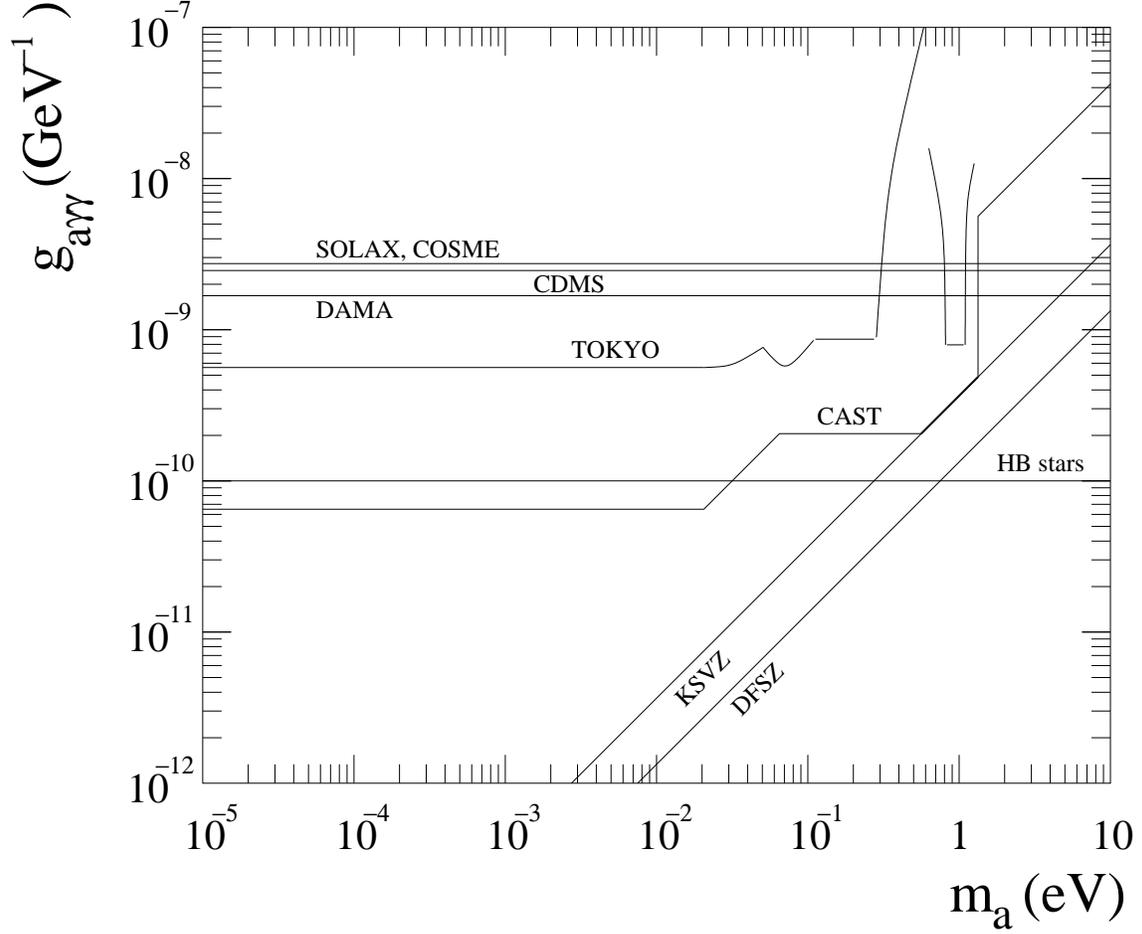}
\vspace{0.2in}
\caption{Limits on the electromagnetic coupling $g_{a\gamma\gamma}$
obtained by the solar axion searches discussed in Section 6, as a
function of axion mass $m_a$.  The relationship between mass and
coupling in the KSVZ and DFSZ axion models, and the limit from
stellar evolution (HB stars), are shown as well.  The Tokyo and 
CAST limits are indicated only in rough outline; for details see 
refs. \cite{Inou08} and \cite{Anas17a}.}
\label{sollim}
\end{figure}

\newpage

\begin{figure}
\includegraphics[angle=360, width=150mm]{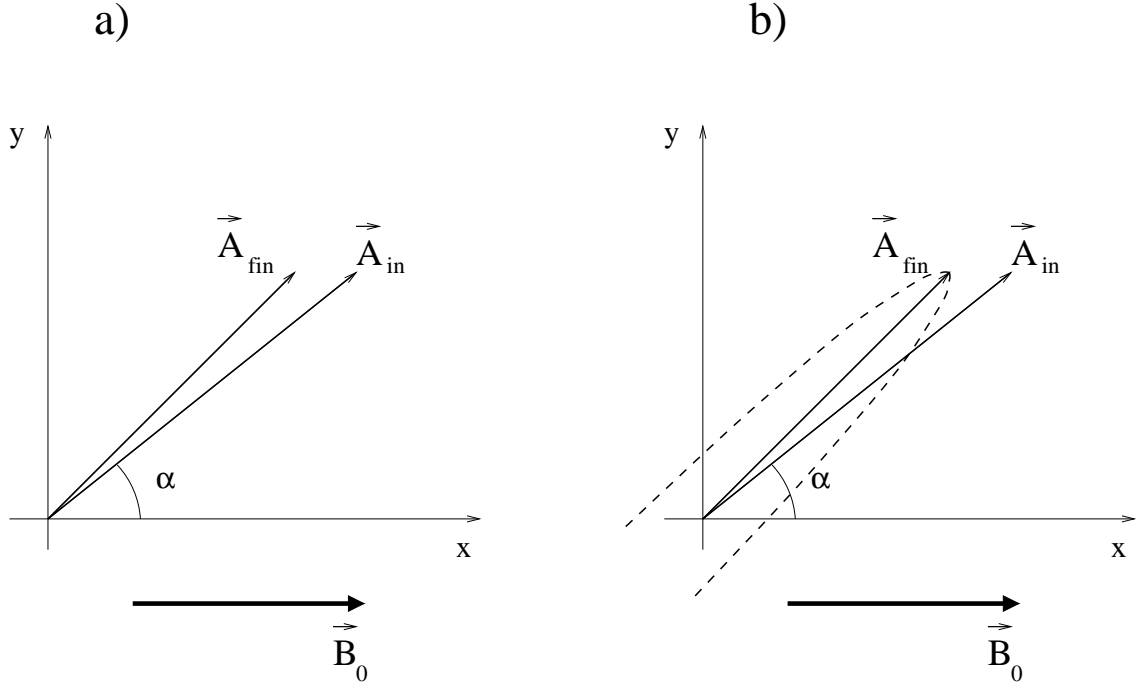}
\vspace{0.3in}
\caption{Axion effects on the propagation of light in a
magnetic field $\vec{B}_0$. Initially the light is linearly
polarized in the direction $\vec{A}_{\rm in}$.   Its
direction of propagation is perpendicular to the plane
of the figure.  (a) The component of light polarized parallel 
to the magnetic field converts partially to axions whereas 
the perpendicular component is unafffected.  This causes 
a rotation of the plane of polarization away from the direction 
of the magnetic field. (b) In addition, a phase difference is 
induced between the parallel and perpendicular components, 
causing light that is initially linearly polarized to acquire 
ellipticity.}
\label{dichrbiref}
\end{figure}


\end{document}